\documentstyle[12pt,epsfig,cite]{article}
\def    \gaga             {$\gamma \gamma$ }
\def    \qq             {$q\bar{q}$ }
\def    \Z              {$Z^0$ }
\def    \J              {{\sc jetset 7.4 }}
\def    \Pt             {$p_T$ } 
\def    \Ptvis         {$\vec{P}^{vis}_\perp$ } 
\def    \Pzvis         {$\vec{P}^{vis}_\parallel$ } 
\def    \gev            {GeV} 

\newcommand\wtF{\widetilde{F}}
\newcommand\smfrac[2]{\mbox{$\frac{#1}{#2}$}}

\newcommand{\bq}{\begin{equation}}
\newcommand{\eq}{\end{equation}}
\newcommand{\ba}{\begin{eqnarray}}
\newcommand{\ea}{\end{eqnarray}}
\newcommand{\ra}{\rightarrow}

\def\u{$\protect\displaystyle\left(\uparrow\right)$}
\def\d{$\protect\displaystyle\left(\downarrow\right)$}

 1
 1
 1

\newcommand{\cbar}{\bar{c}}
\newcommand{\sgaga}{s_{\gamma\gamma}}
\newcommand{\epem}{{\rm e}^+{\rm e}^-}
\newcommand{\NP}{Nucl.\ Phys.\ }
\newcommand{\PL}{Phys.\ Lett.\ }
\newcommand{\PR}{Phys.\ Rev.\ }
\newcommand{\PRL}{Phys.\ Rev.\ Lett.\ }
\newcommand{\ZPH}{Z.\ Phys.\ }
\newcommand{\be}{\begin{equation}}
\newcommand{\ee}{\end{equation}}
\newcommand{\bea}{\begin{eqnarray}}
\newcommand{\ena}{\end{eqnarray}}

\newcommand{\pT}{${p_{T} }\ $}
\newcommand{\alf}{\alpha}
\newcommand{\alfspi}{ {\alpha_s(\mu) \over 2 \pi} }
\newcommand{\dsigd}{  {d \sigma^{D} \over {d {\vec p}_T d \eta}}}
\newcommand{\dsigsf}{{d \sigma^{SF} \over {d {\vec p}_T d \eta}}}
\newcommand{\dsigdf}{{d \sigma^{DF} \over {d {\vec p}_T d \eta}}}
\newcommand{\dsiggg}{ {d \sigma^{\gamma \gamma \rightarrow jet}
            \over {d {\vec p}_T d \eta}}}
\newcommand{\dsigig}{{d \sigma^{i \gamma \rightarrow jet}
            \over {d {\vec p}_T d \eta}}}

\begin{document}

\vspace{5.cm}

\centerline{\Large{\bf $\mathrm{\gamma\gamma}$ Physics}}

\vspace{2.cm}

\centerline{{\it Conveners}: P.~Aurenche, G.A.~Schuler}

\begin{center}
{\it Working group}:  
A.~Bawa, E.~Boudinov,  A.~Buijs, M.~Cacciari, A.~Corsetti, W.~Da
Silva, M.~Dubinin, R.~Engel,  F.~Ern\'e,  J.~Field, A.~Finch,  
J.~Forshaw, R.~Godbole, J.Ph.~Guillet, S.R.~Hou, F.~Kapusta, B.~Kennedy,
M.N.~Kienzle,  M.~Kr\"{a}mer, P.~Kroll,  E.~Laenen, M.~Lehto,  
L.~L\"{o}nnblad, D.~Miller, G.~Pancheri,  J.~Parisi, D.~Perret-Gallix,
J.~Ranft, T.~van~Rhee, S.~Riemersma,  V.~Serbo, M.H.~Seymour, Y.~Shimizu,  
T.~Sj\"ostrand, S.~S\"oldner-Rembold, M.~Stratmann, I.~Tyapkin, 
A.~Vogt, J.~Ward, B.~Wilkens, A.~Wright, N.I.~Zimin
\end{center}
\vspace*{1.0cm}
\setcounter{tocdepth}{1}
\tableofcontents
\clearpage
This chapter is devoted to QCD and, more generally, Strong Interaction
studies in \gaga collisions. For our purposes, LEP2 is a continous
energy \gaga collider with a reach of up to $100$ GeV center of mass
energy for some observables. At low energy, the main studies concern
resonance production and quasi two-body processes which probe the meson
and baryon wave-functions. At high energy, the partonic structure of the
photon  plays a dominant role and, as for hadronic processes, several
tests of  perturbative QCD, using many different observables, are
possible. A specific feature of \gaga collisions is the variability of
the mass of the incoming photons which can be used to tune the
non-perturbative component of the photon.  
\section[ Introduction]{ Introduction}
While LEP1 was dedicated to the study of \Z production and decays the
dominant process at LEP2 will be  $e^+ e^- \rightarrow e^+ e^-  X$ where
the system $X$ is produced in the scattering of two quasi-real photons
by \gaga$\rightarrow X$. It is well known 
that\footnote{taking into account the finite angular acceptance of
any detector}
this cross
section grows like $(\ln s / m^2_{electron})^2$, where $s$ is the
invariant energy squared of the incoming $e^+ e^-$ pair, whereas the
annihilation cross section decreases like $s^{-1}$. Thus LEP2 can be
considered as the highest luminosity as well as the highest energy \gaga
collider presently available. When one of the outgoing electrons is
tagged it is possible to probe the photon "target" at short distance in
deep-inelastic experiments. In fact, concerning the study of the
hadronic structure of the photon  ($i.e.$ its quark and gluon content)
LEP2 is the analogue of both an $e p$  collider and $p p$ collider for
the study of the proton structure.  As in the purely hadronic case the
main processes of interest in this respect will be, besides
deep-inelastic scattering, large $p_T$ phenomena and heavy flavor
production. The high luminosity (500 pb$^{-1}$) and the high energy (in
the following we use $\sqrt {s} = 175$ GeV) available will make it
possible to undertake precision phenomenology and obtain quantitative
tests of perturbative QCD. Furthermore, combining LEP2 data with the
lower energy TRISTAN and LEP1 data and with the high luminosity, high
energy HERA results on photoproduction, a truly quantitative picture of
the hadronic structure of the photon should emerge over a wide
kinematical domain. Let us recall that a precise
knowledge of the photon structure is required if reliable estimates are
to be made of the background to new physics expected at LEP2.

Considering semi-inclusive or exclusive  processes at high energy and
relatively  high momentum transfer it should be possible to probe
diffractive phenomena and shed some light on the nature of the
perturbative Pomeron (the so-called BFKL Pomeron) and the elusive
Odderon (with vacuum quantum numbers but negative C-parity). These
topics have undergone very interesting developments recently in 
connection with HERA results.

Finally, the traditional domain of \gaga physics has been the formation
of resonances and the study of two-body reactions of the type
\gaga$\rightarrow$ meson-meson or  \gaga$\rightarrow$
baryon-$\overline{\mathrm{baryon}}$. In the first case resonances in the
C$=+1$ state, not directly accessible in $e^+ e^-$ annihilation, are
easily produced in a clean environment: heavy resonances like $\eta_c$
and $\chi_c$'s are produced more aboundantly than in previous $e^+ e^-$
colliders. In the second case, the present situation is often unclear
and the LEP2 results, at higher energy, will be helpful to distinguish
between various models and hadron wave-functions. 

One should point out interesting differences between LEP2, as a \gaga
collider, and a hadron-hadron collider.  In particular, in the former
case the initial energy is not fixed: this will turn out to be a major
nuisance in the study of the deep-inelastic structure function of the
photon but it could be an advantage in the study of the  semi-inclusive
channels  (because it could help disentangle perturbative from
non-perturbative effects). Furthermore, using the forward detectors of
the LEP experiments one can vary the ``mass$^2$" of the incoming 
virtual photons.
This will be used to better constrain the non-perturbative component in
the photon, which rapidly decreases with the photon
virtuality,  in the study of deep-inelastic, total cross section or
large $p_T$ processes, for example. More generally, it will help
understand the transition from a non-perturbative to a perturbative
regime in QCD studies.

On the theoretical side, considerable progress has been recently
achieved on the various topics mentioned above. Of particular interest
for data analysis and the study of the event structure of \gaga
collisions is the existence of several general purpose Monte-Carlo codes
({\sc Ariadne}, {\sc Jetset}, {\small HERWIG}, {\sc Phojet}, {\sc Pythia}) 
which are described in the
``\gaga event generator" chapter. These generators are adapted from
hadron-hadron and electron-positron studies and they have been (or are
being) tuned to HERA data thus incorporating all the physics constraints
necessary to reliably describe \gaga reactions. The crucial test of
confronting in detail the models with the LEP1 results on \gaga physics
is still in progress as both data and models are very recent and little
discussion on this point will be given below. In any case, the situation
is much improved compared to only a year ago, when essentially every
experimental group had its own specific event generator, making the
comparison between the various experimental results rather delicate. One
interesting outcome of the recent studies is that the global features of
\gaga scattering are predicted to be rather similar to those of
hadron-hadron scattering at the same energy.
\\
\indent
For the anticipated quantitative studies in perturbative QCD one
obviously needs theoretical predictions at (at least) the
next-to-leading logarithmic order in perturbation theory. All relevant
calculations for \gaga processes have been performed or are being
completed. Depending on the channel under study it will be seen that the
sensitivity of the theoretical predictions under the various unphysical
parameters (scales) is not perfect but, overall, the situation is not
worse than in the purely hadronic channels. 

The plan of the chapter is as follows. We first discuss, in some detail,
the deep-inelastic scattering process on a photon target  ($\gamma^*
\gamma$ process) and its relevance for the determination of the parton
distributions and the $\Lambda_{QCD}$ scale.  We then turn to 
quasi-real \gaga
scattering and discuss the equivalent photon approximation, the
(anti-)tagging conditions which define what we mean by \gaga processes
as well as the background to it. Global features of \gaga events are
described next. Large $p_T$ phenomena and heavy flavor production are
then discussed in the context of next-to-leading QCD phenomenology. The
chapter ends with the discussion of resonance production and exclusive
processes.

Photon-photon physics has been the object of many review articles, see 
$e.g.$ \cite{drgoSG, drgoGS}. 
A look at  \cite{Aachen} shows to what extent the scope
of \gaga physics has extended since the previous LEP2 workshop. 
The topics discussed below are described from a different perspective,
and  with complementary details, in the ``\gaga Event Generators" 
chapter \cite{gggen}.

\section[Structure functions]{Structure functions
\label{ggst1}
{\protect
\footnote{ J.\ Field, B.\ Kennedy, E.\ Laenen, 
M.\ Lehto, L.\ L\"{o}nnblad, D.\ Miller, G.A.\ Schuler, 
M.H.\ Seymour, M.\ Stratmann, I.\ Tyapkin, A.\ Vogt, J.\ Ward, A.\ Wright}}}

%
The measurements of the hadronic structure functions of the photon 
\cite{F2Revs} at LEP1 and lower-energy $e^+ e^-$ colliders 
\cite{F2LEP1,F2prev} can be extended in a number of important ways at 
LEP2. The higher beam energy  will extend the kinematic reach both to 
lower Bjorken-$x$ and to higher scales $Q^2$. Especially, the evolution 
of the real-photon structure function $F_2^{\, \gamma}(x,Q^2)$ can be 
investigated experimentally via single-tag events up to $Q^2 \approx 
500 \mbox{ GeV}^2$; and measurements can be done down to lower values 
of $x$ than ever before, $x \approx 10^{-3}$, entering the small-$x$ 
region where the HERA experiments observe a strong rise of the proton 
structure function $F_2^{\, p}$ \cite{F2HERA}. The increased integrated 
luminosity will be equally important, in principle allowing for a so 
far unachieved statistical accuracy of $F_2^{\, \gamma}$ data of a few 
per cent over a large part of the accessible region. See 
sec.~\ref{avs2} for a more detailed discussion of the kinematical coverage, 
and sec.~\ref{avs4} for a study of the sensitivity of $F_2^{\, 
\gamma}$ at LEP2 to the QCD scale parameter $\Lambda_{\rm QCD}$.

Improved techniques for reducing systematic uncertainties will be needed
to exploit fully these increasing statistics and kinematical coverage, 
and to approach the experimental precision achieved in lepton-hadron 
structure function studies, cf.\ sec.~\ref{avs3} and the report of 
the ``$\gamma\gamma$ event generator'' working group in these 
proceedings. In this context, some of the experiments are improving 
tracking and calorimetry in the forward region to obtain a more complete
coverage for $\gamma \gamma $ events, but no major detector upgrades 
are planned. Masking the detectors against the increased synchrotron 
radiation expected at LEP2 will limit the coverage of structure function
measurements at low $Q^2$ to values above about 3 GeV$^2$. 

There is no prospect of measuring the longitudinal structure function 
$F_L^{\, \gamma}$ at LEP2. However, in sec.~\ref{avs5} a new 
technique is presented which could allow for a measurement of related 
unintegrated structure functions via azimuthal correlations between the 
tagged electron and an outgoing inclusive hadron or jet. Moreover, a 
sufficient number of double-tag events is expected at LEP2 for a study 
of the transition from quasi-real ($P^2 \ll \Lambda_{QCD}^2$) to 
highly-virtual ($P^2 \gg \Lambda_{QCD}^2$) photon structure. Although 
these measurements will remain limited by statistics at LEP2, they can 
considerably improve upon the present experimental information obtained 
by PLUTO \cite {PLUTOv} as discussed in sec.~\ref{avs6}.
\subsection{Kinematical coverage for photon structure measurements}
\label{avs2}
\begin{figure}[htb]
\vspace*{-0.5cm}
\centerline{\epsfig{file=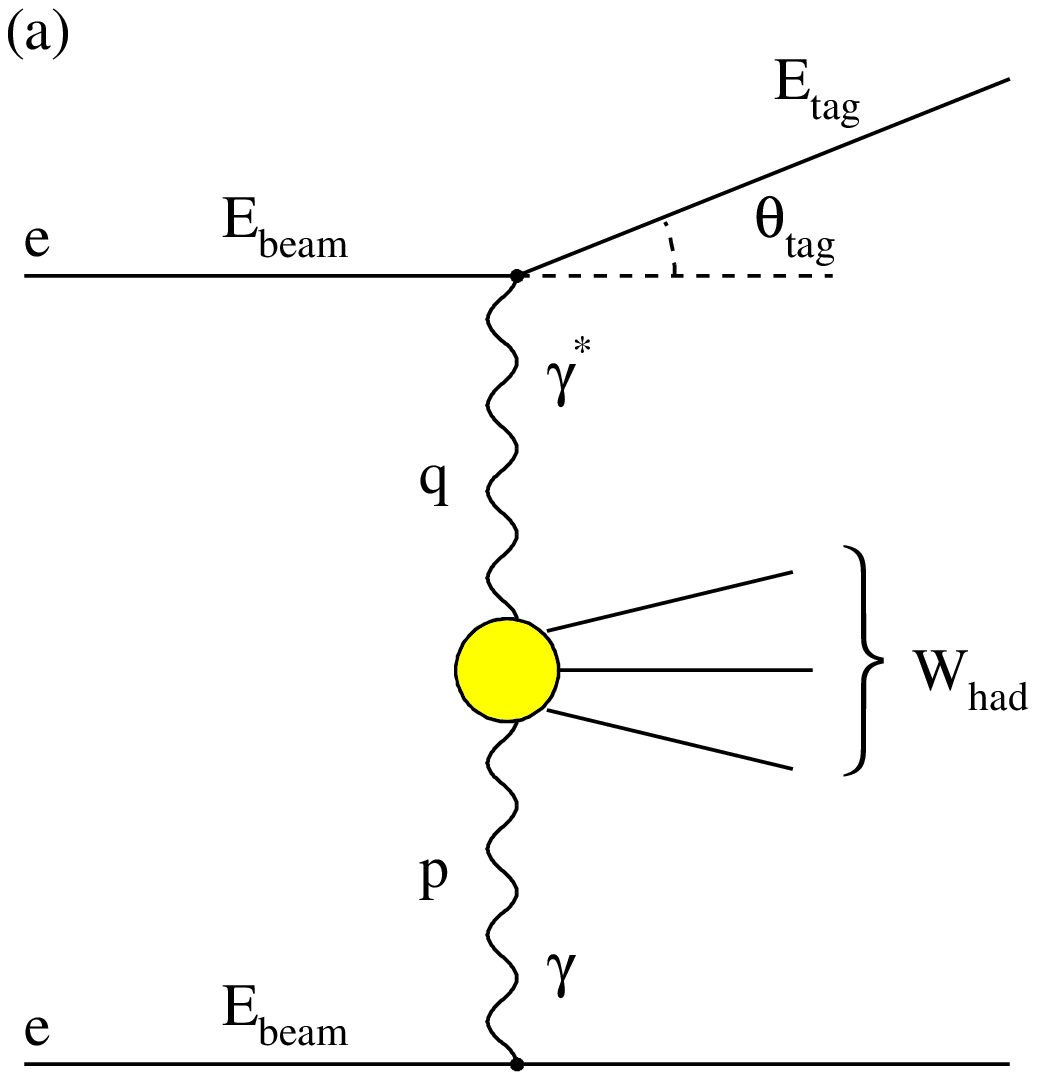,height=9.0cm}\hspace{0.5cm}
\epsfig{file=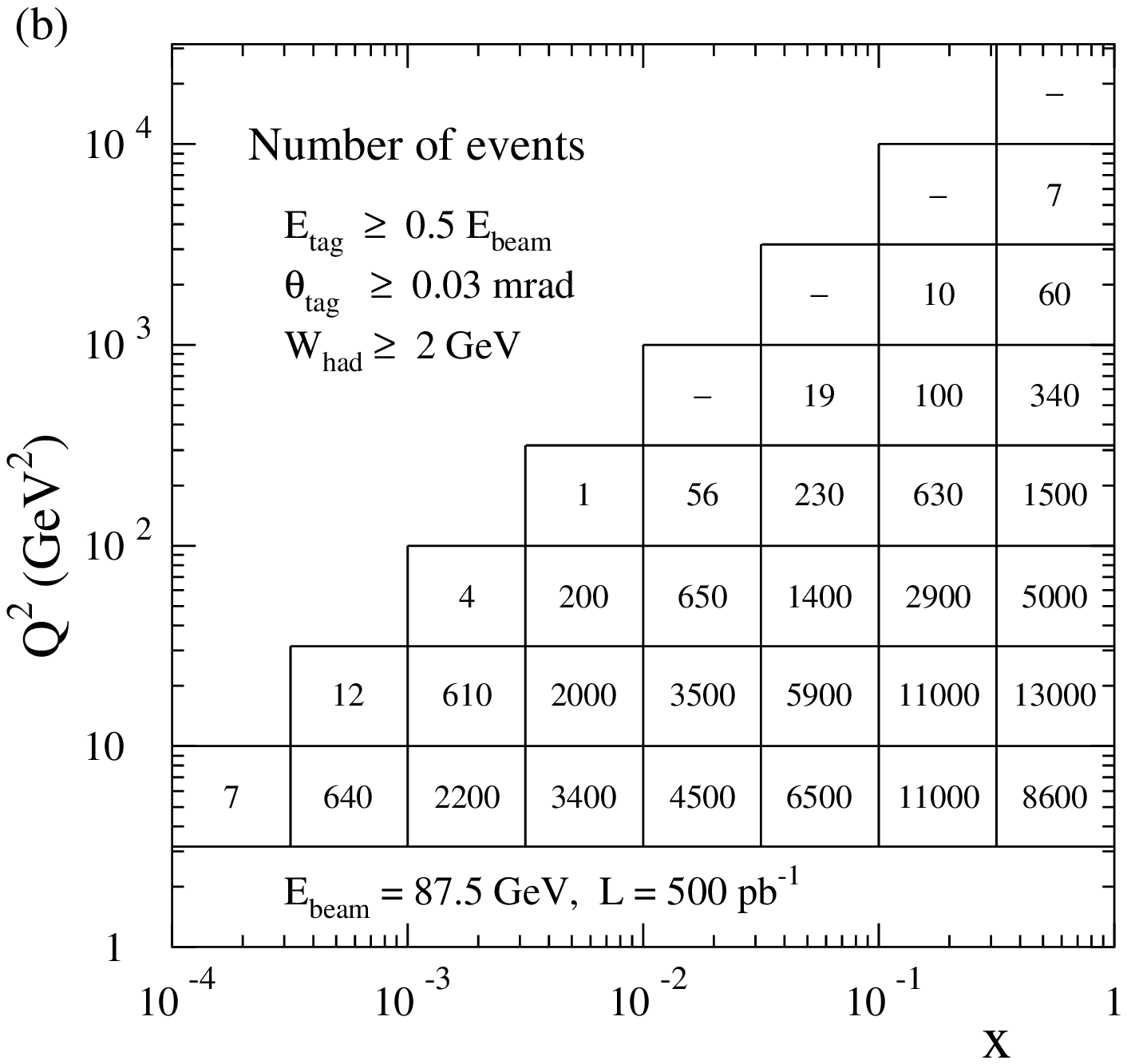,height=9.0cm}}
\caption[]{\em 
 {\bf (a)} The kinematics of a single--tag inclusive $\gamma 
\gamma $ event. {\bf (b)} The expected number of events for the
determination of $F_2^{\, \gamma}$ (including the charm contribution)
at LEP2 (see text for details). The standard antitag
Weizs\"acker--Williams photon spectrum  \cite{avWW} has been used with
$\theta < 30$ mrad. The LO GRV  parametrization of the photon structure
\cite{GRVphot} has been  employed to estimate $F_2^{\, \gamma}$.}
\label{fig:avf1}
\end{figure}

The kinematics of deep--inelastic lepton--photon scattering in $e^+ e^-$
collisions is recalled in Fig.~\ref{fig:avf1}a. Shown is the `single--%
tag' situation, where the electron {\it or} the positron is detected
at $\theta_{\rm tag} > \theta_0$, with a veto against a second tag 
anywhere in the detector. The bulk of the events useful for structure 
function studies are of this type; the generalization to `double--tag' 
events is obvious. At LEP2, the limit of the main detector coverage 
will be about $\theta_0 \simeq 30 $ mrad, slightly higher than at LEP1 
due to the synchrotron radiation shielding already mentioned in the 
introduction. For double--tag events, the very forward calorimeters 
which are used for online high-rate luminosity monitoring will be 
employed (see sec.~\ref{avs6}). 

The cross section for (unpolarized) inclusive lepton--photon scattering 
reads to lowest order in the electromagnetic coupling $\alpha $:
\bq
\label{ave1}
  \frac{d \sigma (e \gamma \rightarrow eX)}{dE_{\rm tag}\, d\cos \theta
  _{\rm tag}} = \frac{4\pi \alpha^2 E_{\rm tag}}{Q^4 y}
  \left[ \{ 1 + (1-y)^2 \} F_2^{\, \gamma}(x,Q^2) - y^2 F_L^{\, \gamma}
  (x,Q^2) \right] \:\: .
\eq
Here $F_{2,L}^{\, \gamma}(x,Q^2)$ denote the structure functions of 
the real photon. The virtuality of the probing photon and the 
invariant mass of the (hadronic) final state are given by
\bq
\label{ave2}
  Q^2 \equiv -q^2 = 2 E_{\rm beam} E_{\rm tag} (1-\cos \theta_{\rm tag})
  \:\: , \hspace{1cm} W^2_{\rm had} = (q+p)^2 \:\: ,
\eq
and we have introduced the usual dimensionless variables
\bq
\label{ave3}
  x = \frac{Q^2}{Q^2+W^2_{\rm had}} \:\: , \hspace{1cm}
  y = 1- \frac{E_{\rm tag}}{E_{\rm beam}} 
      \cos^2 \left( \frac{\theta_{\rm tag}}{2} \right) \:\: .
\eq
Since usually $y^2$ is rather small due to background suppression cuts, 
typically at least $ E_{\rm tag} > 0.5 \, E_{\rm beam}$, only $F_2$ has
been accessible experimentally so far. Under these circumstances $Q^2$ 
is limited to the region $Q^2 > 3 \mbox{ GeV}^2$ for $\theta_{\rm tag}
\geq 30$ mrad at LEP2 energy, which in turn limits the reach towards
small $x$.  

It was demonstrated already at the 1986 LEP2 workshop \cite{Aachen} 
that the longitudinal structure function $F_L$ would be very difficult 
to measure also at LEP2. This statement remains valid. Even in the 
most favoured kinematic region with $y>0.5$, the correction from $F_L$ 
to the main part of the signal due to $F_2$ is only about 14\%. 
Achieving this marginal sensitivity in practice would require a costly 
(some 0.5 MChF per experiment) dedicated detector effort. The point 
is that events with $y>0.5$ must have a low energy for the tagged 
electron. But experience at LEP1 has shown that there are significant 
numbers of off-momentum electrons that give spurious tags. To eliminate 
this background under present detector conditions, the so far published 
analyses for $F_2$ have required something like $E_{\rm tag} > 0.7 \, 
E_{\mathrm{beam}}$ \cite{F2LEP1}. The off-momentum electrons come from 
beam--gas brems\-strahlung in the straight sections, and there is no 
reason to expect that their rate will be significantly reduced at LEP2. 

The cross section (\ref{ave1}) has to be convoluted with the
Weizs\"acker--Williams (WW) spectrum \cite{avWW} for the  photon, of
virtuality $P^2$ (there is a high-$P^2$ tail which has to be corrected
for in  determinations of $F_2^{\, \gamma }$), emitted by  the
antitagged electron. For the explicit form of the WW spectrum see
sec.~\ref{ggepa}. This fact leads to a key systematic
problem in the determination of the photon structure functions: since
$p$ is unknown, $W_{\rm had}$ in eq.~(\ref {ave2}) and hence $x$ in
eq.~(\ref{ave3}) cannot by determined from the outgoing electron alone,
in contrast to the situation in (electromagnetic) lepton--nucleon
deep--inelastic scattering. This brings the hadronic final state into
the game, of which only a part $W_{\rm vis}$ of the invariant mass is
seen in the tracking regions of the detectors. The reconstruction $W_{\rm vis} \rightarrow
W_{\rm had} \equiv W_{\rm true}$ requires a reliable modeling of the
final state fragmentation. More on this issue can be found in 
sec.~\ref{avs3} and in the \gaga event generator report.

Estimated event numbers for the measurement of $F_2^{\, \gamma }(x,Q^2)
$, including the $\gamma^* \gamma \rightarrow c \bar{c} $ and $\gamma^* 
g\rightarrow c \bar{c}$ Bethe-Heitler charm contributions, are given in 
Fig.~\ref{fig:avf1}b in bins in $x$ and $Q^2$. Only simple cuts have 
been applied here: on $E_{\rm tag}$, $\theta_{\rm tag}$ and $W_{\rm 
true}$. The nominal LEP2 integrated luminosity has been 
used\footnote{The reduction in the event number 
due to further experimental final-state 
cuts ($W_{\rm vis}$ instead of $W_{\rm true}$, number of tracks etc.) 
will be approximately compensated by the presence of more than one 
experiment doing the measurements.}. 
Some $ (0.5 \ldots 1) \cdot 10^6 $ events, all in the deep--inelastic
regime  $Q^2 > 3 \mbox{ GeV}^2$, can be expected, a dramatic increase
over the  rates that have been available for $F_2^{\, \gamma }$
determinations  so far \cite{F2LEP1,F2prev}.
 
\begin{figure}[thb]
\begin{center}
\vspace{-0.5cm}
\epsfig{file=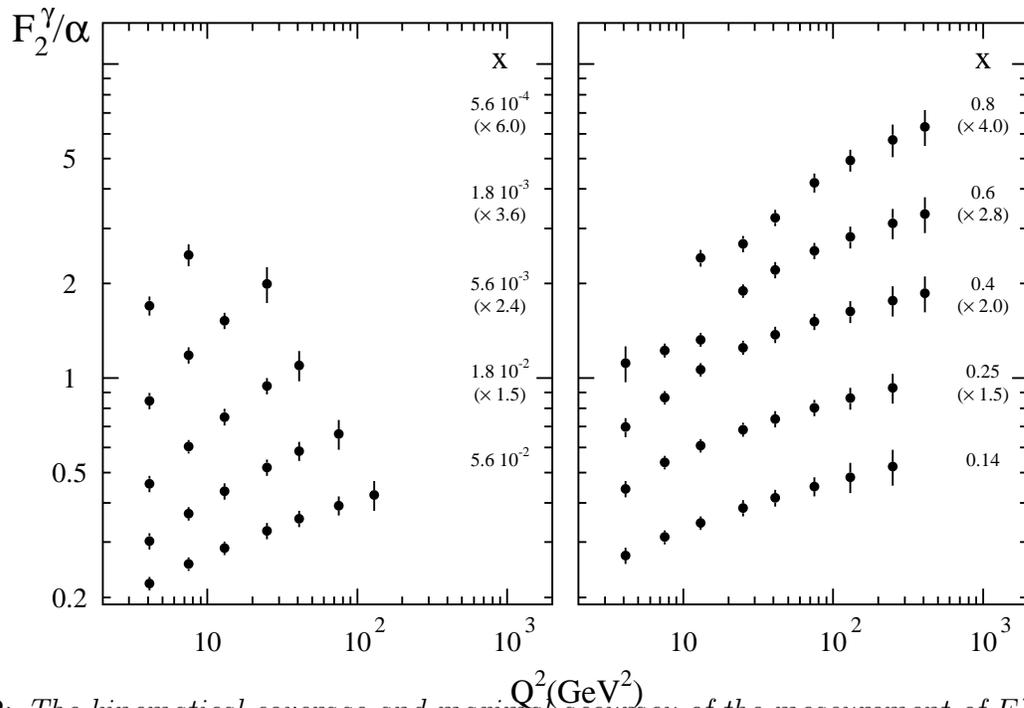,height=9.5cm}
\vspace{-1.25cm}
\end{center}
\caption[]{\em
 The kinematical coverage and maximal accuracy of 
 the measurement of $F_2^{\, \gamma}$ at LEP2, using the event numbers
 of the previous figure for the statistical errors. The assumed 
 systematic uncertainty of at least 5\% has been added quadratically 
 for the error bars shown. The central values have been estimated using 
 the GRV(LO) parametrization \cite{GRVphot}.}
\label{fig:avf2}
\end{figure}

If we put aside the $W_{\rm vis}$ problem for a moment, and assume that 
a systematic error between 5\% and 8\% (depending on the statistical 
accuracy of the bin under consideration) can be achieved, the potential
of LEP2 on $F_2^{\, \gamma}$ is illustrated in Fig.~\ref{fig:avf2}.
Note that in regions where a high-precision measurement is statistically
possible at LEP2, $Q^2 \stackrel{<}{{\scriptstyle \sim}} 100 \mbox{ 
GeV}^2$, such results will dominate our knowledge in the foreseeable 
future. A linear collider with 500 GeV center of mass energy is most 
likely to have no access to this region, since $\theta_{\rm tag} $ 
would be to small to be accessible there.
\subsection{Determining $F_2^{\,\gamma}$ from the experimental 
               information}
\label{avs3}
{\bf Unfolding the photon structure function at small $x$ }
 
\noindent
The rise in the proton structure function $F_2^p$ at small $x$ \cite
{F2HERA} is one of the most important results reported from HERA so far.
It has been suggested that this rise is a signal for so-called BFKL \cite
{LLbfkl} evolution. However, the observed rise has been obtained from 
``conventional" Altarelli-Parisi renormalization group $Q^2$-evolution, 
by starting from a sufficiently low scale $Q_0^2 < 1 \mbox{~GeV}^2$ 
\cite{GRVprot}, see also \cite{Ball}. The small-$x$ coverage of
 $F_2^{\, \gamma}$ at LEP2 cannot compete with that of $F_2^{\, p}$ 
at HERA, hence the LEP2 measurement cannot be expected to shed any new 
light on the origin of this rise. But an important check of 
the universality of the rise is possible, which will also provide 
valuable constraints on the parton densities of the photon at 
small-$x$, where they are dominated by their hadronic 
(vector meson dominance, VMD) component.

So far all measurements of $F_2^{\,\gamma}$ have been at $x$-values
above $0.01$, which is outside the region where a rise would be
expected. At LEP2, we expect to observe a significant number of events
around  and even below $x = 10^{-3}$ (see Fig.~\ref{fig:avf1}b), where a
rise  corresponding to that at HERA certainly should be visible.
Measuring  $F_2^{\,\gamma}$ at such small $x$ values is, however, far
from trivial. While the value of $Q^2$ can be accurately determined to
within a few  per cent, since $\theta_{\rm tag}$ and $E_{\rm tag}$ are
well measured,  $x$ is not directly measurable because the energy of the
target photon  is unknown. The conventional procedure has been to
measure the visible  hadronic mass $W_{\rm vis}$ in each event and
calculate an $x_{\rm vis}$ using eq.~(\ref{ave3}). Then, an unfolding is
made to get the true $x$  distribution. This unfolding requires that the
relationship between  $W_{\rm vis}$ and $W_{\rm true}$ is well described
by the event  generator used.

During this workshop much work has been devoted to get a better
understanding of how this unfolding behaves at small $x$. 
These results are described in more detail in the report from the 
``$\gamma\gamma$ event generator'' working group in these proceedings. 
In that report methods to overcome these problems are also presented,
using additional kinematic variables of the final state and limited
information from the end-caps and the luminosity taggers, where much of
the ``missing" hadronic energy goes. The conclusion is that there is
good  hope that we will be able to measure at sufficiently small $x$ to
detect a rise in $F_2^\gamma$.

\noindent
{\bf QED radiative corrections }
 
\noindent
In analogy to measuring $F_2^p$ at HERA, an accurate measurement of 
$F_2^{\gamma}$ must involve a careful treatment of radiative QED 
corrections to the basic one-photon exchange process described by 
eq.~(\ref{ave1}). Experiments have so far estimated the size of 
radiative corrections by comparing a Monte Carlo event generator for 
$e^+e^- \rightarrow e^+ e^-  \gamma \mu^+\mu^-$ \cite{Berends} (with 
appropriate changes of the muon mass and charge to conform with $q\bar 
q$ production) to one for $e^+e^- \rightarrow e^+e^- q\bar q$ via
$e \gamma \rightarrow e q\bar q$, with the photon energy distribution
given by the Weizs\"{a}cker-Williams spectrum.

A more careful treatment would take into account the hadronic structure 
of the photon and effects of QCD evolution. Such a treatment is given 
in \cite{LS}, in leading logarithmic approximation, which is known to 
be accurate to within a few percent for the proton case. It was found 
that the size of the corrections may vary from as much as 50\% if one 
uses only so-called leptonic kinematic variables, to only a few per cent
using only so-called hadronic variables. As the actual measurement will 
involve a mixture of such variables, a more extended study of the size 
of radiative corrections is needed.
\subsection{The $Q^2$ evolution of $F_2^{\,\gamma}$ and the QCD 
               scale parameter $\Lambda_{QCD}$}
\label{avs4}
At next-to-leading order (NLO) of the QCD improved parton model, the
structure function $F_2^\gamma(x,Q^2)$ is related to the photon's parton
distributions \cite{F2Revs} via 
\begin{equation}
\label{ave4}
  \frac{1}{x} F_{2}^{\,\gamma} =  \sum 2e_{q}^{2} \Big\{ q^{\gamma}
      +  \frac{\alpha_{S}}{2\pi} (C_{q} \ast q^{\gamma} + C_{G}
      \ast G^{\,\gamma})   +  \frac{\alpha}{2\pi} e_{q}^{2} 
      C_{\, \gamma} \Big\} + \frac{1}{x} F_{2,\, h}^{\,\gamma} 
      + {\cal O}\Big( 1/Q^2 \Big) \:\: ,
\end{equation}
where $\ast $ denotes the Mellin convolution. The summation extends 
over the light $u$, $d$ and $s$ quarks. The heavy flavour contribution 
$F_{2,\, h}^{\,\gamma}$ has recently been calculated to second order in 
$\alpha_{S}$ \cite{LSRN} and is discussed in sec.~\ref{gghq1}. 
$C_{q,G}$ are the usual (scheme-dependent) hadronic NLO 
coefficient functions, and for the commonly used $\overline{\mbox{MS}}$ 
factorization there is a `direct' term $C_{\gamma} = 6\, C_{G}$.
Besides the leading--twist contribution written out in (\ref{ave4}), at 
large $x$ (close to and within the resonance region) power-law 
corrections $\propto \mu^2/Q^2 (1-x)$ become important, with $\mu$ 
being some hadronic scale. The $Q^2$-evolution of the quark and gluon 
densities $q^{\gamma}$, $G^{\,\gamma}(x,Q^2)$ is governed by generalized
(inhomogeneous) Altarelli-Parisi evolution equations. For the singlet 
case the solution can be decomposed as
\begin{equation}
\label{ave5}
  \vec{q}^{\,\,\gamma} = \left( \!\! \begin{array}{c} 2 \sum_{q} 
    q^{\gamma} \\ G^{\,\gamma} \end{array} \!\! \right)
    = \vec{q}_{PL}^{\,\,\gamma} + \vec{q}_{had}^{\,\,\gamma} \:\: ,
\end{equation}
where the well-known homogeneous (`hadronic') solution $\vec{q}_{had}^ 
{\, \,\gamma}$ contains the perturbatively uncalculable boundary 
conditions $ \vec{q}^{\,\,\gamma} (Q_{0}^{2})$. The photon-specific 
inhomogeneous (`pointlike', PL) part is given by
\begin{equation}
\label{ave6}
  \vec{q}^{\,\,\gamma}_{PL} = \Big\{ \frac{1}{\alpha_{S}} + \hat{U} 
      \Big\} \ast \Big\{ 1-\big[\alpha_{S}/\alpha_{S}(Q_{0}^{2})\big]
      ^{1+\hat{d}} \Big\} \ast \frac{1}{1+\hat{d}} \ast \vec{a} 
      + \Big\{ 1-\big[\alpha_{S}/\alpha_{S}(Q_{0}^{2})\big]
      ^{\hat{d}} \Big\} \ast \frac{1}{\hat{d}} \ast \vec{b} + 
      {\cal O} (\alpha_{S}) \:\: .
\end{equation}
Here $\vec{a} $, $\vec{b} $, $\hat{d} $ and $\hat{U} $ are combinations 
of the LO and NLO splitting-function matrices. 

At asymptotically large $Q^2$ {\em and} large $x$, eq.~(\ref{ave6})
reduces to the well-known asymptotic solution $\propto 1/\alpha_{S} $,
suggesting a parameter-free extraction of $\Lambda_{\rm QCD}$ from the
photon structure. At energies accessible at present and in the 
foreseeable future, however, the non-asymptotic contributions cannot be
neglected even at large $x$, and $\Lambda_{\rm QCD}$ determinations 
involve a model or a simultaneous free fit of the non-perturbative 
boundary conditions $ q^{\gamma},\, G^{\,\gamma}(Q_0^2)$.
In eq. (\ref{ave6}), $Q^2_0$ is an arbitrary 
reference scale; hence $\vec{q}_{had}^{\,\,\gamma}$ in eq.~(\ref{ave5}) 
will in general contain not only the non-perturbative (coherent) 
hadronic part, but also contributions originating in the pointlike 
photon--quark coupling. However, final state information suggests that 
there is some low scale $Q^2_S$ close to the border of the perturbative 
regime, where (in NLO in some suitable factorization scheme, see 
\cite{GRVth,AFG,avLund}) the parton structure of the photon is purely 
hadronic and given by the fluctuations to virtual vector mesons (VMD) 
\cite{F2prev,FKP,FSexp}.

\begin{figure}[thb]
 \begin{center}
 \vspace{-0.5cm}
 \epsfig{file=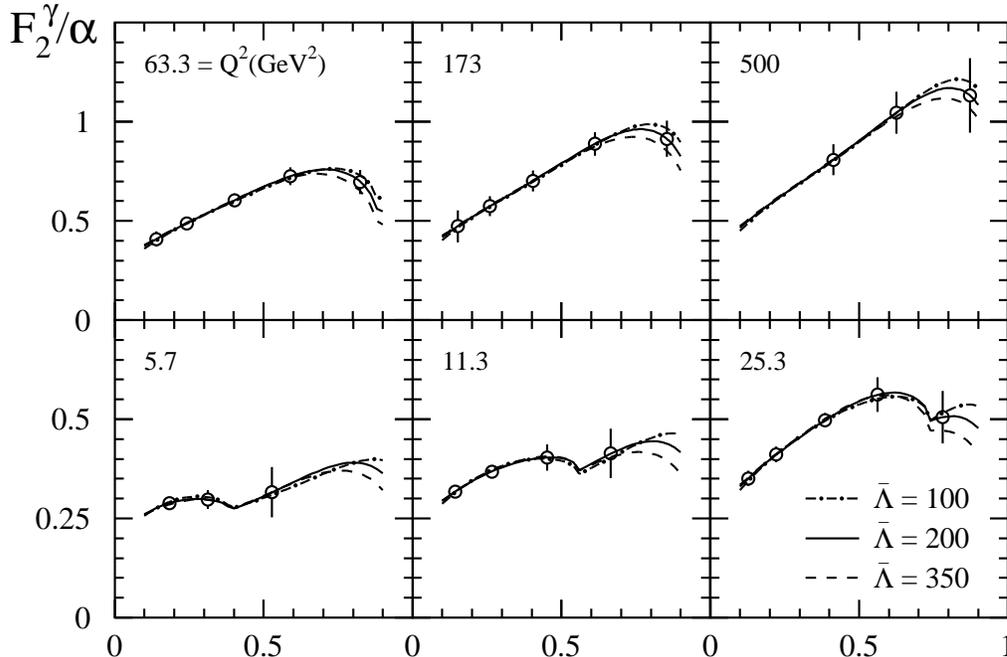,height=9.5cm}
 \vspace{-1.5cm}
 \end{center}
 \caption[]{\em The estimated accuracy of $F_2^{\, \gamma}$ from one 
  experiment at LEP2 using a (linear) Blobel unfolding \cite{Blobel}.
  The (in-)sensitivity to $\Lambda_{\rm QCD}$ is illustrated by the 
  $\pm 1 \sigma $ results described in the text. $\bar{\Lambda}$  
  denotes $\Lambda_{\overline{\rm MS}}^{(4)}$ in MeV.}
\label{fig:avf3}
\end{figure}

In order to estimate the possible sensitivity of $F_2^{\,\gamma}$
to $\Lambda_{\rm QCD}$ at LEP2, deep--inelastic electron--photon 
collisions corresponding to an integrated luminosity of $500\,$pb$^{-1}$
have been generated using the SaS1D distribution functions \cite{SaS}, 
passed through a (fast) detector simulation (DELPHI) and unfolded using 
the Blobel program \cite{Blobel}. Possible systematic errors due to 
the dependence on fragmentation parameters have been neglected. We 
have chosen six bins in $Q^2$ (logarithmically distributed) and let the 
unfolding program choose the number and sizes of $x$ bins. In total 21 
bins have been obtained at $x>0.1$, shown in Fig.~\ref{fig:avf3}. 
Alternatively, we have used the theoretical error estimates and bins 
of sec.~\ref{avs2} as representative for the best possible 
measurement using the combined statistics of two experiments.

Next, fictitious $F_2^\gamma(x,Q^2)$ data have been generated at these 
$(x,Q^2)$ points. The input distributions of a simple toy-model have 
been evolved in NLO, which however yields very similar numbers of 
events as the SaS \cite{SaS} distributions. Thus the relative errors 
can be taken over from SaS. Specifically, the NLO parton distributions 
of the photon have been generated by a (coherent) sum of the three 
vector mesons $\rho^0$, $\omega$, and $\phi$ at $Q_S = 0.6\,$GeV in the 
DIS$_{\gamma}$ scheme \cite{GRVth}.

Finally, $\Lambda_{\rm QCD}$ is fitted together with the shape 
parameters $N_s$, $a_i$ and $b_i$ of the vector--meson valence, sea and 
gluon input distributions 
\begin{equation}
\label{firstinput}
   xv(x) = \kappa N_v x^{a_v} (1-x)^{b_v} \: , \:\:\:
   xS(x) = \kappa N_s x^{a_s} (1-x)^{b_s} \: , \:\:\:
   xG(x) = \kappa N_g x^{a_g} (1-x)^{b_g}
\end{equation}
at $Q_{\rm ref} = 2$ GeV ($N_v$ and $N_g$ are fixed by the charge and 
momentum sum rules of the vector mesons), the overall normalization 
$ \kappa $, the scale $Q_S$, and the charm quark mass entering via 
$F_{2,\, h}^{\,\gamma}$ in eq.~(\ref{ave4}). The variation of 
the parameters is restricted to values reasonable for vector meson 
states, e.g., $ 0.8 \leq b_{v} \leq 1.3 $ (the ``data" were generated
with the counting-rule value of 1.0). Using the Blobel-unfolded 
``results" of Fig.~\ref{fig:avf3}, one finds an experimental $1\sigma $ 
accuracy of
\begin{equation}
  \Lambda_{\overline{\rm MS}}^{(4)} = \left( 200^{+150}_{-100} 
  \right)\,\mathrm{MeV} \:\:\: \Rightarrow \:\:\: \alpha_{s}(M_Z) = 
  0.109^{+0.010}_{-0.011} \:\: .
\end{equation}
The sensitivity is dominated by the large-$x$ region \cite{Frazer} 
as obvious from Fig.~\ref{fig:avf3}. 
If we use the ``data" of sec.~\ref{avs2} 
instead, which need about the statistics of two experiments and some 
progress in the unfolding, especially at large $x$, we obtain
\begin{equation}
  \Lambda_{\overline{\rm MS}}^{(4)} = \left( 200^{+85}_{-65} 
  \right)\,\mathrm{MeV} \:\:\:\: \Rightarrow \:\:\: \alpha_{s}(M_Z) = 
  0.109 \pm 0.006 \:\: .
\end{equation}
Hence, even under the most optimistic assumptions, the experimental 
error on $\Lambda_{\rm QCD}$ as determined from $F_2^{\,\gamma}$ at 
LEP2 is a factor of two bigger than from fixed-target $ep\, / \mu p$ 
DIS.  The theoretical scale--variation uncertainty of $F_2^{\,
\gamma}$  has been studied in \cite{LSRN} for fixed parton 
distributions. The resulting theoretical error on $\Lambda_{\rm QCD}$
is expected to be of similar size as in $ep$ DIS. It should be 
mentioned that without the VMD-based restrictions on the parameter ranges 
the kinematical coverage and accuracy of LEP2 is not sufficient for any 
sensitive $\Lambda_{\rm QCD} $ determination. 
\subsection{Azimuthal correlations, a substitute for the 
               longitudinal structure function}
\label{avs5}
It has been mentioned already that $F_L^{\, \gamma }$ will not be 
measurable at LEP2. However, $F_L$ is not the only structure function 
that contains additional information. If, instead of just measuring 
the total cross-section $e\gamma\to eX$, one triggers on a final-state 
`particle' $a$ (either a hadron or a jet), more structure functions 
become accessible. The cross section as a function of the direction of 
$a$ can be written in terms of the unintegrated structure 
functions $\wtF_P$, $P=T,\, L,\, A,\,$ and $B$, as
\begin{equation}
  \frac{d\sigma (e\gamma\to eaX)}{dx\;dy\;d\Omega_a/4\pi} =
  \frac{2\pi\alpha^2}{Q^2}\frac{1+(1-y)^2}{xy}
  \biggl[ \Bigl(2x\wtF_T+\epsilon(y)\wtF_L\Bigr)
         -\rho(y)\wtF_A\cos\phi_a
         +\smfrac12\epsilon(y)\wtF_B\cos2\phi_a \biggr] \:\: .
\end{equation}
Here $\Omega_a$ represents the direction of $a$ in the $\gamma\gamma^*$
rest-frame, and $\phi_a$ is its azimuth around the $\gamma\gamma^*$ 
axis, relative to the electron plane. The functions $\epsilon(y)$ and 
$\rho(y)$ are both $1+{\cal{O}}(y^2)$, and can be approximated by 1 
throughout the accessible region of phase space. The standard 
structure functions $F_2$ and $F_L$ are related to the corresponding
$\wtF_P$ by integration over $\Omega_a$. 

For leptonic final states $\wtF_P$ are uniquely given by perturbation
theory, while for hadronic final states these quantities involve a
convolution over the parton densities of the photon:
\begin{equation}
  \wtF_P(x,z) = \sum_{i=\gamma,q,g} \int_x^1 \frac{dx_p}{x_p}
  \;\frac{x}{x_p}f_{i/\gamma}\Big( \frac{x}{x_p}\Big) \;
  \wtF_P^i(x_p,z) \:\: ,
\end{equation}
where $f_{\gamma/\gamma}(x)=\delta(1-x)$ and $z = (p_a\cdot p_i)/
(q\cdot p_i) = \smfrac12(1+\beta\cos\theta)$, with $\beta$ and
$\theta$ denoting the velocity and direction of~$a$ in the $i\gamma^*$ 
rest-frame, respectively. One can find many incorrect formulae for 
$\wtF_P^i$ in the literature. We have performed an independent 
calculation and confirm the leading-order results given in 
\cite{John:Sheffield}, obtaining 
\begin{eqnarray}
  \wtF_T^\gamma(x_p,z) &=& e_q^4\frac{\alpha}{2\pi}
    (x_p^2+(1-x_p)^2)\frac{z^2+(1-z)^2}{2z(1-z)}  \nonumber \\
  \wtF_B^\gamma(x_p,z)  =  \wtF_L^\gamma(x_p,z) &=&
    e_q^4\frac{4\alpha}{\pi}x_p^2(1-x_p)  \nonumber \\
  \wtF_A^\gamma(x_p,z) &=& e_q^4\frac{4\alpha}{\pi}
    x_p(1-2x_p)(1-2z)\sqrt{\frac{x_p(1-x_p)}{4z(1-z)}}  \nonumber \\
  \wtF_P^g(x_p,z) &=& \frac{T_R\alpha_s}{e_q^2\alpha}\wtF_P^\gamma, \\
  \wtF_T^q(x_p,z) &=& e_q^2\frac{C_F\alpha_s}{4\pi}
    \left[\frac{x_p^2+z^2}{(1-x_p)(1-z)}+2(x_pz+1)\right]  \nonumber \\
  \wtF_B^q(x_p,z)  =  \wtF_L^q(x_p,z) &=&
    e_q^2\frac{2C_F\alpha_s}{\pi}x_p^2z  \nonumber \\ 
  \wtF_A^q(x_p,z) &=& e_q^2\frac{4C_F\alpha_s}{\pi}
    x_p(x_pz+(1-x_p)(1-z))\sqrt{\frac{x_pz}{4(1-x_p)(1-z)}}\:\: ,
  \nonumber 
\end{eqnarray}
up to terms of order $m_q^2x/(1-x)Q^2$.
In the quark case, the azimuth is that of the outgoing quark~-- the
equivalent expressions for the outgoing gluon are identical but with
$z$ replaced by $1-z$ and $\wtF_A^q$ negated. The photon and gluon
cases are identical for either outgoing parton.

Note that $\wtF_B^i=\wtF_L^i$ for all parton types, so a measurement 
of $\langle\cos2\phi_a\rangle$ gives the same information about the 
parton content of the photon as $F_L^{\,\gamma},$ despite the fact 
that they arise from different spin states of the virtual photon 
(purely longitudinal for $F_L$ and transverse-transverse interference 
for $F_B$).\footnote{In
  fact this has only been proved at lowest order. Different
  definitions of $\wtF_B$ (eg.\ single-particle vs.\ jet inclusive) 
  will have different higher-order corrections, and it is currently
  unknown whether any obey the same relationship at higher orders.}
This is a consequence of the fact that the struck parton is a fermion.
In the leptonic case, the two outgoing particles can be distinguished,
and the above distributions directly measured. In the hadronic case,
however, quark, antiquark and gluon jets cannot be distinguished, and
one must sum over all assignments. Since each event consists of two
jets with $z$ and $1-z$, all three sub-processes give equal and 
opposite $\cos\phi_a$-dependence for the two jets, and $\langle\cos\phi_a
\rangle$ defined naively is identically zero. But if we instead use 
only the more forward of the two jets (i.e., the one with larger $z,$ 
which is more often the quark in the $q\to qg$ case), the constant and
$\cos2\phi_a$ terms remain unchanged, but also the $\cos\phi_a$ term 
becomes nontrivial, being always negative for quarks and taking either
sign for the other two processes, depending on~$x_p$.

The measurement of azimuthal correlations involves reconstructing the
hadronic final state to a much greater degree than does the measurement
of the total cross-section. For this reason, a number of additional 
problems occur, including:
 How well do jet (or inclusive-particle) momenta mirror the underlying
 parton momenta?
 How well can the jets be reconstructed experimentally?
 How much are the azimuthal correlations smeared by the fact that the
 $\gamma\gamma^*$ rest-frame is not exactly known, or by target photon
 mass effects?
 How much artificial (de)correlation is induced by the fact that any
 cuts made in the lab frame are azimuth-dependent in the $\gamma
 \gamma^*$ rest-frame?
A detailed detector-level study would be needed to answer these points,
as discussed for CELLO in \cite{CELLO}, but this has not yet been done
for LEP2 energies. However, azimuthal correlations have been measured
on a proton target at HERA \cite{Chris:thesis}, and in the leptonic
final-state on a photon target at PETRA \cite{PETRAazi} and
LEP1 \cite{Leonardi:Sheffield}, and
all of the above problems have been addressed, and overcome, in one or 
other of these analyses. It thus seems hopeful that the measurement 
can be done at LEP2.

In the absence of a detailed experimental investigation, the group has 
performed a brief generator-level study using {\small HERWIG}, version 5.8d 
\cite{HERWIG}. The jet reconstruction and cuts are loosely based
on those used by the H1 collaboration \cite{Chris:thesis}, adapted for
our assumed LEP2 detector coverage. The difficulties in measuring $x$
discussed in the context of $F_2$ are left aside~-- it is assumed that
$x$ and $Q^2$ are perfectly known. Events passing the canonical cuts
are selected ($E_{tag}>0.5\, E_{\mathrm{beam}}, \; \theta_{tag}>30\, \mathrm
{mrad}$ and $W_{\mathrm{vis}}>2\, \mathrm{GeV}$ within $|\cos\theta|<
0.97$). All particles within the central region are boosted to the 
Breit frame of the virtual photon and target electron beam\footnote{This
 frame has the same transverse boost as their rest-frame but a 
 different longitudinal boost, which results in improved jet properties
 if a frame-dependent jet algorithm is used \cite{CDW:ktDIS}.},
and jets are reconstructed using the $k_\perp$ jet algorithm 
\cite{CDW:ktDIS}, with a cutoff of 2 GeV. No attempt was made to 
optimize this value. As discussed in \cite{Mike:simple}, the maximum 
correlations are achieved for $p_t=Q/2$.

\begin{figure}
  \vspace{0.2cm}
  \hspace*{-0.4cm}
  \centerline{\epsfig{figure=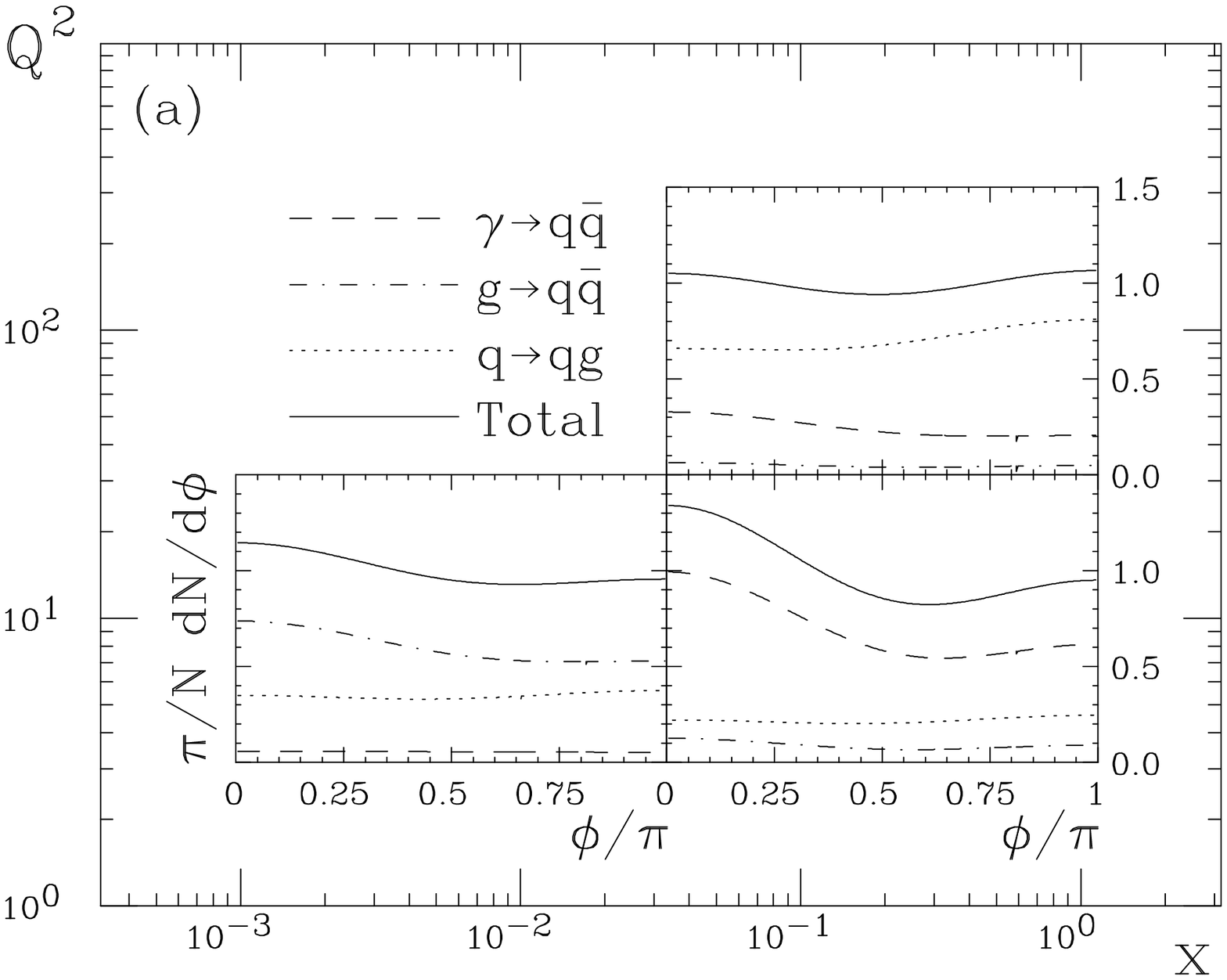,height=5.3cm,angle=90}
  \hspace{3.2cm}\epsfig{figure=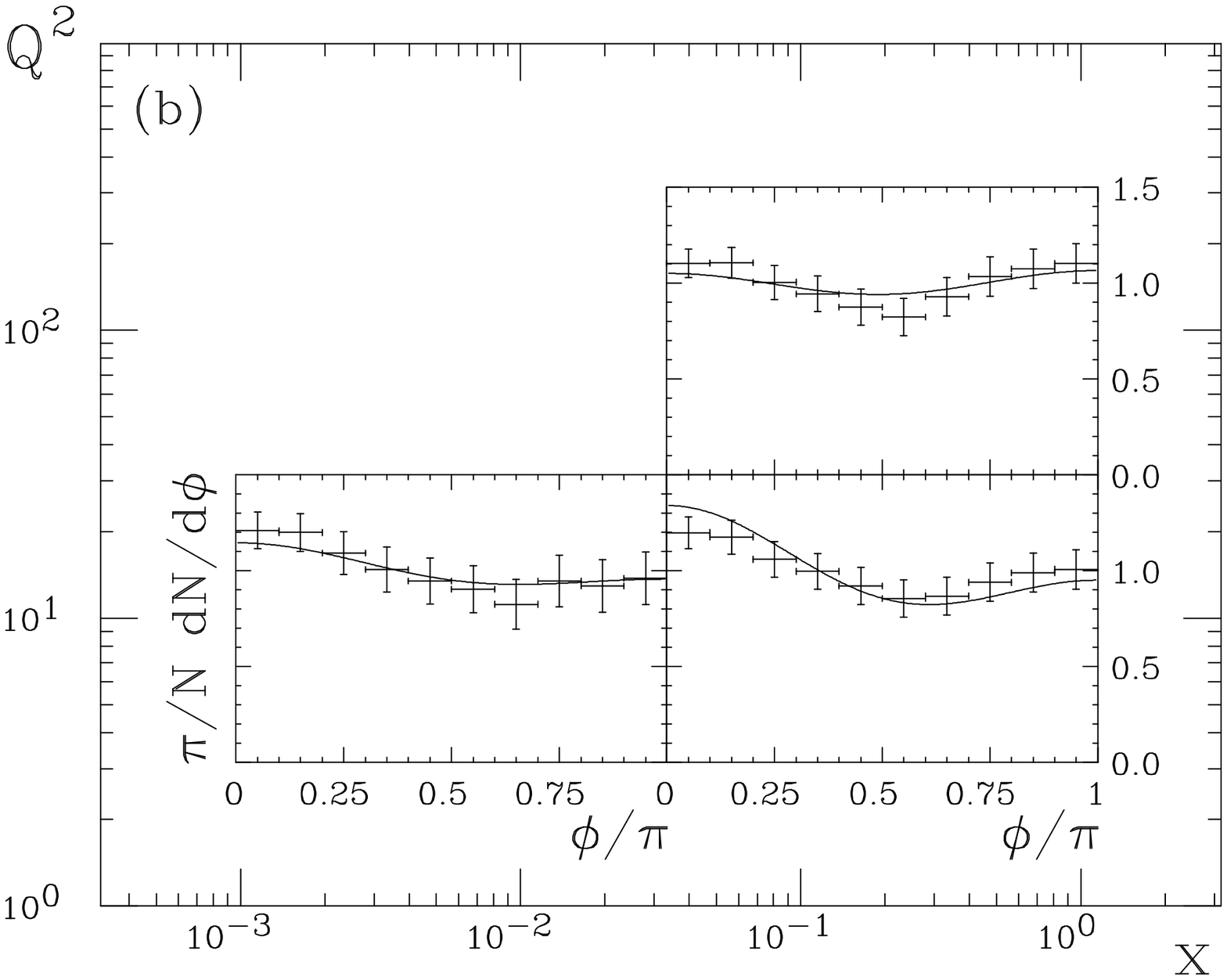,height=5.3cm,angle=90}}
  \vspace{-0.7cm}
  \caption[]{\em
   Distribution of azimuthal angle of the more forward jet in
   two-jet events. Canonical cuts are made, plus a 2 GeV jet-$p_t$ cut.
   In {\bf (a)} parton--level results are shown, {\bf (b)} depicts 
   generator-level results (data points with statistical errors 
   corresponding to 500 pb${}^{-1}$), in comparison with the total 
   parton-level prediction of (a).}
  \label{fig:azidist}
\end{figure}

In Fig.~\ref{fig:azidist}a we show parton-level results in three $x$
and $Q^2$ bins, broken down into the contributions from the different
parton types $\gamma,\;q$ and~$g$.
By `parton--level' we mean at leading order, with a direct 
correspondence made between partons and jets, and with the photon 
assumed collinear with the incoming electron beam. The SaS1D 
parametrization of $f_{i/\gamma}$ \cite{SaS} has been used. We see 
that the $\cos2\phi_a$-dependence arises predominantly from the gluon- 
and photon-induced sub-processes, and that a different initial-state 
parton dominates within each bin.
In Fig.~\ref{fig:azidist}b generator-level results are presented, 
including parton showering, hadronization, jet reconstruction and 
target--mass smearing. The error bars shown correspond to the 
statistical errors one could expect from 500pb${}^{-1}$ of data.
While the correlations are somewhat smeared, a signal still persists. 
It is possible that one could improve the statistics by including 
lower $p_t$ scattering, provided a hadronic plane could be cleanly 
defined. Hence it seems likely that the azimuth--dependent structure 
functions of the photon can be measured at LEP2, providing additional
constraints on the parton content of the photon.
\subsection{Virtual photon structure}
\label{avs6}
Effects of a non-zero virtuality of the target photon, $P^2 \neq 0$, 
have attracted considerable interest recently \cite{SaS,borzu,afg1,%
GRSphot}. The non-perturbative hadronic (VMD) contribution to the photon 
structure is expected to go away with increasing $P^2$, allowing for a 
purely perturbative prediction for $F_2^{\,\gamma}(x,Q^2;P^2)$ at 
sufficiently high $P^2$ \cite{uematsu}. The fall-off of the 
non-perturbative part with increasing $P^2$ is theoretically uncertain 
and model dependent \cite{GRSphot,SaS}, hence experimental clarification
is required. An improved understanding of this transition is also 
relevant for the experimental extraction of the real--photon structure 
functions, since corrections of the order of 10\% have to be made to 
take into account the finite range of $P^2$ in single-tag events.
However, the present experimental knowledge is very poor: the 
older\footnote{Note that actually $F_{eff} \equiv F_2 + 3 F_L/2$ was 
measured by PLUTO.}
measurement of $F_2^{\,\gamma}(x,Q^2;P^2)$ by the PLUTO 
collaboration \cite{PLUTOv} suffers from low statistics and a rather 
limited kinematical coverage: $Q^2 = 5{\mbox{ GeV}}^2$, $P^2 \le 0.8 
{\mbox{ GeV}}^2$, and $x \stackrel{>}{\sim} 0.1$. 
Some new data are however being
collected at HERA \cite{utley}.

Especially because of its higher energy, but also due to its increased
integrated luminosity, LEP2 can provide much improved information from 
double-tagged events. The rate of such events in the main forward 
luminometers ($\theta > 30 $ mrad for the electron {\it and} the 
positron) is very small. The more important double-tagged results at 
LEP2 are expected to come from events with first tags in the main 
forward luminometers ($\theta _1 > 30 $ mrad) and second tags in the 
very forward calorimeters which are used for online high-rate 
luminosity monitoring in each of the four LEP experiments. These small 
detectors (currently being upgraded in ALEPH and L3) are situated at 
7 -- 8 meters from the interaction point, beyond the minibeta 
quadrupole magnets. The defocussing effect of the quadrupoles distorts 
the acceptance of the detectors, but lepton tags can be reconstructed 
in the range $ 5 \stackrel{<}{{\scriptstyle \sim}} \theta_2 \stackrel{<}
{{\scriptstyle \sim}} 15 $ mrad, yielding $ 0.1 \stackrel{<}
{{\scriptstyle \sim}} P^2 \stackrel{<}{{\scriptstyle \sim}} 
1.0 \mbox{ GeV}^2$ (the exact coverage varies between experiments).
\begin{figure}[htb]
\begin{center}
\vspace*{-0.6cm}
\epsfig{file=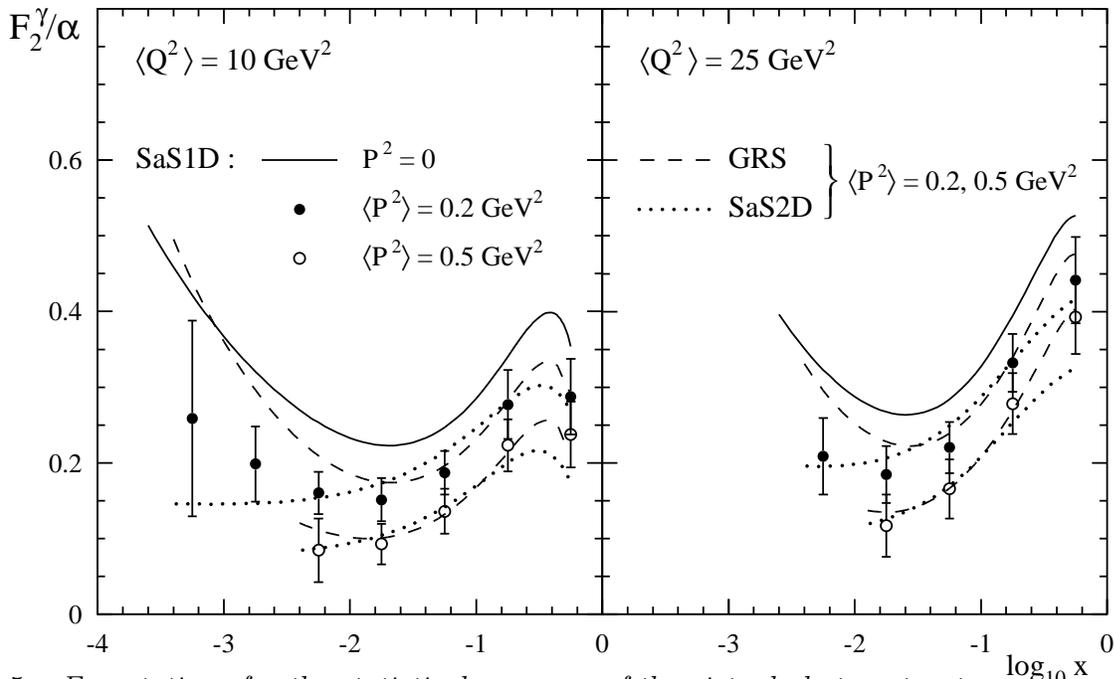,height=9.4cm}
\end{center}
\vspace{-1.2cm}
\caption[]{ \em
  Expectations for the statistical accuracy of the virtual 
  photon structure measurement at LEP2 in two different $P^2$ and $Q^2$ 
  bins, using the SaS1D \cite{SaS} distributions. The SaS1D prediction 
  for the real photon and the $P^2 \neq 0$ results for the GRS \cite 
  {GRSphot} and SaS2D \cite{SaS} distributions are shown as lines for 
  comparison. The upper (lower) curves for GRS and SaS2D refer to 
  $P^2= 0.2$ (0.5) GeV$^2$, respectively.}
\label{fig:awf1}
\end{figure}
 
For 500 pb$^{-1}$ of data collected at LEP2, it is expected that about 
800 double-tagged events of this type will be seen within the range $ 3 
\times 10^{-4} < x < 1$ and $3 < Q^2 < 1000$ GeV$^2$. The invariant
mass of each event can be reconstructed from the two tagged leptons and 
hence the correlation between the measured value ($W_{\rm vis}$) and 
the true value is good -- there is no loss of correlation at high $W$ 
(unlike for single--tagged events, see sec.~\ref{avs3}), and the 
structure function can be measured more easily down to low $x$.

Fig.~\ref{fig:awf1} shows the virtual photon structure function 
$F_2^{\gamma}(x,Q^2;P^2)$ as predicted by the SaS1D, SaS2D \cite{SaS} 
and GRS \cite{GRSphot} models in two bins for $P^2$ and $Q^2$. The error
bars indicate the statistical error expected on each point for a bin 
width leading to two points per decade in $x$, using the SaS1D parton 
distributions \cite{SaS}. The SaS2D \cite{SaS} and GRS \cite{GRSphot} 
distributions, both showing a rather different small-$x$ behaviour as 
compared to SaS1D, lead to similar results for the expected statistical 
errors. A measurement of $F_2^{\gamma}(x,Q^2;P^2)$, as distinct from 
the real ($P^2=0$) photon structure function (shown as solid curves for 
SaS1D), should be possible at LEP2 and could be compared to the results 
from PLUTO \cite {PLUTOv} in the region of overlap, 
as well as to different model predictions \cite{GRSphot,SaS}.
Additional information will be obtained in single tag events where 
high $p_T$ jets are produced (see sec.~\ref{ggpt1}).
\subsection{Summary}
\label{avs8}
Due to its high beam energy and increased integrated luminosity, LEP2 
will be a unique place for studying the hadronic structure functions of 
the photon. Some $ (0.5 \ldots 1) \cdot 10^6 $ single--tag events for
deep--inelastic ($Q^2 > 3 \mbox{ GeV}^2$) electron--photon scattering
can be expected, for the first time allowing for a determination of 
$F_2^{\,\gamma}$ (and hence of the photon's quark content) with 
statistically high precision up to scales $Q^2$ of a few 100 GeV$^2$ 
and down to Bjorken-$x$ values as low as about $10^{-3}$. Depending on 
how well systematic uncertainties can be controlled, a determination of 
the QCD scale $\Lambda_{\rm QCD}$ may become possible from $F_2^{\,
\gamma}$. But even under optimistic assumptions, the experimental error 
on $\alpha_{s}(M_Z)$ will be about a factor of two bigger than that one
obtained from electron--nucleon deep--inelastic scattering.
It seems likely that supplementary information on the parton densities
can be obtained by measuring azimuthal correlations between the tagged 
electron and a final state hadron or jet. 
About $10^3$ 
double--tag events are expected to be seen with $0.1 \stackrel{<}
{{\scriptstyle \sim}} P^2 \stackrel{<}{{\scriptstyle \sim}} 1.0 
\mbox{ GeV}^2$ and $Q^2$ as above, allowing for a study of the virtual 
photon structure function over a much wider kinematical range than so 
far.
It should be noted that the data taken at LEP2 on photon structure 
functions will dominate our knowledge in most of the accessible range  
discussed above for the foreseeable future, since a 500 GeV linear 
collider will most likely have no access to $Q^2 < 100 \mbox{ GeV}^2$.

\input feynman
%
%
\section[The equivalent photon approximation]
{The equivalent photon approximation }
\label{ggepa}
%
%
The cross section for a \gaga process is related to the cross section
at the $e^+ e^-$ level, which is measured in the laboratory, by
the formula
%
%
\begin{equation}
  d \sigma(e^+e^-\rightarrow e^+e^- X) = 
  \sigma(\gamma_1 \gamma_2 \rightarrow X)
   \frac{ d^2 n_1 }{ d z_1 d P_1^2}
   \frac{ d^2 n_2 }{ d z_2 d P_2^2}
   d z_1 d z_2
   d P_1^2 d P_2^2
\label{eq:epa}
\end{equation}
where $z_i$ is the scaled photon energy in the laboratory frame and
$P_i^2$ is  the photon invariant mass. This is the equivalent photon
approximation (EPA)~\cite{Budnev} where 
the longitudinal polarization component as
well as the mass of the incoming photons are neglected in $\sigma
(\gamma \gamma \rightarrow X)$. The $P_i^2$ integration can be carried
out to give the photon ``density" in the $e^\pm$ 
(the photon flux)~\cite{avWW}
%
%
\begin{equation}
f_{\gamma/e}(z,P_{min},P_{max}) = \int^{P^2_{max}}_{P^2_{min}}
 \frac{d^2 n} {d z d P^2} d P^2 \\
 = \frac{\alpha}{2 \pi}
\left[ \frac{1+(1-z)^2}{z} 
\ln \frac{P^2_{max}}{P^2_{min}}
- 2 m_e^2 z \left( 
\frac{1}{P_{min}^2}
-\frac{1}{P_{max}^2}\right)
\right].
\label{eq:wewi}
\end{equation}
For untagged experiments $P_{min}$ is the kinematic limit 
\begin{equation}
 P^2_{min} = \frac{m^2_e z^2}{1-z}
\end{equation}
and $P_{max} \simeq E_{beam}$. The quality of the approximation is not 
guaranteed in this case as the EPA is derived under the hypothesis that
$P^2 \ll E^2_{beam}$ which is not always satisfied here. In most 
of the following we use
antitagging conditions where the $e^\pm$ are confined to small angles 
$\theta < \theta_{max}$ (typically $\theta_{max}= 30$ mrad) so that 
\begin{equation}
P^2_{max} = (1-z) E^2_{beam} \theta^2_{max}.
\end{equation}
Using antitagging conditions rather than untagged conditions reduces
somewhat the cross sections (about $30\ \%$ in the case of heavy flavor
production and a factor 2 for large $p_T$ jet production) but improves
the reliability of  the theoretical calculations based on the EPA.
Finally for tagged conditions both $P^2_{min}$ and $P^2_{max}$ are set by
the detector configuration.

Even in the case of antitagging it is always worthwhile, whenever
possible, to  check the validity of the EPA  for each of the considered
process. For large $p_T$ jets and heavy flavor production  a possible
check consists in comparing, with the appropriate choice of cuts and
kinematics, the lowest order matrix element calculation of $e^+ e^-
\rightarrow e^+ e^- q \bar{q}$ to the approximate one. Good agreement is
found provided the ``constant" (non-logarithmic) term is kept in eq.
(\ref{eq:wewi}). If the constant term is ignored the cross sections are
over-estimated by roughly $10 \%$ when both $e^+$ and $e^-$ are
antitagged. The same is true for minimum bias type physics. 
Special
attention should be given to processes which involve the photon
structure function, $i.\ e.$, the so-called resolved processes. Under
certain conditions it may
be necesary to take into account the effect of the virtuality of the
quasi-real photons initiating the process. Indeed as written in eq.
(\ref{ave5}) the structure function has two components
\begin{equation}
 F_{i/\gamma}(x,Q^2;P^2) = F^{PL}_{i/\gamma}(x,Q^2;P^2) 
+ F^{had}_{i/\gamma}(x,Q^2;P^2)   
\end{equation}
with a different dependence in $P^2$, the virtuality of the quasi-real
photon. In particular $F^{had}_{i/\gamma}$ is roughly suppressed by the
usual $VMD$ form factor $(m^2_\rho / (m^2_\rho+P^2))^2$. In the tagged
case ($P^2 \simeq .5$ GeV$^2$) or in the case with relatively large
antitagging angles this factor should be taken into account when
carrying out the integration over the virtuality in eq. (\ref{eq:wewi}).
The result is a relative reduction of the ``$had$" component of the
photon compared to the case when the photon is assumed to be real. This
reduction obviously affects the rate of the observable cross section. If
the aim is to obtain predictions at a $10\%$ accuracy this effect should
certainly be studied in further details.

Turning now to resonance production (see sec. \ref{ggex1}), 
eq. (\ref{eq:epa}) simplifies since one of the $z_i$ integration can
be performed with the constraint $z_1 z_2 = \tau = M^2/s_{e^+e^-}$
where $M$ is the resonance mass. It is then customary to define 
luminosity functions (see $e.g.$ \cite{Field})
\begin{equation}
{d {\cal L} \over d M} = {2 \tau \over M} \int d z_1 d z_2
f_{\gamma/e}(z_1) f_{\gamma/e}(z_2) \delta(z_1 z_2 - \tau)
\end{equation}
so that
\begin{equation}
  d \sigma(e^+e^-\rightarrow e^+e^- X) = \int d M 
{d {\cal L} \over d M}  \sigma(\gamma \gamma \rightarrow X).
\end{equation}
This luminosity curve makes it easy, in principle, to determine the
counting rate for resonance production knowing the width of the
resonance in the \gaga channel. An important point however concerns the
acceptance cuts of the detector which reduce the  observed rates
compared to the theoretical predictions. Such cuts are taken into
account in sec. \ref{ggex1}.  Detailed studies of luminosity curves were
done for the previous LEP2  and we do not repeat them here \cite {Aachen}.
%
\section[Tagging conditions, cuts and background to \gaga processes]
{Tagging conditions, cuts and background to \gaga processes
\label{ggtag}
{\protect\footnote{
 P.~Aurenche, J.Ph.~Guillet, F.~Kapusta, D.~Perret-Gallix, N.I.~Zimin}}}
In this section we discuss what we call \gaga events 
(as opposed to $\gamma^* \gamma$) $i.e.$
events where the electron and positron of the diagram in 
Fig. \ref{fig:avf1} escape detection ($i.e.$ essentially 
``go down the beam pipe"). This require a precise definition of the
tagging conditions as well as the cuts necessary to reduce the
background to \gaga physics.
Background sources come from all processes not initiated by \gaga 
interactions but exhibiting similar features such that they can be taken,
mistakingly, for genuine \gaga events.
%
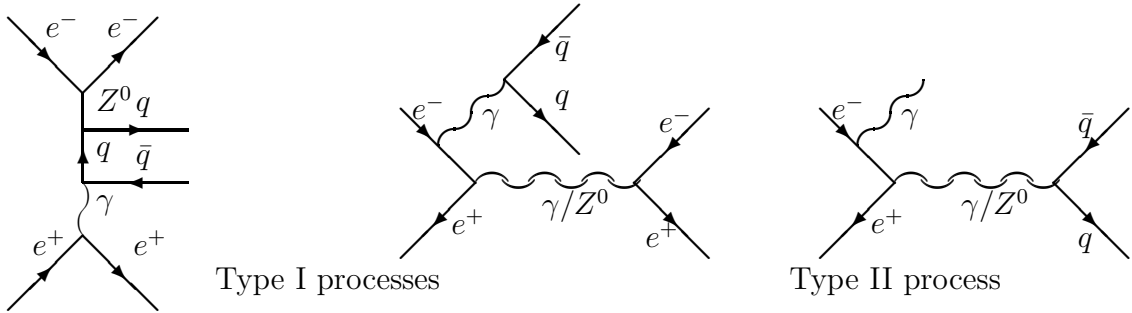
\begin{figure}[hbtp]
  \begin{picture}(10000,10000)(0,0)
   \THICKLINES
   \drawline\photon[\N\REG](5000,2000)[2]
   \global\advance \pmidx by +500
   \put(\pmidx,\pmidy){$\gamma$}
   \drawline\fermion[\SW\REG](\photonfrontx,\photonfronty)[4000]
   \drawarrow[\NE\ATBASE](\pmidx,\pmidy)
   \global\advance \pmidx by -500
   \global\advance \pmidy by +700
   \put(\pmidx,\pmidy){$e^+$}
   \drawline\fermion[\SE\REG](\photonfrontx,\photonfronty)[4000]
   \drawarrow[\SE\ATBASE](\pmidx,\pmidy)
   \global\advance \pmidy by +700
   \global\advance \pmidx by +500
   \put(\pmidx,\pmidy){$e^+$}
   \drawline\fermion[\E\REG](\photonbackx,\photonbacky)[4000]
   \drawarrow[\W\ATBASE](\pmidx,\pmidy)
   \global\advance \pmidy by 700
   \put(\pmidx,\pmidy){$\bar{q}$}
   \drawline\fermion[\N\REG](\photonbackx,\photonbacky)[2000]
   \drawarrow[\N\ATBASE](\pmidx,\pmidy)
   \global\advance \pmidx by 500
   \put(\pmidx,\pmidy){$q$}
   \drawline\fermion[\E\REG](\fermionbackx,\fermionbacky)[4000]
   \drawarrow[\E\ATBASE](\pmidx,\pmidy)
   \global\advance \pmidy by 700
   \put(\pmidx,\pmidy){$q$}
   \drawline\scalar[\N\REG](\fermionfrontx,\fermionfronty)[1]
   \global\advance \pmidx by 500
   \put(\pmidx,\pmidy){$Z^0$}
   \drawline\fermion[\NW\REG](\scalarbackx,\scalarbacky)[4000]
   \drawarrow[\SE\ATBASE](\pmidx,\pmidy)
   \global\advance \pmidy by +700
   \put(\pmidx,\pmidy){$e^-$}
   \drawline\fermion[\NE\REG](\scalarbackx,\scalarbacky)[4000]
   \drawarrow[\NE\ATBASE](\pmidx,\pmidy)
   \global\advance \pmidy by +700
   \global\advance \pmidx by -500
   \put(\pmidx,\pmidy){$e^-$}
\put(10000,0){Type I processes}
  \end{picture}
 \hfill
  \begin{picture}(10000,10000)(0,0)
   \THICKLINES
   \drawline\photon[\E\REG](4000,4000)[6]
   \global\advance \pmidx by -500
   \global\advance \pmidy by -1200
   \put(\pmidx,\pmidy){$\gamma/Z^0$}
   \drawline\fermion[\NW\REG](\photonfrontx,\photonfronty)[4000]
   \global\Xone=\pmidx
   \global\Yone=\pmidy
   \global\advance \pmidx by -500
   \global\advance \pmidy by +500
   \drawarrow[\SE\ATBASE](\pmidx,\pmidy)
   \global\advance \pmidx by -500
   \global\advance \pmidy by +500
   \put(\pmidx,\pmidy){$e^-$}
   \drawline\fermion[\SW\REG](\photonfrontx,\photonfronty)[4000]
   \drawarrow[\SW\ATBASE](\pmidx,\pmidy)
   \global\advance \pmidx by +500
   \global\advance \pmidy by -500
   \put(\pmidx,\pmidy){$e^+$}
   \drawline\fermion[\NE\REG](\photonbackx,\photonbacky)[4000]
   \drawarrow[\SW\ATBASE](\pmidx,\pmidy)
   \global\advance \pmidx by -500
   \global\advance \pmidy by +500
   \put(\pmidx,\pmidy){$e^-$}
   \drawline\fermion[\SE\REG](\photonbackx,\photonbacky)[4000]
   \drawarrow[\SE\ATBASE](\pmidx,\pmidy)
   \global\advance \pmidx by -1000
   \global\advance \pmidy by -1000
   \put(\pmidx,\pmidy){$e^+$}
   \drawline\photon[\NE\REG](\Xone,\Yone)[4]
   \global\advance \pmidx by 400
   \global\advance \pmidy by -400
   \put(\pmidx,\pmidy){$\gamma$}
   \drawline\fermion[\NE\REG](\photonbackx,\photonbacky)[4000]
   \drawarrow[\SW\ATBASE](\pmidx,\pmidy)
   \global\advance \pmidx by +500
   \global\advance \pmidy by -500
   \put(\pmidx,\pmidy){$\bar{q}$}
   \drawline\fermion[\SE\REG](\photonbackx,\photonbacky)[4000]
   \drawarrow[\SE\ATBASE](\pmidx,\pmidy)
   \global\advance \pmidx by +500
   \global\advance \pmidy by +500
   \put(\pmidx,\pmidy){$q$}
  \end{picture}
 \hfill
  \begin{picture}(10000,10000)(0,0)
   \bigphotons
   \THICKLINES
   \drawline\photon[\E\REG](4000,4000)[6]
   \global\advance \pmidx by -500
   \global\advance \pmidy by -1200
   \put(\pmidx,\pmidy){$\gamma/Z^0$}
   \drawline\fermion[\NW\REG](\photonfrontx,\photonfronty)[4000]
   \global\Xone=\pmidx
   \global\Yone=\pmidy
   \global\advance \pmidx by -500
   \global\advance \pmidy by +500
   \drawarrow[\SE\ATBASE](\pmidx,\pmidy)
   \global\advance \pmidx by -500
   \global\advance \pmidy by +500
   \put(\pmidx,\pmidy){$e^-$}
   \drawline\fermion[\SW\REG](\photonfrontx,\photonfronty)[4000]
   \drawarrow[\SW\ATBASE](\pmidx,\pmidy)
   \global\advance \pmidx by +500
   \global\advance \pmidy by -500
   \put(\pmidx,\pmidy){$e^+$}
   \drawline\fermion[\NE\REG](\photonbackx,\photonbacky)[4000]
   \drawarrow[\SW\ATBASE](\pmidx,\pmidy)
   \global\advance \pmidx by -500
   \global\advance \pmidy by +500
   \put(\pmidx,\pmidy){$\bar{q}$}
   \drawline\fermion[\SE\REG](\photonbackx,\photonbacky)[4000]
   \drawarrow[\SE\ATBASE](\pmidx,\pmidy)
   \global\advance \pmidx by -500
   \global\advance \pmidy by -1000
   \put(\pmidx,\pmidy){$q$}
   \drawline\photon[\NE\REG](\Xone,\Yone)[4]
   \global\advance \pmidx by 400
   \global\advance \pmidy by -400
   \put(\pmidx,\pmidy){$\gamma$}
\put(0,0){Type II process}
  \end{picture}
~~~~~~~~~~~~~~~ \hfill
\caption{\it Two types of background processes for \gaga physics}
\label{fig:gf1}
\end{figure}
Background events are, essentially, of two types (Fig. \ref{fig:gf1}).
Type I events have similar final states, although being produced by
different processes:  for example, the t-channel $\gamma  Z$ exchange
diagram or the initial state photon splitting in a \qq pair. The final
state is therefore two electrons and a \qq pair as in the signal.  Type
II background processes can arise  from detector acceptance and
resolution when some particles are lost down the beam pipe or in
detector cracks  or are misidentified. In this  category, one can
mention s-channel initial state radiation: when the photon is  lost in
the beam pipe 
and the boosted \qq pair detected, this event can be interpreted as a no
tag or antitag \gaga event. Four fermions channels where two of the
fermions are lost in the beam pipe region are also part of this category
of \gaga no tag events.  A series of dedicated cuts based on kinematic
constraints must be set to reject most of the background (largest
purity), although keeping most of  the signal (largest efficiency). 
\par
The LEP2 total cross sections are orders of magnitude smaller than those
found at the \Z peak. However, initial state radiations by emitting a
high energy photon, can shift down the centre of mass energy to the \Z
peak energy. In the following we concentrate on the study of this
background as the Type I backgrounds can be argued away easily: the
$\gamma  Z$ exchange processes are suppressed due to the Z propagator
while the other process leads typically to a very small hadronic mass at
large rapidity and leptons at large angle (which would be detected) or a
large missing energy.
\subsection{Tagging conditions and acceptances}
Typical detector acceptances and thresholds have been selected in order
to match an ``average" LEP experiment. All momenta and angles are
expressed in the laboratory frame. The events are selected by
antitagging in the following way: both scattered electrons or photons
have a polar angle $\theta \le 30$ mrad or an energy $E \le 1$ GeV. We
define now the acceptance cuts on the final state particle system\\ 
{\bf cut 0}: no further constraint applied to the final state with the
exception of the electron or photon tagging conditions.\\ 
{\bf cut 1}: Only
charged particles with $20^\circ \le \theta \le 160^\circ$, $p > 0.4$
GeV/c  and neutral particles with $10^\circ \le \theta \le 170^\circ$,
$E > 1$ GeV are accepted for the analysis. At least four charged
particles have to survive the cuts mentioned above to accept the
event.\\
The visible energy is calculated from
the invariant mass of the four-momentum obtained by
summing the four-momenta of all particles satisfying cut 1.
For jet search, a cone algorithm with a cone radius $R = \sqrt{ (\Delta\eta)^2
+(\Delta\phi)^2} = 1$ has been used. Only particles passing cut 1 enter
the jet analysis.
%
\subsection{Radiative return to the \Z cross-section}
Background simulation have been performed at $\sqrt{s}=175 $ \gev\
using \J \cite{jetset}.
\begin{figure}[hbtp]
\begin{minipage}{.495\linewidth}
\epsfig{figure=g2.epsi,height=8cm,angle=0}
\caption{\it Initial state radiated photon energy.}
\label{fig:g2}
\end{minipage}
\hfill
\begin{minipage}{.495\linewidth}
\epsfig{figure=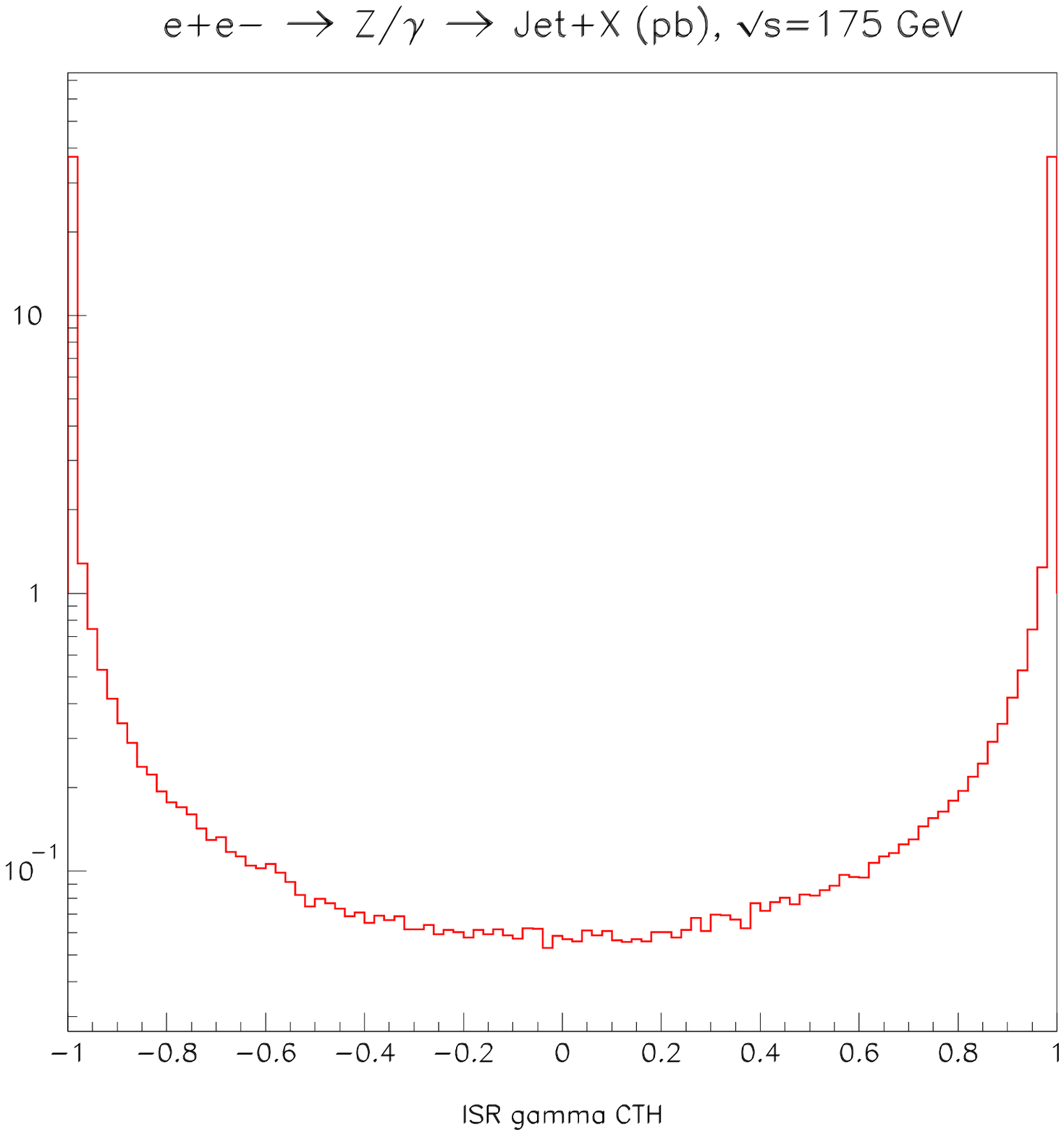,height=8cm,angle=0}
\caption{\it $\cos{\theta}$ distribution of initial state radiation photon.}
\label{fig:g3}
\end{minipage}
\vfill
\end{figure}
We show in Fig. \ref{fig:g2} the initial state radiated photon energy.
The low energy peak reflects the $1/E_\gamma$ behaviour of the
bremsstrahlung process, the high energy peak comes from the
$\sigma_o(\hat{s}) \propto 1/\hat{s}$  ($\hat{s}=(1-x_\gamma)s$)
singularity of the Born term as found in the hard radiative
cross-section formula:
\be
\frac{d\sigma}{dx_\gamma}= \frac{\alpha_{em}}{\pi}(\ln{\frac{s}{m_e^2}}
-1) \frac{1+(1-x_\gamma)^2}{x_\gamma}\sigma_0(\hat{s})
\ee
where $x_\gamma$ is the fraction of the beam energy carried by the real
photon and $\hat{s}$ is the invariant mass squared of the virtual
photon. The large peak, close to 64 \gev, is precisely due to the
so-called return to the \Z for $ E_\gamma = (s-M^2_Z)/2\sqrt{s}$. 
The $\cos{\theta}$ distribution of the photon is shown in Fig. \ref{fig:g3} 
where the forward and backward peaks reflect the cross section divergence
for collinear photon production.
If such a photon remains undetected, the boosted \qq pair system may appear as
a untagged \gaga event. Even worse, the photon can be identified as an
electron in the forward tagging detectors, this event would then be
selected as a one tag \gaga process. \par
\begin{figure}[hbtp]
\begin{minipage}{.495\linewidth}
\epsfig{figure=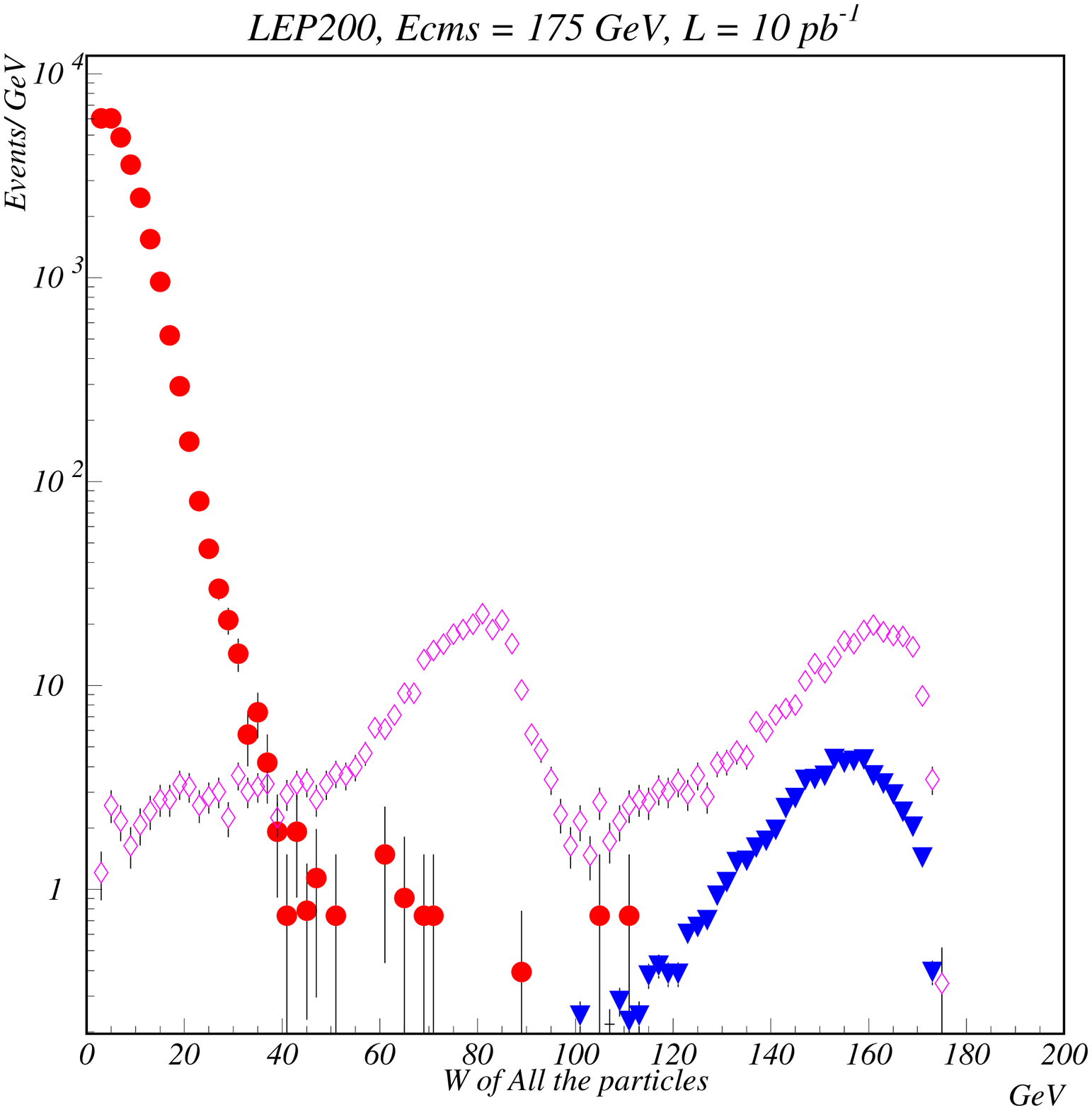,height=8cm,angle=0}
\caption{\it Visible invariant mass of 
$\gamma\gamma$(solid $\ \ $ dots), Z$\gamma$(open diamonds) 
and W$^{+}$W$^{-}$(solid $\ \ $ triangles).} 
\label{fig:g5}
\end{minipage}
\hfill
\begin{minipage}{.495\linewidth}
\vspace{0.1cm}
\epsfig{figure=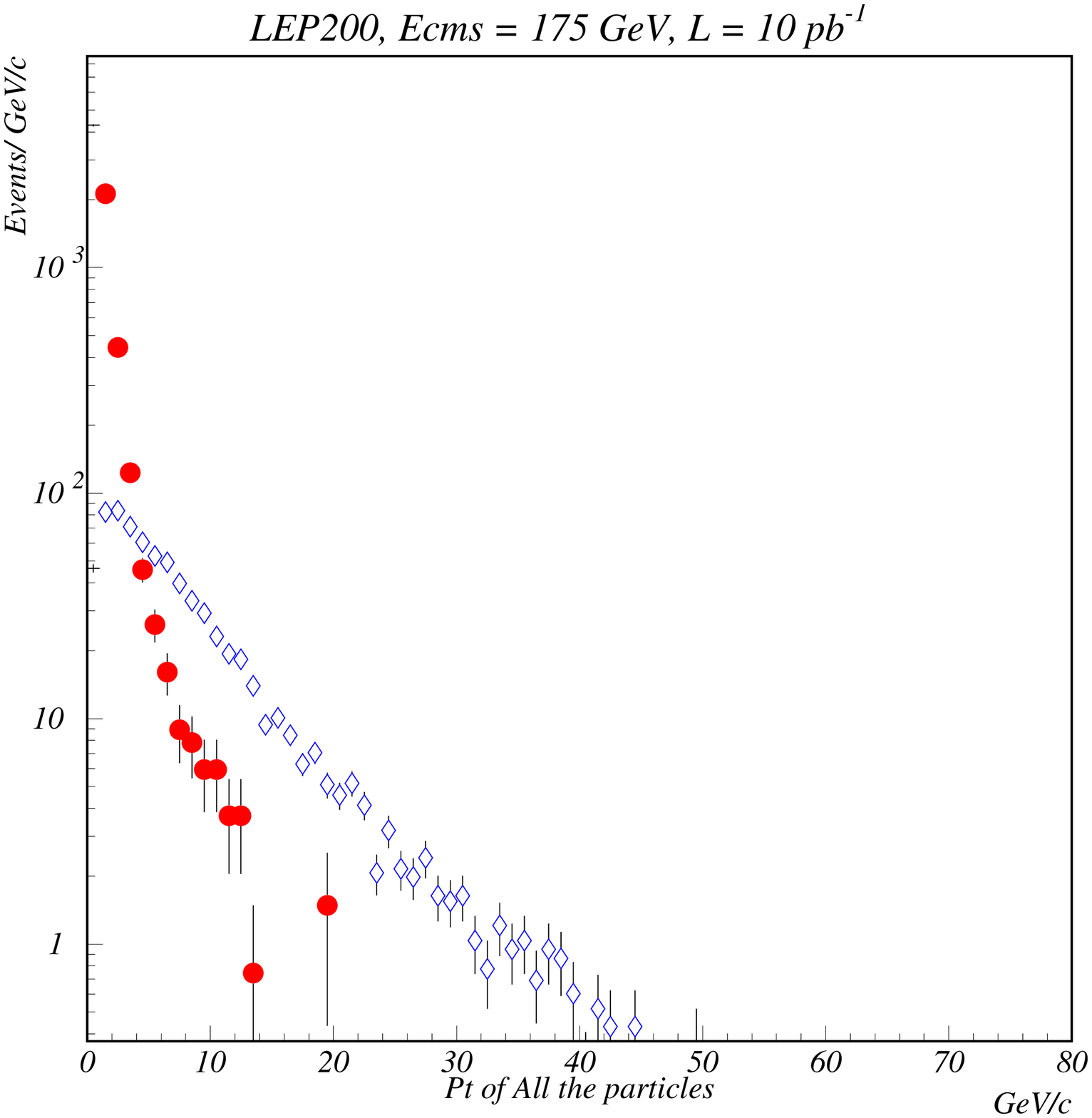,height=8cm,angle=0}
\caption{\it $\vec{P}^{vis}_\perp$ distributions for the signal and 
the background. The symbols have the same meaning as in the previous
figure.}
\label{fig:g6}
\end{minipage}
\vfill
\end{figure}

The next three figures display clearly the differences between the
signal and the radiative \Z production background. 
In these studies the charged hadrons 
angular acceptance  has been increased to 10$^{0}$ with a realistic 
track reconstruction efficiency as expected in DELPHI. 
The two-photon events are generated with TWOGAM and are compared to PYTHIA 
Z$\gamma$ and W$^{+}$W$^{-} \rightarrow \mbox{all decays}$ production 
contributions allowing initial state radiation. As expected
the signal is characterised by a rapidly falling  $W_{\rm vis} $
distribution (due to the Weisz\"acker-Williams convolution) while
the background exhibits a clear peak slightly below the \Z mass 
and the W$^{+}$W$^{-}$ channel shows up at very large invariant mass.
Similarly the $\vec{P}^{vis}_\perp$,  
$\vec{P}^{vis}_\parallel$ spectra from the signal are confined 
to low values while they have a long tail for the background 
(figs. \ref{fig:g5}, \ref{fig:g6}). 
The above features clearly dictate the following cuts to
reject the
background still retaining most of the genuine \gaga events:\\
{\bf cut 2}: the vector sum of the transverse
 momenta of all accepted particles satisfies
$\vec{P}^{vis}_\perp = \sum \vec{p}_\perp \le 10$ GeV/c.\\
{\bf cut 3}: the vector sum of the longitudinal
 momenta of all accepted particles satisfies
$\vec{P}^{vis}_\parallel = \sum \vec{p}_\parallel \le 20$ GeV/c.\\
{\bf cut 4}: the invariant mass calculated from the four momenta of
accepted particles satisfies $W_{\rm vis} \le 50$ GeV. \\
\begin{figure}[hbtp]
\begin{minipage}{.495\linewidth}
\epsfig{figure=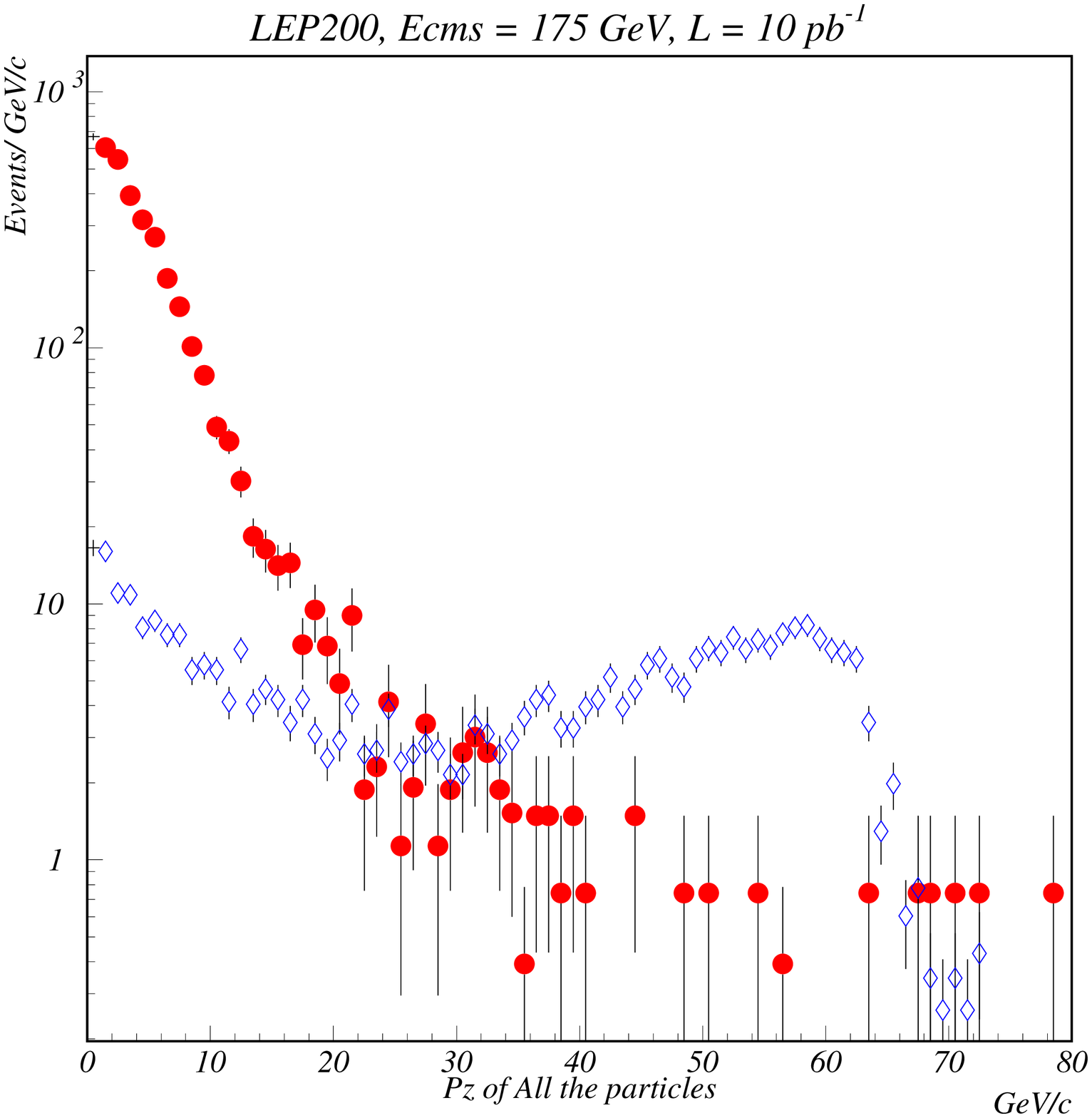,height=8cm,angle=0}
\caption{\it $\vec{P}^{vis}_\parallel$ distributions for the signal and 
the background.}
\label{fig:g4}
\end{minipage}
\hfill
\begin{minipage}{.495\linewidth}
\vspace{0.1cm}
\epsfig{figure=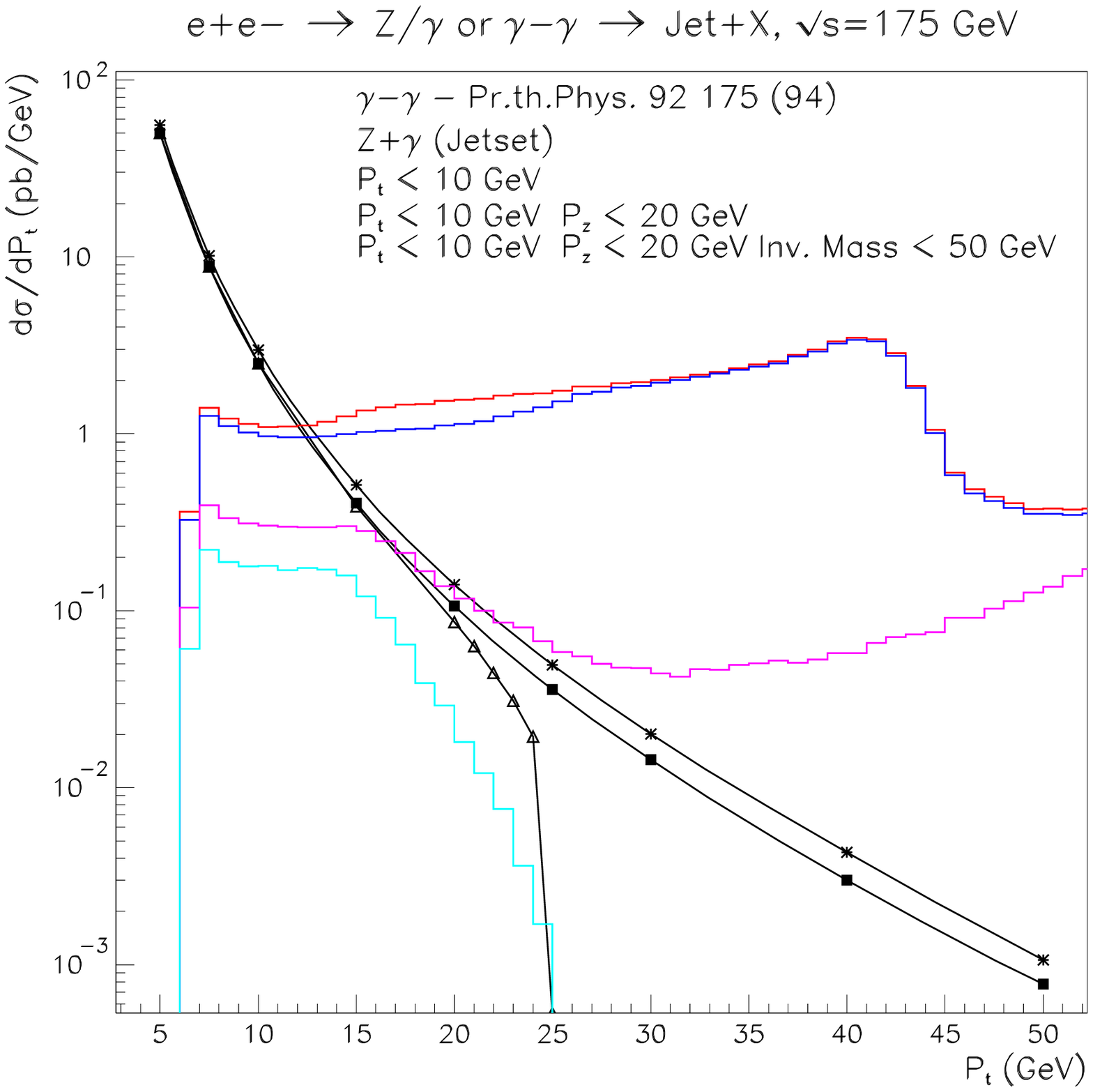,height=8cm,angle=0}
\caption{\it Signal and background \Pt distribution for various cuts.
The cross section is integrated over the jet rapidity range
$|\eta|=1$. See the text for explanations. }
\label{fig:g1}
\end{minipage}
\vfill
\end{figure}
%
Clearly these cuts will not affect the bulk of the \gaga events
($e.g.$ the total charm production cross section is hardly affected),
however they will certainly reduce the rate of rare events in the signal 
characterized by a large invariant mass: this is the case of large $p_T$ 
jet production since one has in general $W \ge 2 p_T$. 
This is illustrated in Fig. \ref{fig:g1} where  
the effect of the cuts on jet searches is discussed. 
Histograms represent the contribution of the \Z return background. 
The upper line
is for events satisfying cut 1 above, the next lower ones
are for cut 1+2, cut 1+2+3 and cut 1+2+3+4 respectively.
The solid lines are the results of an analytic calculation
in the leading-logarithm approximation at the partonic level
where the cuts are approximately implemented:
the top curve is for cut 1+2 (cut 2 is ineffective since
$\vec{P}^{vis}_\perp = 0$ by definition) and the lower two
curves are as above.
The \Pt threshold above which the background rate is larger than the signal
rate goes from 13 \gev\ (\Ptvis cut) up to 16 \gev\ (\Ptvis and \Pzvis).
The final visible invariant mass cut has a strong effect on the background but
reject also most of the signal over \Pt $> 20-25$ \gev. A more detailed 
study of the effects of the cuts on the signal 
is discussed in the ``Large-$p_T$ processes" section.
 

\section[Soft and semihard physics, and event structure]{Soft and 
semihard physics, and event structure
\label{ggso1}
{\protect 
\footnote{ A .\ Corsetti, R.\ Engel, F.\ Ern\'e, A.\ Finch, J.\ Field,
J.\ Forshaw, R.~Godbole, F.\ Kapusta,  G.\ Pancheri, J.\ Ranft, 
G.A.\ Schuler, V.\ Serbo, T.\ Sj\"ostrand, N.I.\ Zimin}}} 
%
%
%
%
%
%
Studies of minimum-bias physics and semihard interactions 
in two-photon events offer a good opportunity to investigate 
the high-energy behaviour of scattering amplitudes 
and the transition from perturbative to non-perturbative QCD. 
%
\subsection{Cross section predictions and general characteristics}
The photon, in its high-energy interactions with hadrons,
behaves very much like a hadron, however with cross sections
reduced strongly relative to pure hadronic cross sections. 
Similarly to a hadron, the photon both undergoes soft hadronic interactions
and has resolved hard interaction between its hadronic constituents and
the hadronic constituents of the target. Additionally, the photon has
a direct pointlike interaction with the hadronic constituents of
the target.

Even at high energies, many features of hadronic 
interactions of photons are dominated  by soft multiparticle production. 
Correspondingly, distributions measured in photoproduction are similar to
those obtained in purely hadronic interactions (provided, of course, 
these are taken at the same center-of-mass energy). 
\setlength{\unitlength}{1pt}
\begin{figure}[hbt]
\begin{picture}(484,198)(0,0)
\put(-10,0){
\setlength{\unitlength}{0.240900pt}
\ifx\plotpoint\undefined\newsavebox{\plotpoint}\fi
\sbox{\plotpoint}{\rule[-0.200pt]{0.400pt}{0.400pt}}%
\begin{picture}(944,826)(0,0)
\font\gnuplot=cmr10 at 10pt
\gnuplot
\sbox{\plotpoint}{\rule[-0.200pt]{0.400pt}{0.400pt}}%
\put(220.0,113.0){\rule[-0.200pt]{4.818pt}{0.400pt}}
\put(198,113){\makebox(0,0)[r]{$0.5$}}
\put(860.0,113.0){\rule[-0.200pt]{4.818pt}{0.400pt}}
\put(220.0,190.0){\rule[-0.200pt]{4.818pt}{0.400pt}}
\put(198,190){\makebox(0,0)[r]{$1$}}
\put(860.0,190.0){\rule[-0.200pt]{4.818pt}{0.400pt}}
\put(220.0,266.0){\rule[-0.200pt]{4.818pt}{0.400pt}}
\put(198,266){\makebox(0,0)[r]{$1.5$}}
\put(860.0,266.0){\rule[-0.200pt]{4.818pt}{0.400pt}}
\put(220.0,343.0){\rule[-0.200pt]{4.818pt}{0.400pt}}
\put(198,343){\makebox(0,0)[r]{$2$}}
\put(860.0,343.0){\rule[-0.200pt]{4.818pt}{0.400pt}}
\put(220.0,420.0){\rule[-0.200pt]{4.818pt}{0.400pt}}
\put(198,420){\makebox(0,0)[r]{$2.5$}}
\put(860.0,420.0){\rule[-0.200pt]{4.818pt}{0.400pt}}
\put(220.0,496.0){\rule[-0.200pt]{4.818pt}{0.400pt}}
\put(198,496){\makebox(0,0)[r]{$3$}}
\put(860.0,496.0){\rule[-0.200pt]{4.818pt}{0.400pt}}
\put(220.0,573.0){\rule[-0.200pt]{4.818pt}{0.400pt}}
\put(198,573){\makebox(0,0)[r]{$3.5$}}
\put(860.0,573.0){\rule[-0.200pt]{4.818pt}{0.400pt}}
\put(220.0,650.0){\rule[-0.200pt]{4.818pt}{0.400pt}}
\put(198,650){\makebox(0,0)[r]{$4$}}
\put(860.0,650.0){\rule[-0.200pt]{4.818pt}{0.400pt}}
\put(220.0,726.0){\rule[-0.200pt]{4.818pt}{0.400pt}}
\put(198,726){\makebox(0,0)[r]{$4.5$}}
\put(860.0,726.0){\rule[-0.200pt]{4.818pt}{0.400pt}}
\put(220.0,803.0){\rule[-0.200pt]{4.818pt}{0.400pt}}
\put(198,803){\makebox(0,0)[r]{$5$}}
\put(860.0,803.0){\rule[-0.200pt]{4.818pt}{0.400pt}}
\put(220.0,113.0){\rule[-0.200pt]{0.400pt}{4.818pt}}
\put(220,68){\makebox(0,0){$10$}}
\put(220.0,783.0){\rule[-0.200pt]{0.400pt}{4.818pt}}
\put(306.0,113.0){\rule[-0.200pt]{0.400pt}{2.409pt}}
\put(306.0,793.0){\rule[-0.200pt]{0.400pt}{2.409pt}}
\put(357.0,113.0){\rule[-0.200pt]{0.400pt}{2.409pt}}
\put(357.0,793.0){\rule[-0.200pt]{0.400pt}{2.409pt}}
\put(393.0,113.0){\rule[-0.200pt]{0.400pt}{2.409pt}}
\put(393.0,793.0){\rule[-0.200pt]{0.400pt}{2.409pt}}
\put(420.0,113.0){\rule[-0.200pt]{0.400pt}{2.409pt}}
\put(420.0,793.0){\rule[-0.200pt]{0.400pt}{2.409pt}}
\put(443.0,113.0){\rule[-0.200pt]{0.400pt}{2.409pt}}
\put(443.0,793.0){\rule[-0.200pt]{0.400pt}{2.409pt}}
\put(462.0,113.0){\rule[-0.200pt]{0.400pt}{2.409pt}}
\put(462.0,793.0){\rule[-0.200pt]{0.400pt}{2.409pt}}
\put(479.0,113.0){\rule[-0.200pt]{0.400pt}{2.409pt}}
\put(479.0,793.0){\rule[-0.200pt]{0.400pt}{2.409pt}}
\put(494.0,113.0){\rule[-0.200pt]{0.400pt}{2.409pt}}
\put(494.0,793.0){\rule[-0.200pt]{0.400pt}{2.409pt}}
\put(507.0,113.0){\rule[-0.200pt]{0.400pt}{4.818pt}}
\put(507,68){\makebox(0,0){$100$}}
\put(507.0,783.0){\rule[-0.200pt]{0.400pt}{4.818pt}}
\put(593.0,113.0){\rule[-0.200pt]{0.400pt}{2.409pt}}
\put(593.0,793.0){\rule[-0.200pt]{0.400pt}{2.409pt}}
\put(644.0,113.0){\rule[-0.200pt]{0.400pt}{2.409pt}}
\put(644.0,793.0){\rule[-0.200pt]{0.400pt}{2.409pt}}
\put(680.0,113.0){\rule[-0.200pt]{0.400pt}{2.409pt}}
\put(680.0,793.0){\rule[-0.200pt]{0.400pt}{2.409pt}}
\put(707.0,113.0){\rule[-0.200pt]{0.400pt}{2.409pt}}
\put(707.0,793.0){\rule[-0.200pt]{0.400pt}{2.409pt}}
\put(730.0,113.0){\rule[-0.200pt]{0.400pt}{2.409pt}}
\put(730.0,793.0){\rule[-0.200pt]{0.400pt}{2.409pt}}
\put(749.0,113.0){\rule[-0.200pt]{0.400pt}{2.409pt}}
\put(749.0,793.0){\rule[-0.200pt]{0.400pt}{2.409pt}}
\put(766.0,113.0){\rule[-0.200pt]{0.400pt}{2.409pt}}
\put(766.0,793.0){\rule[-0.200pt]{0.400pt}{2.409pt}}
\put(781.0,113.0){\rule[-0.200pt]{0.400pt}{2.409pt}}
\put(781.0,793.0){\rule[-0.200pt]{0.400pt}{2.409pt}}
\put(794.0,113.0){\rule[-0.200pt]{0.400pt}{4.818pt}}
\put(794,68){\makebox(0,0){$1000$}}
\put(794.0,783.0){\rule[-0.200pt]{0.400pt}{4.818pt}}
\put(880.0,113.0){\rule[-0.200pt]{0.400pt}{2.409pt}}
\put(880.0,793.0){\rule[-0.200pt]{0.400pt}{2.409pt}}
\put(220.0,113.0){\rule[-0.200pt]{158.994pt}{0.400pt}}
\put(880.0,113.0){\rule[-0.200pt]{0.400pt}{166.221pt}}
\put(220.0,803.0){\rule[-0.200pt]{158.994pt}{0.400pt}}
\put(45,458){\makebox(0,0){\large $\frac{dE_t}{d\eta}$  }}
\put(550,23){\makebox(0,0){\large W  [GeV] }}
\put(220.0,113.0){\rule[-0.200pt]{0.400pt}{166.221pt}}
\put(750,738){\makebox(0,0)[r]{p--p, NA--22}}
\put(794,738){\raisebox{-.8pt}{\makebox(0,0){$\Diamond$}}}
\put(318,182){\raisebox{-.8pt}{\makebox(0,0){$\Diamond$}}}
\put(750,693){\makebox(0,0)[r]{p--p, AFS}}
\put(794,693){\raisebox{-.8pt}{\makebox(0,0){$\Box$}}}
\put(361,197){\raisebox{-.8pt}{\makebox(0,0){$\Box$}}}
\put(447,217){\raisebox{-.8pt}{\makebox(0,0){$\Box$}}}
\put(750,648){\makebox(0,0)[r]{$\bar p-p$, UA1}}
\put(794,648){\circle{24}}
\put(593,312){\circle{24}}
\put(631,335){\circle{24}}
\put(673,366){\circle{24}}
\put(707,389){\circle{24}}
\put(736,404){\circle{24}}
\put(766,435){\circle{24}}
\put(781,443){\circle{24}}
\put(750,603){\makebox(0,0)[r]{$\gamma$--p, H1}}
\put(794,603){\circle*{24}}
\put(583,305){\circle*{24}}
\put(750,558){\makebox(0,0)[r]{$\gamma$--$\gamma$, PHOJET}}
\put(772.0,558.0){\rule[-0.200pt]{15.899pt}{0.400pt}}
\put(220,205){\usebox{\plotpoint}}
\multiput(220.00,205.58)(1.396,0.497){59}{\rule{1.210pt}{0.120pt}}
\multiput(220.00,204.17)(83.489,31.000){2}{\rule{0.605pt}{0.400pt}}
\multiput(306.00,236.58)(1.244,0.498){89}{\rule{1.091pt}{0.120pt}}
\multiput(306.00,235.17)(111.735,46.000){2}{\rule{0.546pt}{0.400pt}}
\multiput(420.00,282.58)(0.912,0.499){187}{\rule{0.828pt}{0.120pt}}
\multiput(420.00,281.17)(171.281,95.000){2}{\rule{0.414pt}{0.400pt}}
\multiput(593.00,377.58)(0.879,0.499){211}{\rule{0.803pt}{0.120pt}}
\multiput(593.00,376.17)(186.334,107.000){2}{\rule{0.401pt}{0.400pt}}
\put(794,558){\circle*{12}}
\put(220,205){\circle*{12}}
\put(306,236){\circle*{12}}
\put(420,282){\circle*{12}}
\put(593,377){\circle*{12}}
\put(781,484){\circle*{12}}
\end{picture}}
\put(230,0){\epsfig{file=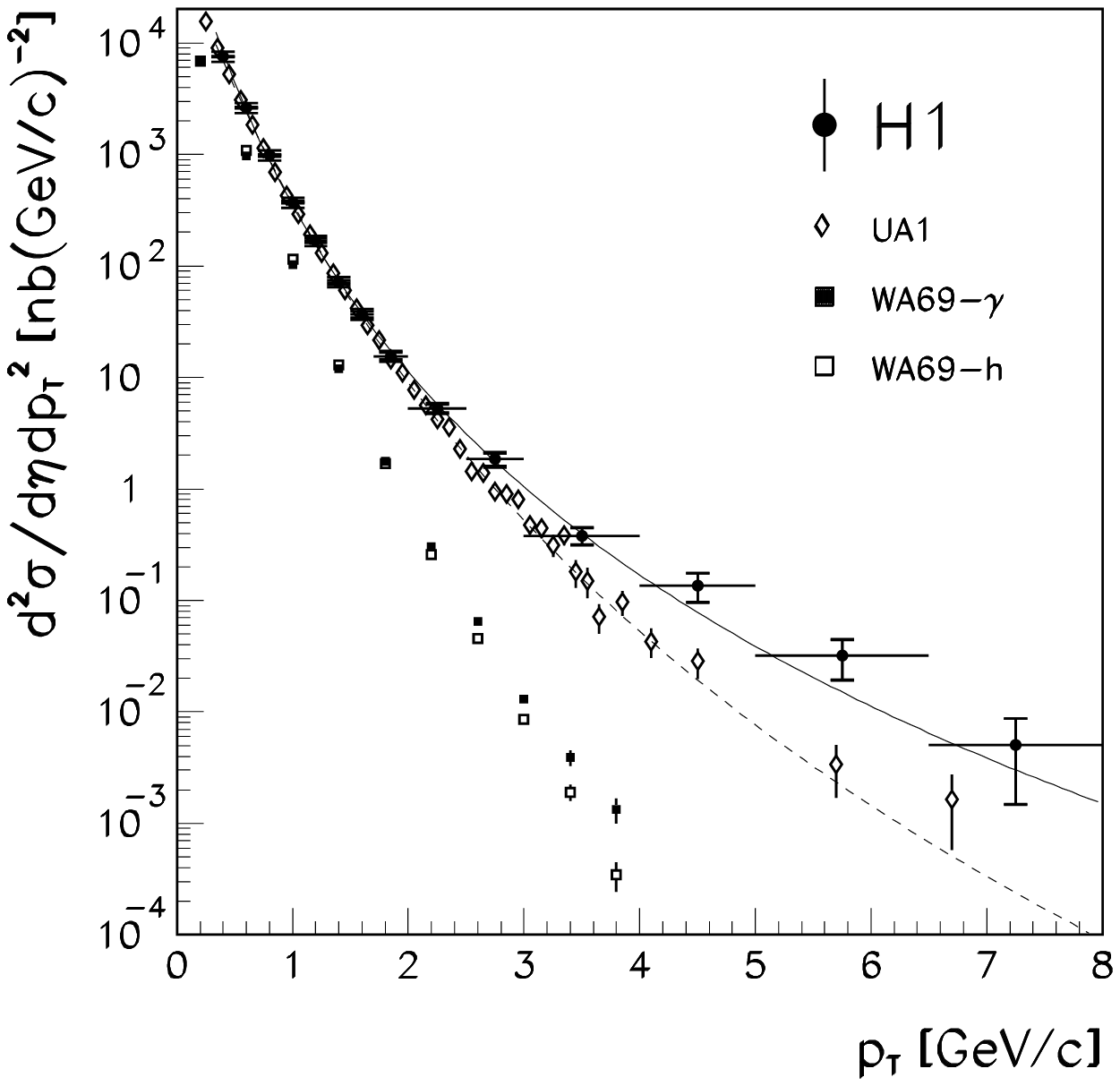,height=7.0cm,width=7.0cm}}
\end{picture}
\vspace*{-0.5cm}
\caption[]{ \em
(a,left)
The transverse energy (in GeV)
per unit of pseudorapidity in the central
region (i.e.\ at $\eta$ = 0)
as a function of the hadronic CM energy $W$. Data from
photoproduction are compared with data from hadron-hadron
collisions; from \protect\cite{Aid95a}. For $\gamma\gamma$ collisions the
\protect{\sc Phojet} predictions are shown.
%
(b right) 
Charged hadron differential cross sections for photon-hadron
scattering compared to the shape of hadron-hadron scattering:
the $\protect \gamma$p data from H1 at $\sqrt s$ = 200 GeV exceed in
the high $\protect p_{\perp}$ region the 
\protect$\bar{p}p$ data from UA1 at the same
energy. The results from a fixed target experiment (WA69) at
$\sqrt s$ = 18 GeV show a similar difference between 
$\protect \gamma$p and hadron-hadron data; from \protect\cite{erdmann}.}
\label{eteta}
\end{figure}
This is nicely illustrated  in Fig.~\ref{eteta} for the central
transverse energy density  and the one-particle inclusive $p_T$
spectrum. It is only at high $p_T$  that photon-induced reactions differ
because of the photon's pointlike  interactions and its correspondingly
harder parton  distribution functions (PDF).  Based on these
observations we can safely predict that  minimum-bias physics of
$\gamma\gamma$ interactions will follow that of  $\gamma p$ or $pp$
interactions. At high $p_T$, the spectra should become  harder when
going from $pp$ to $\gamma p$ to $\gamma\gamma$ interactions. 

In view of what we said above, any model that aims at an complete
description of $\gamma\gamma$ interactions should better successfully 
describe the wealth of data taken in hadronic collisions, notably 
at $p\bar{p}$ colliders. Reactions to be modelled include elastic 
scattering, diffractive dissociation and hard, perturbatively calculable 
interactions. On top of that, unitarity constraints 
have to be incorporated implying, in general, the existence 
of multiple (soft and hard) interactions (for a review and references 
see e.g.~\cite{Schuler:1994a}). The SaS model
(Schuler and Sj\"ostrand) \cite{Schuler:1994b} has been implemented
in  {\sc Pythia} \cite{jetset} while the  DPM (Dual-Parton-Model) 
\cite{Capella:1994}
has been extended (Engel and Ranft) to $\gamma p$ \cite{Engel95a}
and \gaga reactions \cite{Engel95d} in {\sc Phojet}.
Minimum-bias physics in $pp$, $\gamma p$, and $\gamma\gamma$ 
collisions is currently being improved in {\small HERWIG} by the inclusion 
of multiple hard scatterings \cite{Butter:1995}.

These event generators and the physics of the corresponding
models are described in detail in the ``Event generators" chapter. 
Here we discuss only some differences among the three models
which have been used in Fig.~\ref{ggxtot}.

To extend the description of $pp$ interactions to $\gamma p$ (and 
$\gamma\gamma$) ones it is convenient to represent the physical photon
as the superposition \cite{Schuler:1993}
\begin{equation}
|\gamma \rangle =  
   \sqrt{Z_3}\, |\gamma_B \rangle 
  +  P^\gamma_{had}\, |\gamma_{had} \rangle
\equiv  
   \sqrt{Z_3}\, |\gamma_B \rangle 
  +  \frac{e}{f_{low}}\, |q\bar{q}_{low} \rangle
  +  \frac{e}{f_{high}}\, |q\bar{q}_{high} \rangle
\ ,
\label{gammadeco}
\end{equation}
where the (properly normalized) first term describes the pointlike 
interaction of the photon. The spectrum of hadronic fluctuations of 
the photon is split into a low- and a high-mass part, separated by some
scale $Q_0$. Both contributions can, in general, undergo soft and 
hard interactions. The soft interactions are mediated by Pomeron/Reggeon 
exchange whose amplitudes can be inferred from the ones of $pp$ 
interactions assuming photon--hadron duality. Hard interactions 
are those that contain at least one hard scale and can be expressed 
in terms of the ``minijet'' cross section $\sigma_{jet}(s;p_{Tmin})$, 
i.e.\ the cross section for perturbatively calculated partonic $2 
\rightarrow 2$ scatterings above a $p_T$ cutoff $p_{Tmin}$. 
Again, unitarization leads to multiple (soft and hard) scatterings. 

In a most na\"\i ve scenario, the probability $P^\gamma_{had} = P(\gamma 
\rightarrow q\bar{q})$ is taken to be constant. At high energies, 
this cannot, however, be correct \cite{Schuler:1993} since the
contribution from high-mass hadronic fluctuations becomes important. 
These are perturbatively calculable and lead to a logarithmic 
increase of $P^\gamma_{had}$ with the hard scale $Q \sim p_T$, 
$(e/f_{high})^2 \propto \ln(Q^2/Q_0^2)$. In the SaS 
approach \cite{Schuler93b}, 
most parameters, in particular the coupling of the low-mass part 
of (\ref{gammadeco}) are determined using VMD-type arguments. 
The only two additional parameters in the extension from $pp$ to 
$\gamma p$ collisions, namely $Q_0$ and the $p_T$ cutoff 
for the hard cross section originating from the high-mass part 
($p_{Tmin}^{anom}$)
were fixed by low-energy $\gamma p$ data, prior to the HERA data. 
Elastic and diffractive cross sections as well as minimum-bias distributions 
were succesfully predicted \cite{Schuler:1994a}. The prediction for the 
$\gamma\gamma$ total cross section is shown in Fig.~\ref{ggxtot}. 
\setlength{\unitlength}{1pt}
\begin{figure}[hbt]
\begin{picture}(484,198)(0,0)
\put(-10,0){\input{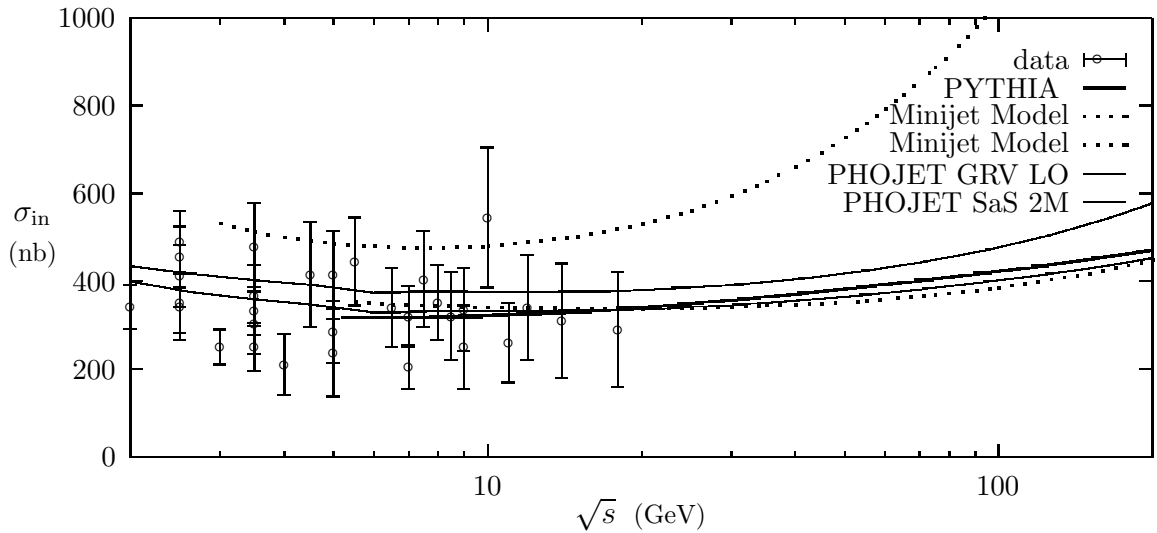}}
\end{picture}
\vspace*{-0.5cm}
\caption[]{ \em
Inelastic  photon-photon cross sections calculated in the SaS approach 
\protect\cite{Schuler93b} implemented in \protect{\sc Pythia},
the DPM model implemented in \protect{\sc Phojet} \protect\cite{Engel95a}, 
and an eikonalized minijet model \protect\cite{Corsetti95}
compared with data. The two curves
from  \protect{\sc Phojet} were calculated
using the GRV LO photon structure
function \protect\cite{GRVphot} (upper curve)
and the SaS 2M photon structure function
\protect\cite{SaS} (lower curve).
The two curves according to the unitarized minijet model are the
highest and the lowest prediction presented in
\protect\cite{Corsetti95} for values of the parameters compatible
with the photoproduction data.
}
\label{ggxtot}
\end{figure}

The DPM approach extended to $\gamma p$ collisions \cite{Engel95a} 
(in the  {\sc Phojet} event generator)  differs in 
several important 
aspects  from the SaS approach. 
The 
unitarization requirements
are obeyed  by strictly sticking to the eikonal approach. This leads 
to\footnote{ This also leads to   
another difference compared to {\sc Pythia}, which is already present for 
purely hadronic collisions, namely the generation of events containing 
multiple soft interactions in combination with any number 
(including zero) of hard interactions.}
multiple partonic scatterings also for high-mass photonic states. 
Furthermore the  
probabilities  $e^2/f_{low}^2$ and $e^2/f_{high}^2$ as well
as the Pomeron and Reggeon  coupling constants and effective intercepts
have been determined by  fits to data on the total photoproduction cross
section and the cross  section for quasi-elastic $\rho^0$ production.
Once these parameters are  fixed, $\gamma\gamma$ collisions can be
predicted without further  new parameters \cite{Engel95d}. The predicted
rise of $\sigma_{tot}^{\gamma\gamma}$  is shown in Fig.~\ref{ggxtot}. It
is governed by the small-$x$ behaviour  of the PDF of the photon. 

The eikonalized mini-jet model is well described in the literature 
(see e.g.\ \cite{drgoSG,Schuler:1994a}). In addition to the above-mentioned 
parameters such as $P^\gamma_{had}$ and $p_{Tmin}$ the predictions of 
this model depend also on $\rho(b)$, the distributions of the photonic 
partons in the impact-parameter space. The
new feature in the calculation in \cite{Corsetti95} is
that  for $\rho (b)$ 
they use the Fourier transform of the partonic transverse
momentum distribution instead of the Fourier transform of the pionic
form factor which is normally the case. The former has recently been
measured \cite{zeusgod} and has the form, in agreement with the
expectations of perturbative QCD,
$d N_{\gamma} / d k_t^2 = 1 / ( k_t^2 + k_0^2 )$
with $k_0 = 0.66 \pm 0.22 $ GeV. Interestingly, the normal usage of pionic 
form factor corresponds to $k_0 = 0.735$.  Predictions of the model are
shown by the dotted lines in Fig.~\ref{ggxtot}. 
 
\subsection{Production of hadrons and jets}
%
In order to illustrate characteristic differences and similarities
between $\gamma\gamma$, $\gamma p$, and $pp$ collisions we first 
show comparisons at fixed CM energy. Since elastic
hadron-hadron collisions usually are excluded in studies of 
inclusive secondary distributions, we also exclude the analogous ones for 
the photon-induced reactions, i.e.\ the quasi-elastic
diffractive channels $\gamma \gamma \rightarrow V + V'$, 
$\gamma p \rightarrow V + p$ ($V = \rho , \omega ,\phi$) 
but we include  all other diffractive processes.
 
\begin{figure}[hbt]
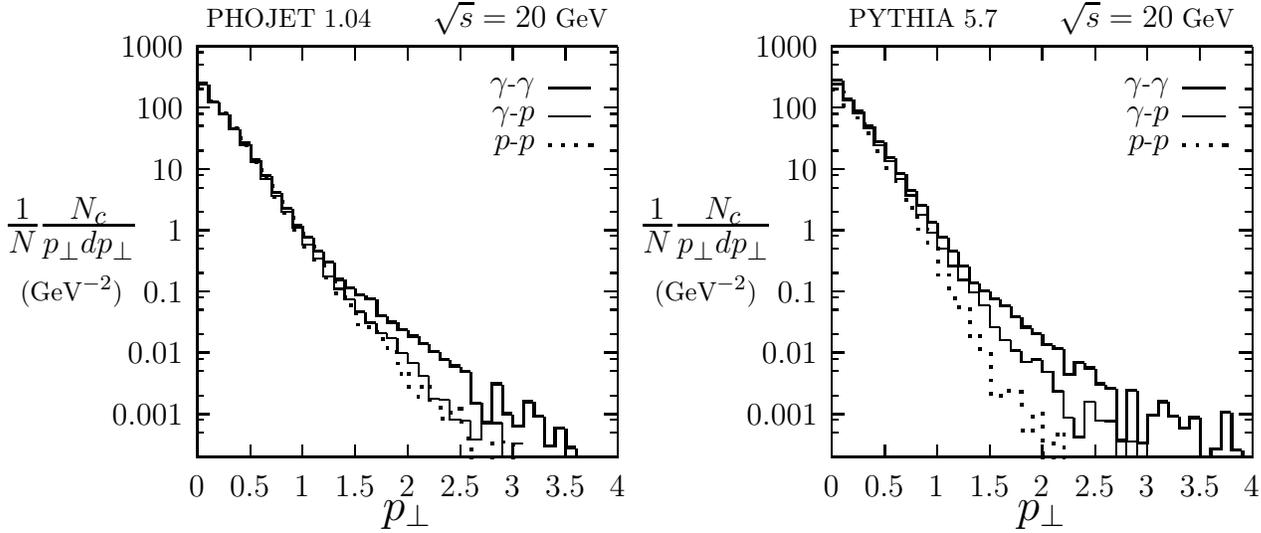

\begin{picture}(484,198)(0,0)
\put(-10,0){\input{dndpt20dd.pic}}
\put(230,0){\input{dndpt20ll.pic}}
\end{picture}
\vspace*{-0.8cm}
\caption{ \em
Comparison at the collision energy $\protect\sqrt s$ = 20 GeV of
 the transverse momentum distribution in invariant form for all
  charged hadrons produced in
  $pp$, $\gamma p$ and $\gamma\gamma$
 collisions. The calculation was done with
 {\sc Phojet} (left) and {\sc Pythia} for inelastic collisions.
\label{dndpt20} }
\end{figure}
First we show the transverse momentum distribution in Fig.~\ref{dndpt20}. 
Both {\sc Pythia} and {\sc Phojet} show a similar behaviour and
agree very well with the behaviour of the data in Fig.~\ref{eteta}b.
In fact the distributions from both models for $\gamma\gamma$ interactions
are very similar, differences between the two models appear mainly
for $p p$ collisions. These differences are probably due to the
use of different parton distribution functions and cutoffs for minijets.
\begin{table}[htb]
\begin{center}
\begin{tabular}{|c|c|c|c|c|c|c|}
\hline
      $\sqrt s$     & 10 & 10 & 10 & 20 & 20 & 20
\\ \hline
Quantity      &$pp$  & $\gamma p$ &$\gamma\gamma$ &
$pp$ & $\gamma p$ &$\gamma\gamma$
\\ \hline
 $n_{ch}$             &6.39   &6.38    &6.80     &9.17   &9.15 & 9.64
\\ \hline
 $n_{\pi^-}$          &1.94   &2.28   &2.76     &3.10   &3.40 & 3.92
\\ \hline
$n_{\bar p}$          &0.06    &0.10     &0.14    &0.122   &0.17 & 0.22
\\ \hline
$<p_{\perp  ch}>_{centr. \eta}$ [GeV/c] &0.36   &0.36   &0.39
   &0.37 &0.38    &0.40
\\ \hline
\end{tabular}
\end{center}
\caption{
\label{GGRJtable1} \em
Comparison of average quantities characterizing hadron
production in nondiffractive
$pp$, $\gamma p$ and $\gamma\gamma$
collisions according to {\sc Phojet}
at CM energies between 10 and 20 GeV.}
\end{table}
As expected, at low $p_T$, the distributions of $\gamma\gamma$, $\gamma p$, 
and $pp$ collisions are very similar, while the 
fraction of hard interactions in minimum bias interactions rises from
$pp$ to $\gamma p$ to $\gamma\gamma$ collisions. 
The reason for this is the direct
photon interaction and the fact that the photon structure
function is considerably harder than the proton one. 
In \gaga collisions it is easy to observe
already with moderate statistics hadrons with transverse
momenta approaching to the kinematic limit.

However, the differences in the hard scatterings hardly 
influence such average properties of the collision as average
multiplicities or even average transverse momenta. This can be
seen from Table \ref{GGRJtable1}, where we collect some average quantities
characterizing  nondiffractive
$pp$, $\gamma p$ and \gaga
collisions in {\sc Phojet} at CM energies of 10 and 20 GeV. The total
and charged multiplicities at all energies are rather similar to
each other in all channels. The differences in the
multiplicities of non-leading hadrons like $\pi^-$ and $\bar
p$ are more significant and we find them at all energies rising
from $p p$ to $\gamma p$ to $\gamma\gamma$ collisions.
Also the average transverse momenta rise as expected from
$pp$ to $\gamma p$ to $\gamma\gamma$.

\begin{figure}[htb]
\begin{picture}(484,170)(0,0)
\put(70,0){\input{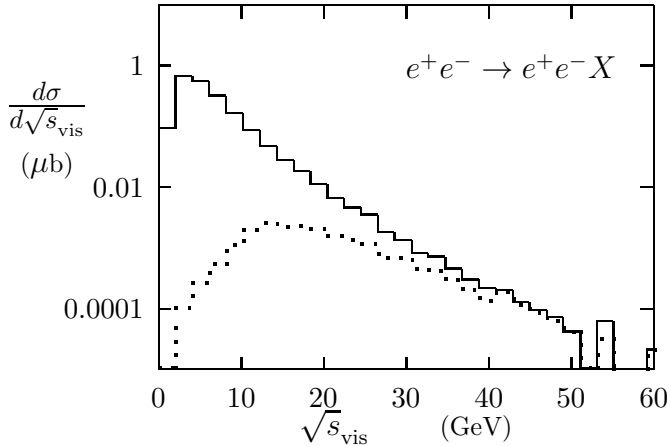}}
\end{picture}
\vspace*{-0.8cm}
\caption{ \em
Cross section at $\protect\sqrt{s} = 175\,$GeV
as function of the visible $\gamma\gamma$ CM
energy with cuts 1+2+3 on the final state system (full line)
and for events with jets after the application of cut 1 (dotted line),
(calculated with {\sc Phojet}; the cuts are defined in 
sec.~\protect\ref{ggtag}).}
\label{evislep} 
\end{figure}

%
%
Next we consider an example of 
hadron and jet production in $e^+e^-$ collisions. 
In Fig.~\ref{evislep} 
the $e^+ e^- \rightarrow e^+ e^- X$ cross section is
shown as a function of the visible photon-photon energy.
Also shown in this Figure is the cross section for events with
jets ($p_T^{jet} > 5\ \mbox{GeV/c}$). We predict that nearly all 
events have jets at large $W_{vis}$.

\begin{figure}[htb]
\begin{picture}(484,227)(0,0)
\put(-10,0){\epsfig{file=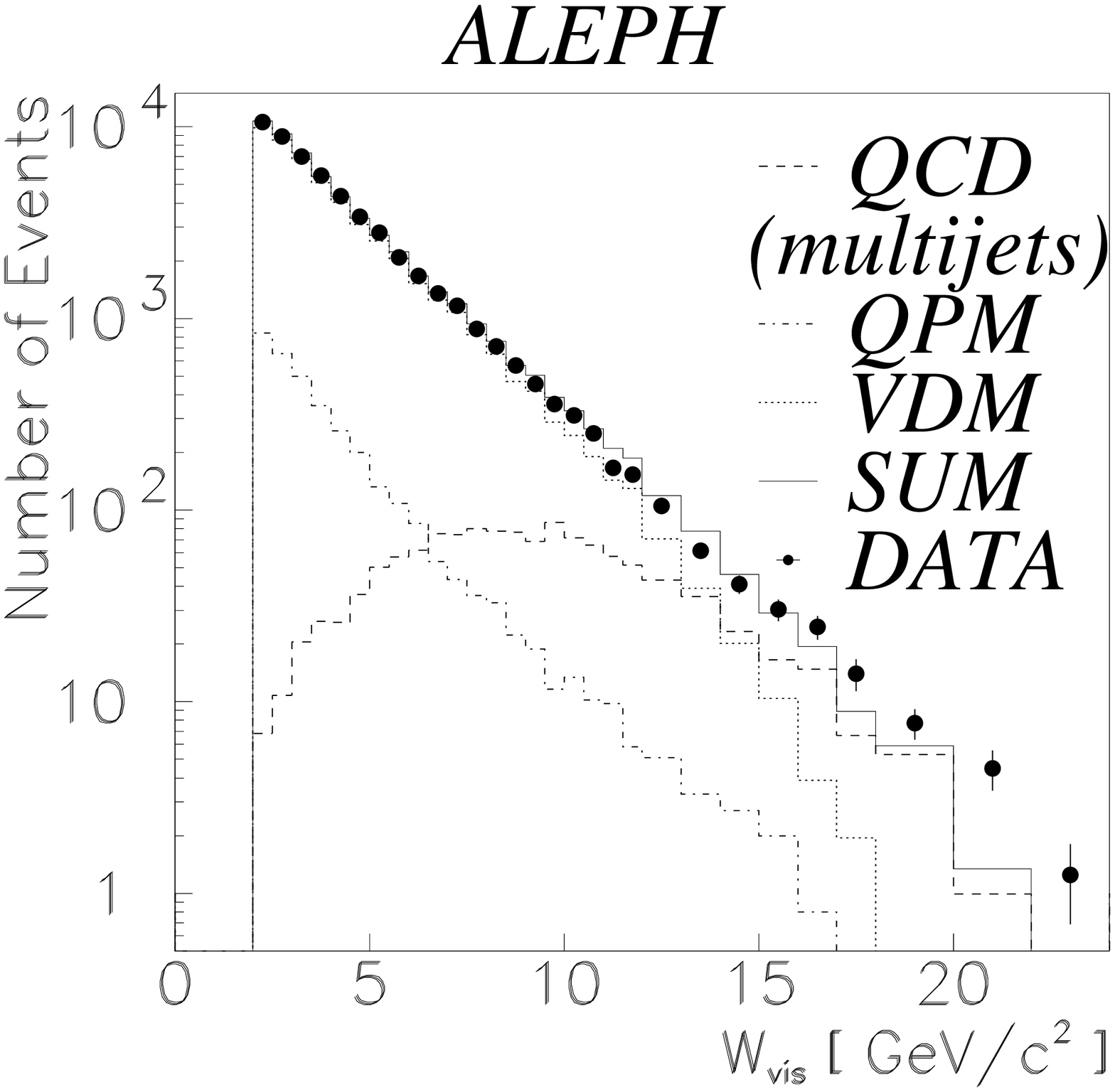,height=8.0cm}}
\put(230,0){\epsfig{file=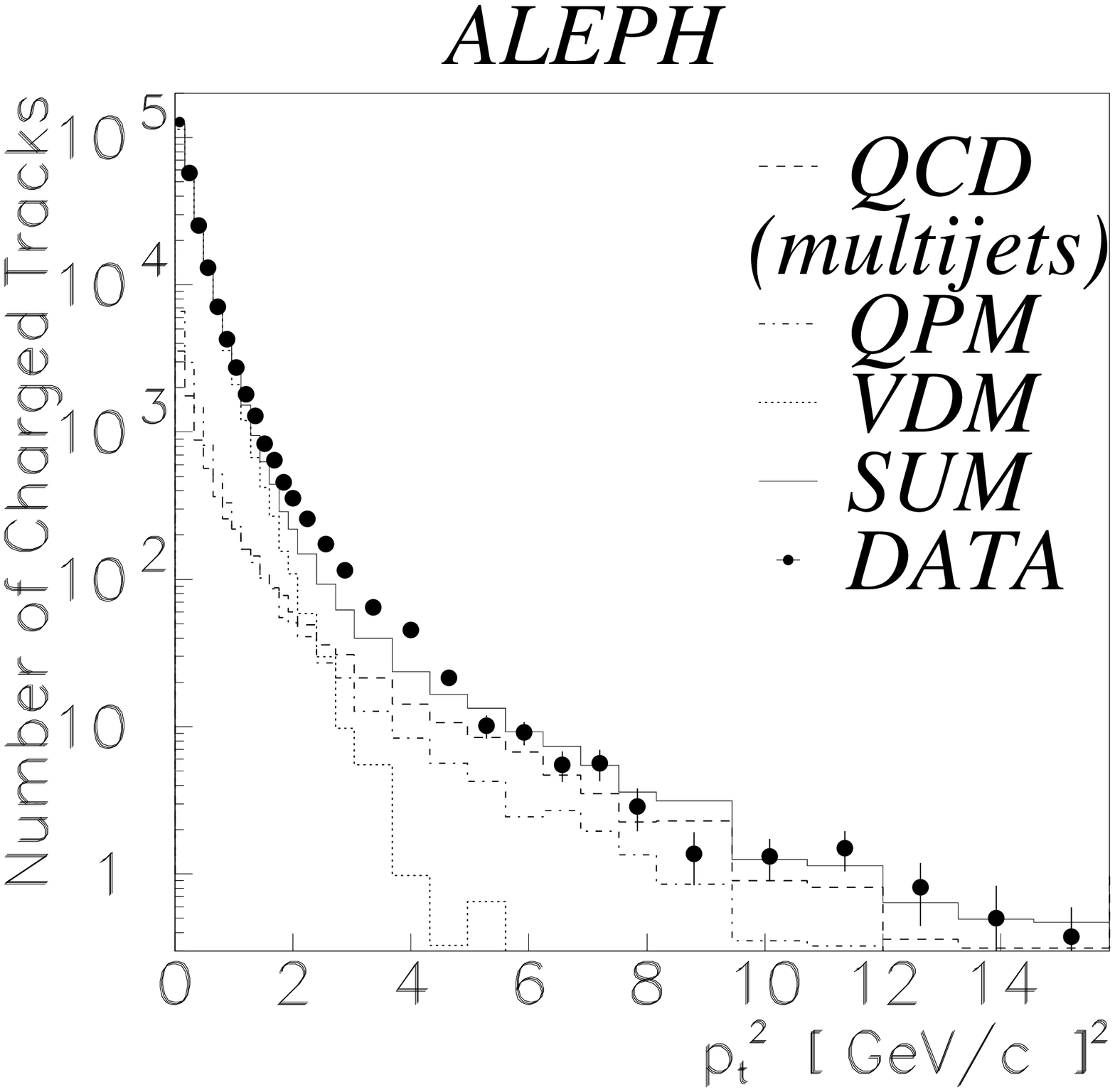,height=8.0cm}}
\end{picture}
\vspace*{-1.2cm}
\caption[]{\em 
(a, left) Visible energy distribution.
(b, right) Transverse momentum distribution.}
\label{fig:softW}
\end{figure}

Studies at LEP1 have started recently showing 
that measurements of minimum bias events are indeed possible. 
For example, Fig. \ref{fig:softW} shows that small $p_T$ values
are accessible. 
No detailed comparison between the multipurpose Monte-Carlo
generators such as {\sc Phojet} and {\sc Pythia} 
has yet been done with the existing 
LEP1 data. Instead, we show here the Aleph \cite{Aleph_paper} results 
on the $\mathrm{W}_{vis}$ distribution and on the charged track
$p_T$ distribution for  minimum bias untagged events compared
to a specific in-house $\gamma\gamma$ event generator.
The data has been modelled by a sum of four contributions 
(Fig.~\ref{fig:softW}). The bulk of
the data is described by a VMD model which includes only limited $p_T$
with respect to the $\gamma\gamma$ direction, and has been tuned to the
data. At high $p_T$ and $\mathrm{W}_{vis}$ the data require additional 
contributions from QCD. These are separated into the direct process
(labelled QPM), and the sum of single- and double-resolved processes
(labelled QCD--multijets).

\subsection{Measurement of the \gaga hadronic
cross section}
%
In untagged two-photon events the photons are nearly on-shell.
The measurement of $\sigma_{tot}^{\gamma\gamma}$
is not so much limited by statistical accuracy
as by a precise description of competing processes,
such as beam-gas or beam-wall scattering, annihilation reactions, and a
careful simulation of the process itself.
The invariant mass from beam-gas scattering is effectively limited
to $\approx 15\,$GeV by the kinematics of
electron scattering off slowly moving nucleons. It can be suppressed by
requiring the event vertex to be at the interaction point.
The contribution from the annihilation process ``$e^+e^-\rightarrow
\hbox{ hadrons}$'' can be reduced by requiring the average rapidity of
the observed hadrons to be non-zero, while in the ``$e^+e^-\rightarrow
\hbox{hadrons}+\gamma$'' process the average rapidity of the hadrons is
kinematically related to the invariant mass of the hadrons. Furthermore,
the different s-dependence of various processes helps their
identification. \\ 
The event characteristics are dominated by the Vector
Dominance process, which accounts for about 70 to 80\% of the events
according to present models. Much information from two-photon events
disappears along the holes in the detectors around the beams. As a
hardware solution to this problem is not envisaged by the LEP
experiments, one has to cope with partial event information; 30 to 50\%
of the hadronic energy is lost in the forward region, and the fraction
increases with the \gaga invariant mass and LEP beam energy. Some useful
information can be collected from small-angle electromagnetic detectors.
At low \gaga invariant mass the measurements are limited by trigger
requirements. Unfolding of the true \gaga mass distribution from the
data with the help of event simulation is imperative, and it gives rise
to model dependence in the cross section. Results are expected up to
about 80 GeV,
somewhat below the $Z^0$ mass.
\begin{figure}[htb]
\begin{picture}(484,250)(0,0)
\put(60,-90){\epsfig{file=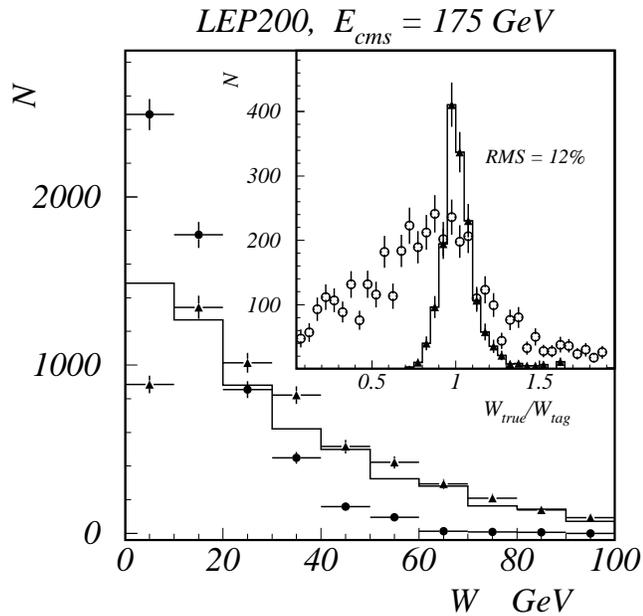,height=10.0cm}}
\end{picture}
\vspace*{-1.5cm}
\caption{\em Distribution in the number of events according to the
various hadronic mass determinations.}
\vspace*{-0.4cm}
\label{fig:zimin3} 
\end{figure}

Interesting studies can be performed at LEP2 with double tag
$\gamma\gamma$ events using Very Small Angle Tagger (VSAT) detectors
due to the high enough cross section for the polar angle
region (2~-~15~mrad) covered. 
The DELPHI collaboration had already obtained some new 
results studying the single
tag events in their VSAT \cite{DELVSAT}
(similar detectors are currently being used in OPAL or 
upgraded in ALEPH and L3). 
The double tag mode is attractive due to the possibility of a
direct measurement of both the hadronic invariant mass produced
and the absolute momentum transfers squared for both photons.
Taking into account the experimental constraints and  the efficiency 
of the hadronic system registration, for 500 pb$^{-1}$ of data,
we expect about of 6000 VSAT double tagged events
per experiment. \\
In Fig. \ref{fig:zimin3} the ${\gamma\gamma}$ invariant mass 
distributions
are illustrated for the visible $W_{vis}$ (black circles), 
the true $W_{true}$ (solid line) and $W_{tag}$ (triangles) 
reconstructed from the tag measurements.
Also displayed is the ratio $W_{true}/W_{tag}$ 
shown for two regions of $W_{true}$, above 40 GeV (triangles 
and solid line) and below (white circles).  
It appears that for $W$ above 40 GeV, there is
a good agreement between the $W_{tag}$ and $W_{true}$. This 
means that one expects reliable results because there is little
need to apply unfolding procedures in that region.
The conclusion from this picture is that the 
extraction of the total ${\gamma\gamma}$ cross-section 
$\sigma_{\gamma\gamma}$ is possible in a wide region indeed: 
even above 80-90 GeV the statistics available is greater
than 100 events.   

Summarizing the above we can say that  LEP2 will open
a new opportunity to obtain reliable values of total
$\sigma_{\gamma\gamma}$ for a ${\gamma\gamma}$ central energy up to 
$80\,$GeV  ($100\,$GeV) in the untagged (double-tag) case. This will 
increase five times the presently accessible range
\cite{TPCTWG} (see Fig. \ref{ggxtot}). The extrapolation of 
the cross section for virtual photons in the double-tag case to 
$Q^2 = P^2 = 0$ should be safe since the average virtualies 
($\sim 0.5\,$GeV$^2$) are much lower than where HERA sees a strong 
$x$ dependence in $F_2^p(x,Q^2)$. The uncertainty in the extrapolation will 
further be reduced by future HERA measurements at low $Q^2$. 
Finally, double-tag events provide a non-trivial check of 
hadronization models (see also sec.~3 in \cite{gggen}). 

\subsection{Semihard quasidiffractive processes}
%
We consider now the 
exclusive or semi-exclusive production of neutral
mesons $M$
\begin{equation}
\gamma \gamma \to M \; M', \hspace*{3cm}
\gamma \gamma \to M + X
\label{1}
\end{equation}
in the semihard region
\begin{equation}
W^2 \gg p_\bot^2 =|t| \gg \mu^2,\;\; W^2=(p_{1\gamma} +
p_{2 \gamma})^2, \  \ t=(p_{1\gamma} - p_M)^2, \ \
\mu = 0.3 \, \mbox{GeV}.
\label{3}
\end{equation}
Here $M$ is a vector ($V = \rho^0, \omega, \phi, ... \Psi $), or
pseudoscalar ($P= \pi^0, \eta, \eta'$), or tensor ($T= a_2, f_2,
f^\prime$) neutral meson and $X$ is a hadron system with not too large
invariant mass $M_X^2 \stackrel{<}{\sim} |t|$. Such processes are
discussed in a number of papers (see, for example, 
\cite{CZ,GPS87,FR}).


The condition $|t| \ll W^2$ determines these processes as 
quasi-Regge ones. 
In the production of the vector meson $V$ vacuum quantum numbers are 
transferred from the photon to the meson. The energy dependence 
of these processes is determined by the Pomeron singularity. In
the production of pseudoscalar $P$ or tensor $T$ mesons
the corresponding singularity is the odderon.
In perturbative QCD (pQCD), the Pomeron (PP) and the odderon (PO)
have the same status,
however, the current data do not indicate unambiguously the odderon
contribution to the total cross sections. 
%

In the lowest order of pQCD the processes (\ref{1}) are 
described by diagrams with two quark exchange. Their contribution 
to the cross section $d\sigma / d t$ decreases as $W^{-4}$ with increasing $W$ 
\cite{CZ}.
However, the contribution of the diagrams with the gluon 
exchange does not decrease with $W$ while $t$ is fixed. For the 
production of vector mesons $V$ the lowest nontrivial diagrams of 
pQCD corresponds to the two-gluon exchange and for the production of
$P$ and $T$ mesons to the three-gluon exchange.
The current lowest order calculations are performed in this approximation
\cite{CZ,GPS87}. In Tab.~\ref{tab1} the expected event rates are
given. 

It should be emphasized that the lowest order calculations (LO) provide
a lower limit of the expected event rates, as, at high enough energies,
higher-order terms in the perturbative series such as $\alpha_s
(p^2_\bot) \ln{(W^2/p^2_\bot)}$ become large and lead to a considerable
increase of the cross sections.
%
For the PP case, this effect has been calculated in the leading 
logarithm approximation (LLA) (see Refs. \cite{BL} and 
references therein). 
Defining $ z = 3 \alpha_s/(2 \pi) \ln W^2/W_0^2$
the expression for the differential cross section 
can be written \cite{FR}, as a power series in $z$, $i.e.$
\begin{equation}
\frac{d\sigma}{dt} = \frac{d\sigma_{2\rm gluon}}{dt}( 1 + \sum_{n=1}^{\infty}
 c_n z^n) = \frac{d\sigma_{2\rm gluon}}{dt}\;|K(W^2, t)|^2.
\label{EQ-LLA}
\end{equation}
The value of $z$ is very sensitive to the choice of parameters: for
example,  taking $\alpha_s=.2$ (small), $W_0^2 = 4(p_T^2 + m_V^2)=16$
GeV$^2$ (large) and $W_{min} = 15$ GeV one obtains the conservative
estimate $z=.25$. LEP2 can get statistics in the region  $z=.25 - .5$.
If $z \ge .5$, 
the perturbative
series needs to be summed to all orders and the striking power
behaviour of the cross sections emerges, $i.e.$
\begin{equation}
\frac{d \sigma}{dt} \sim f(W,t) \left( \frac{W^2}{W_0^2} \right)^{2 \omega_0},
\label{4}
\end{equation}
where $1+\omega_0$ is the ``intercept'' of the BFKL Pomeron
and the $W$ dependence in $f(W,t)$ is weak.
A corresponding strong enhancement over the two-gluon exchange results
is thus anticipated.
For large values of $z$ we have the LLA result
$2 \omega_0= (12 / \pi) \alpha_s (t) \ln{4} $  
leading eventually to a violation of the Froissart bound. At large 
enough $W^2$ 
this growth should be stopped by unitarity 
to satisfy approximately $K(W^2, t) < 25$\ \cite{CZ}.
The PO estimations in \cite{BL} show that 
there is an increasing function of a parameter like $z$ also for the PO.

In the case of $J/\Psi$ production there exists a prediction \cite{FR}
which takes into account the coupling of the reggeised gluons to the 
$c \bar c$-$J/\Psi$ system.
For $p_T^2 = -t \ll M_{\psi}^2$ and large $z$ the cross section is given by
\begin{equation}
\frac{d\sigma(J/\Psi J/\Psi)}{dp_T^2} = 16 \pi^2 \alpha^2
(\alpha_s C_F)^4 \frac{\pi^3}{4} \frac{\exp(16z\ln 2)}{(7 \pi \zeta(3) z)^3}
\left( \frac{c_{\psi} f_{\psi} }{M_{\psi}^2}\right)^4
\ln^4 \frac{M_{\psi}^2}{p_T^2},
\label{EQ-FR}
\end{equation}
where $c_{\psi} = 3/4$ and $f_{\psi} = 0.38$ GeV.

\begin{table}[htbp]
\renewcommand{\arraystretch}{1.1}
\begin{tabular}{|c|c|c|c|c|c|c|c|}
\hline
process & $W_{\rm min}$ (GeV & $|t|_{\rm min}$ (GeV$^2$) & $\Theta_{\rm min}$
(mrad) & LO & LLA Eq.~(\ref{4}) & LLA Eq.~(\ref{EQ-FR}) \\ \hline
$\rho \rho$ &  15  &  9  &  -   &  0.8 & 180/140  &  \\ 
            &  15  &  4  &  87  &  9.2 &          &  \\
            &  15  &  4  & 175  &  2.4 &          &  \\
\hline
$\rho X$    &  15  &  9  &  -   & 100  & 23000/18000  &  \\
            &  25  & 16  &  -   &  13  & 2300/2100  &  \\
            &  15  &  4  &  87  &  840 &            &   \\
            &  15  &  9  &  87  &  240 &            &   \\
\hline
$J/\Psi J/\Psi$ &  15 &  -   &  -   & 10    & 2500/2000  &   \\
                 & 15  &  4  &  87  &  1.4  &           & 570   \\
                 & 15  &  4  & 175  &  0.5  &           & 320    \\
                 & 25  &  4  &  87  &  0.5  &           & 440  \\
                 & 25  &  4  & 175  &  0.1  &           & 220   \\ 
\hline
$J/\Psi X$      &  15 &  -   &  -   & 120   & 26000/21000 &   \\
\hline
$\pi^0 X$       &  10 &  5   &  -   & 180   &             &   \\
\hline
\end{tabular}
\caption{\em Number of expected events for LEP II with
an integrated $e^+e^-$ luminosity of 500 pb$^{-1}$. 
The calculation was done for
anti-tagged $e^+e^-$ events with $\Theta_e \le 30$ mrad. \label{tab1}}
\vspace*{-2mm}
\end{table}
In Tab.~\ref{tab1} we present the number of expected events for LEP2
based on the two regimes (LO and LLA) described above. Since one does
not know at which energy the asymptotic predictions become valid we
expect the number of experimentally observed events to be bracketed by
the two sets of predictions. Further uncertainties are associated to the
choice of $\alpha_s$ and $W_0$ and the fact that the enhanced rates are
obtained using the large $z$ approximation in solving the BFKL equation.
More modest enhancements can be expected for the $z$ values typical of
LEP2 (see the numerical studies in  \cite{FR}) .
The values for the  $\gamma\gamma$ cross
sections in lowest order are the following: for the $J/\Psi$  production
we use the total cross sections obtained in \cite{GPS87},  for $\rho^0$
and $\pi^0$ the cross section corresponds to the region $-t \ge
|t_{min}|$. 
In the calculations, the photon-photon energy was restricted to
$W_{\gamma\gamma} \ge W_{min}$. Furthermore, results are given for the cross 
sections demanding the meson to emerge at angles larger than $\Theta_{min}$.
For the estimation of the effects of
BFKL Pomeron we use the $K$ factor calculated for the process
$\gamma\gamma \to \gamma\gamma$ via two $c\bar c$ pairs \cite{BL}. To show the
strong dependence of the results on the limitation of $K$, two sets
of numbers are given (last row), for $K \le 25$ and for $K \le 20$. 
In addition, the event rates according to Eq.~(\ref{EQ-FR}) (neglecting
unitarity corrections) are given.
Note that the numbers obtained with the constraint on $\Theta_{min}$ are
calculated with a fixed value of $\alpha_s = 0.32$ whereas for the
other numbers running $\alpha_s$ was used.

In conclusion, it appears that the study of diffractive phenomena
in the semi-hard domain may yield interesting information on the 
nature of the Pomeron.


\section[Large-$p_T$ processes and NLO phenomenology]
{Large-$p_T$ processes and NLO phenomenology
\label{ggpt1}
{\protect 
\footnote{ P.\ Aurenche, M.\ Dubinin, R.\ Engel, J.P.\ Guillet, J.\ Ranft,
Y.\ Shimizu, M.H.\ Seymour}}}
 
As it is the case in the study of the proton structure the production
of jets or hadrons
at large $p_T$ in $\gamma \gamma$ collisions is complementary to
the deep-inelastic scattering of a real photon: indeed the latter
reaction essentially probes the quark distribution while large
$p_T$ phenomena are also sensitive to the gluon distribution.
In order to reliably understand the hadronic structure of the photon
both types of processes (as well as heavy flavor production, see next
section) should be studied and compared to the
theory. Taken together, they will allow the determination of the
parton distributions in the photon.
For this purpose, we discuss below the jet inclusive  cross
section at the next-to-leading order (NLO) in QCD and
related processes in the case of anti-tag or single tag experiments. The
possibility to identify the gluon jet in three-jet events is also
considered.
 
\subsection{The structure of the hard process}
 
\begin{figure}[htb]
\begin{center}
\vspace*{-1.0cm}
\mbox{\epsfxsize=10.cm\epsfysize=10.cm\epsffile{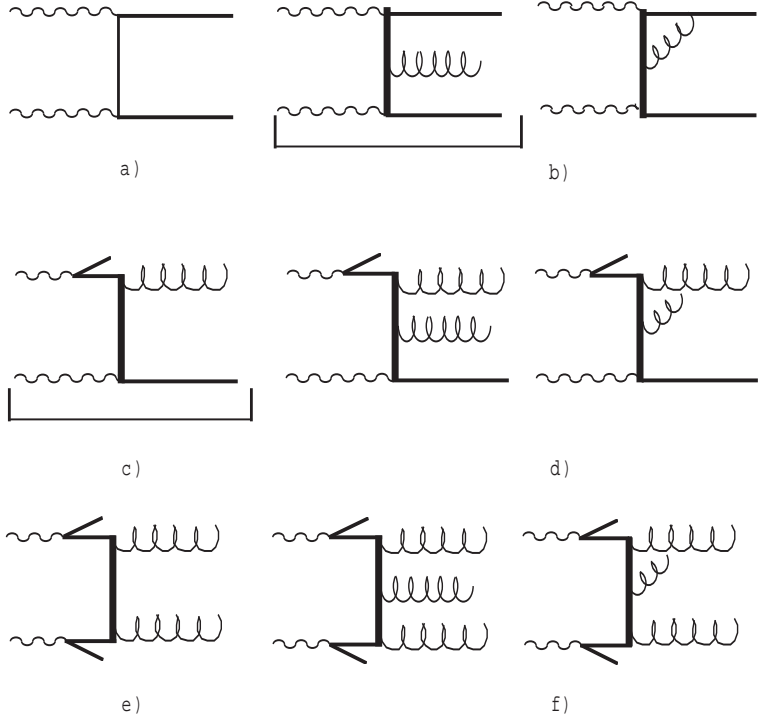}}
\vspace*{-1.05cm}
\end{center}
\caption{\label{fig:ptone}
{\em Some diagrams contributing to jet production in $\gamma \gamma$
collisions in the NLO approximation.} }
\end{figure}

Large $p_T$ processes involving real photons
are rather complex. This arises from the fact that the photon
couples to the hard sub-process either directly or through its quark or
gluon contents.
\noindent
The cross section for the production of a jet of a given
\pT and pseudorapidity $\eta$ can therefore be decomposed
as
(D: direct, SF: structure function, DF: double structure functions)
\be
\frac{d\sigma}{d\vec{p}_T d \eta} \ = \ \dsigd + \ \dsigsf + \ \dsigdf
\label{eq:section}
\ee
where each term is now being specified. In the NLO approximation
\cite{afg1} the ``direct" cross section takes the form
\be
\dsigd (R) =  \dsiggg \ + \ \alfspi K^{D} (R;M) .
\label{eq:dir}
\ee
with the corresponding diagrammatic decomposition shown in
Fig.~\ref{fig:ptone}a and~b.
\begin{figure}[htb]
\begin{picture}(484,198)(0,0)
\put(-10,0){\epsfig{file=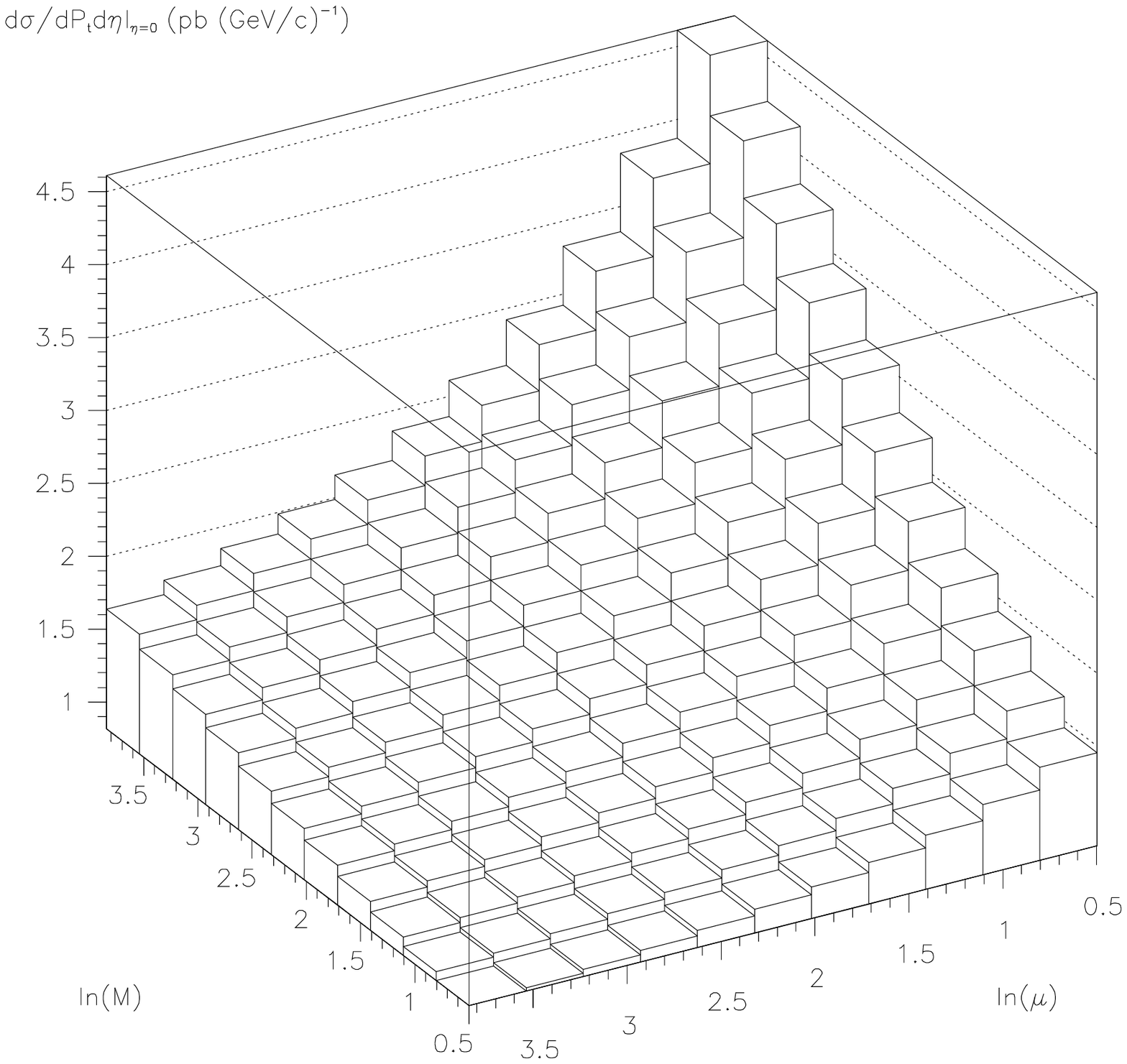,height=7.0cm}}
\put(230,0){\epsfig{file=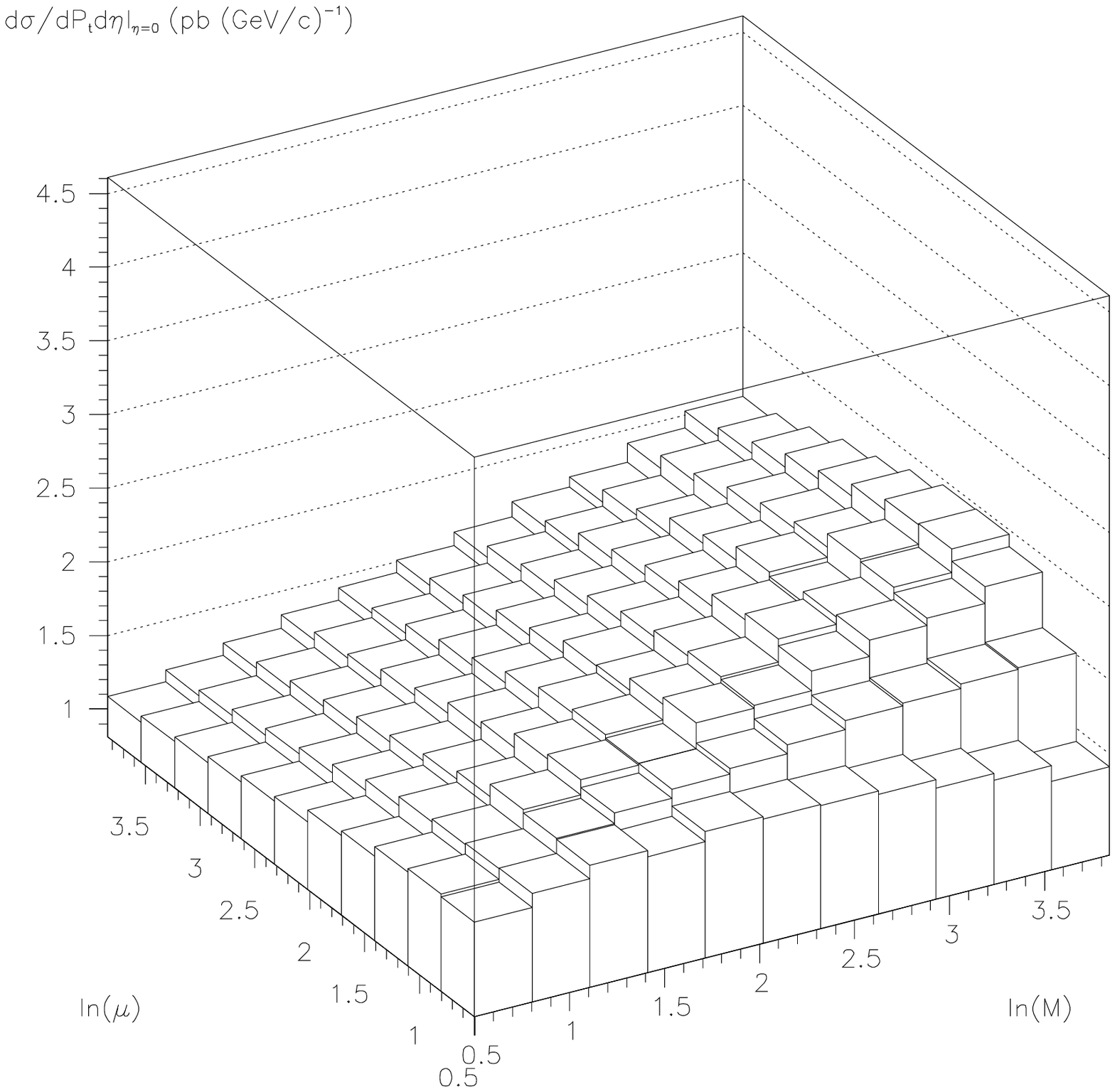,height=7.0cm}}
\end{picture}
\vspace*{-0.9cm}
\caption{\label{fig:pttwo}
{\em Variation of the $e^+ e^- \rightarrow e^+ e^- jet$ cross section under
changes of scales: a) leading logarithmic approximation; b)
next-to-leading logarithmic approximation.} }
\end{figure}
The parameter $R$ specifies the jet cone size, while $\mu$
and $M$ are the renormalization and factorization scales respectively.
When one photon couples directly and the other one through its
structure function, 
it leads to
\bea
\dsigsf (R) = \sum_{i=q,G} \int dx_{1} F_{i/\gamma}(x_{1},M)
\alfspi \left( \dsigig + \alfspi K^{SF}_{i \gamma}(R;M, \mu) \right)
+ 1 \rightarrow 2
\label{eq:sf}
\ena
where some diagrams representative of the ${\cal{O}}(\alf_s)$ and
${\cal{O}} (\alf_s^2)$ terms on the right hand side are shown in
Fig.~\ref{fig:ptone}c and~d, respectively.
The underlined diagrams in Fig.~\ref{fig:ptone}b and~c are in fact the
same but they contribute to different regions of phase space. When the
final state quark is not collinear to the initial photon (as in
Fig.~\ref{fig:ptone}b))  the exchanged propagator has a large virtuality
(shown by the fat line)  and the corresponding contribution is
associated to the hard subprocess $K^D$. When the final quark is almost
collinear to the initial photon (as in Fig.~\ref{fig:ptone}c))  the
virtuality of the exchanged propagator is small and the interaction is
soft (long range). Roughly speaking the factorization scale $M$
separates the hard region from the soft region and changing this
arbitrary scale shifts contributions from $d\sigma^D$ to $d\sigma^{SF}$
but clearly should not affect the sum $d\sigma^D + d \sigma^{SF}$
\cite{afg2}.  A
similar compensation occurs between $d\sigma^{SF}$ and $d\sigma^{DF}$
(see Fig.~\ref{fig:ptone}d and~e).
In conclusion, only the sum eq. (\ref{eq:section}) has a physical
meaning and it is not legitimate to associate $d \sigma^{SF}$ and $d
\sigma^{DF}$ to experimentally measured ``once resolved" and ``twice
resolved" components.
To illustrate quantitatively the variation of the theoretical
predictions under changes of $M$ and $\mu$ we consider the production of
jets at $p_T=10$ GeV/c and $\eta=0$:
Fig.~\ref{fig:pttwo}a is obtained when setting arbitrarily $K^D = K^{SF}
= K^{DF} =0$ (the so-called leading order (LO) predictions) while
Fig.~\ref{fig:pttwo}b  takes into account the full expressions: the gain
in stability is remarkable despite the fact that no saddle-point or
extremum is found \cite{afg1}. In the following we always use for
definiteness $M=\mu=p_T$.
\subsection{Transverse momentum and rapidity distributions}
In order to compare theory and experiment one must convolute the above
cross sections with the Weizs\"acker-Williams spectrum of quasi-real
photons emitted by the electrons, taking into account the experimental
conditions. The usual antitagging conditions defined in  sec.
\ref{ggtag} are used here ($\theta_{e} < 30$ mrad or $E_e < 1\ GeV$).
The appropriate Weizs\"acker-Williams spectrum  is given in eq.
(\ref{eq:wewi}).
\begin{figure}[htb]
\begin{picture}(484,198)(0,0)
\put(-10,0){\epsfig{file=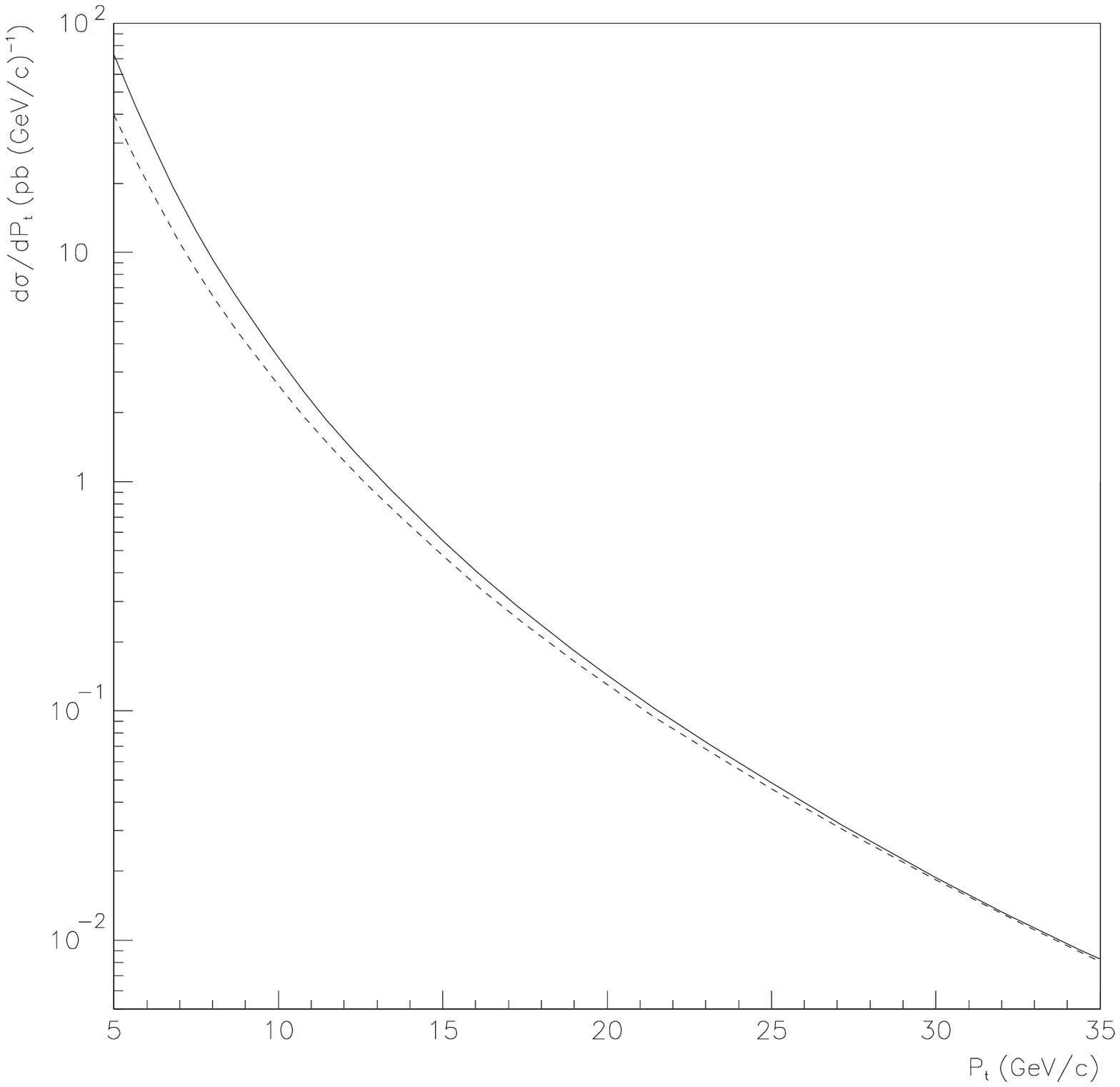,height=7.0cm}}
\put(230,0){\epsfig{file=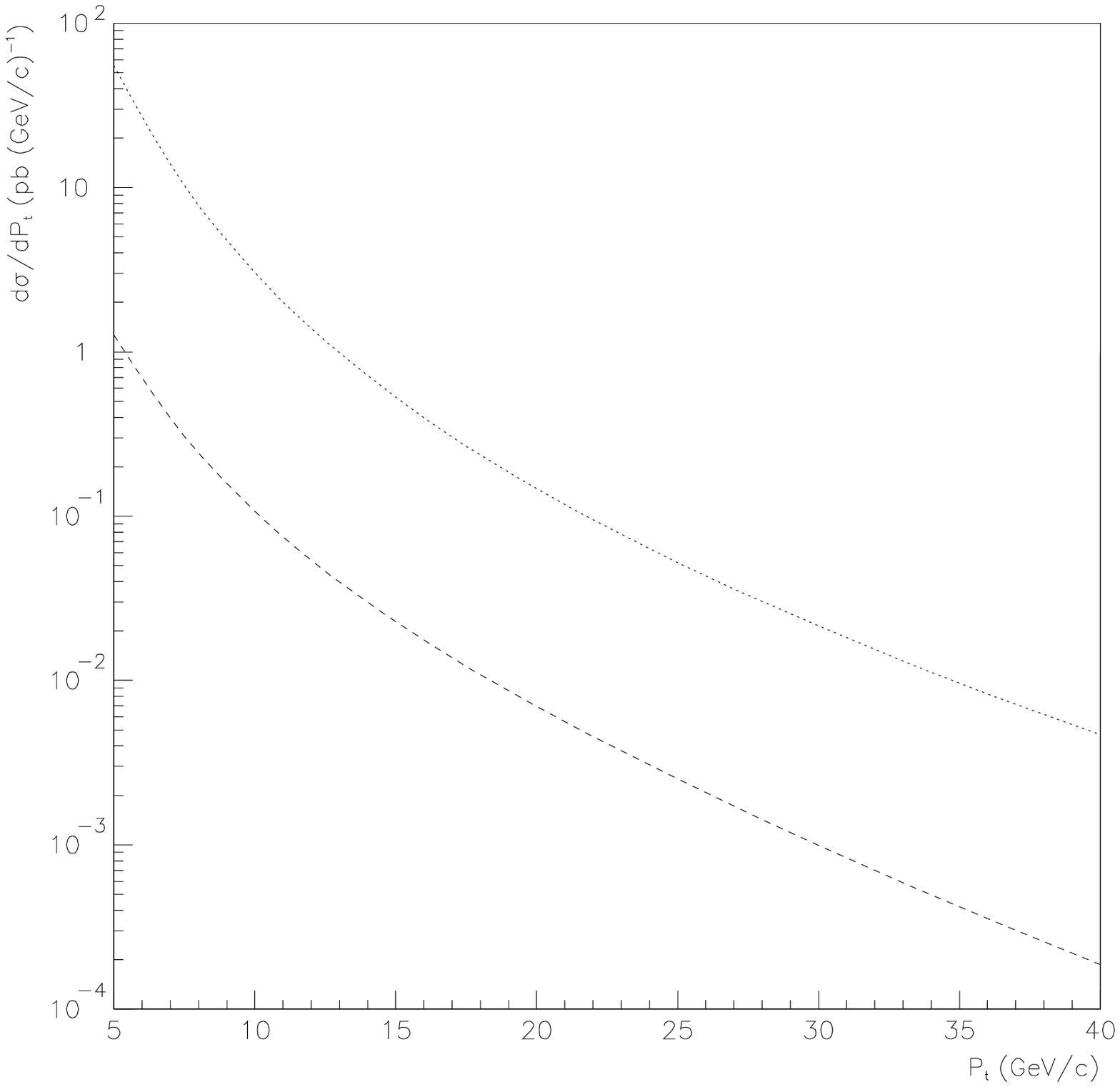,height=7.0cm}}
\end{picture}
\vspace*{-1.cm}
\caption{\em Jet pt distributions.  Left: The NLO cross section using
the full structure functions (solid line) compared to the case when
only the perturbative part is kept (dashed line). Right:  the same
as above (solid line) in the LO approximation and the cross section
when one of the photon is tagged (dashed line).}
\label{pttag}
\end{figure}
We display in Fig. \ref{pttag} the jet $p_T$ distribution, the jet
pseudo-rapidity being integrated in the interval $|\eta| <1$. The
complete NLO expressions are used. The solid
curve is the prediction  when using the full AFG parton distributions
(with a VMD input at $Q^2_0 = .5\ \mbox{GeV}^2$) \cite{AFG},
while the dashed curve is obtained when one artificially sets the
VMD component equal to $0$. More precisely, in the latter
case, the quark and gluon densities are vanishing at
$Q^2_0$ and they are generated by the evolution
equations at larger $Q^2$.
It is clearly seen that the VMD component is quite
important at $p_T=5 \ GeV/c$ (about $40\%$) but rapidly decreases as
$p_T$ increases since it is less than $10\%$ at $p_T=20 \ GeV/c$. One
may note that the VMD component of the quark is also tested in photon
structure function studies but large  $p_T$ processes are also very
sensitive to the VMD component of the gluon which dominates for $x <.5$.
With the luminosity expected at LEP2 the error bars in the data will be
sufficiently small to constrain the size of the VMD input.
 
As mentioned in the structure function  section above it will be
possible to probe the virtual photon structure in deep-inelastic type
experiments. We show here that it can also be probed in jet studies.
Using the tagging conditions $.2 < Q^2\ (\mbox{GeV}^2) < .8$ and
$z_{\gamma} < .5$, typical of LEP2, on one photon
and the usual anti-tagging condition on the other, one still obtains an
appreciable  jet cross section as shown on Fig. \ref{pttag}: about 100
events with $p_T=10 \pm 1 \mbox{GeV/c}$ are expected. Let us remark
that,  thanks to the usual VMD form-factor, the VMD contribution is
rapidly reduced when considering tagged electron (it is reduced by
roughly $75\% $ when going from a real photon to a photon of virtuality
$Q^2=.5\ \mbox{GeV}^2$).
\begin{figure}[htb]
\begin{center}
\vspace*{-1.0cm}
\mbox{\epsfxsize=10.cm\epsfysize=10.cm\epsffile{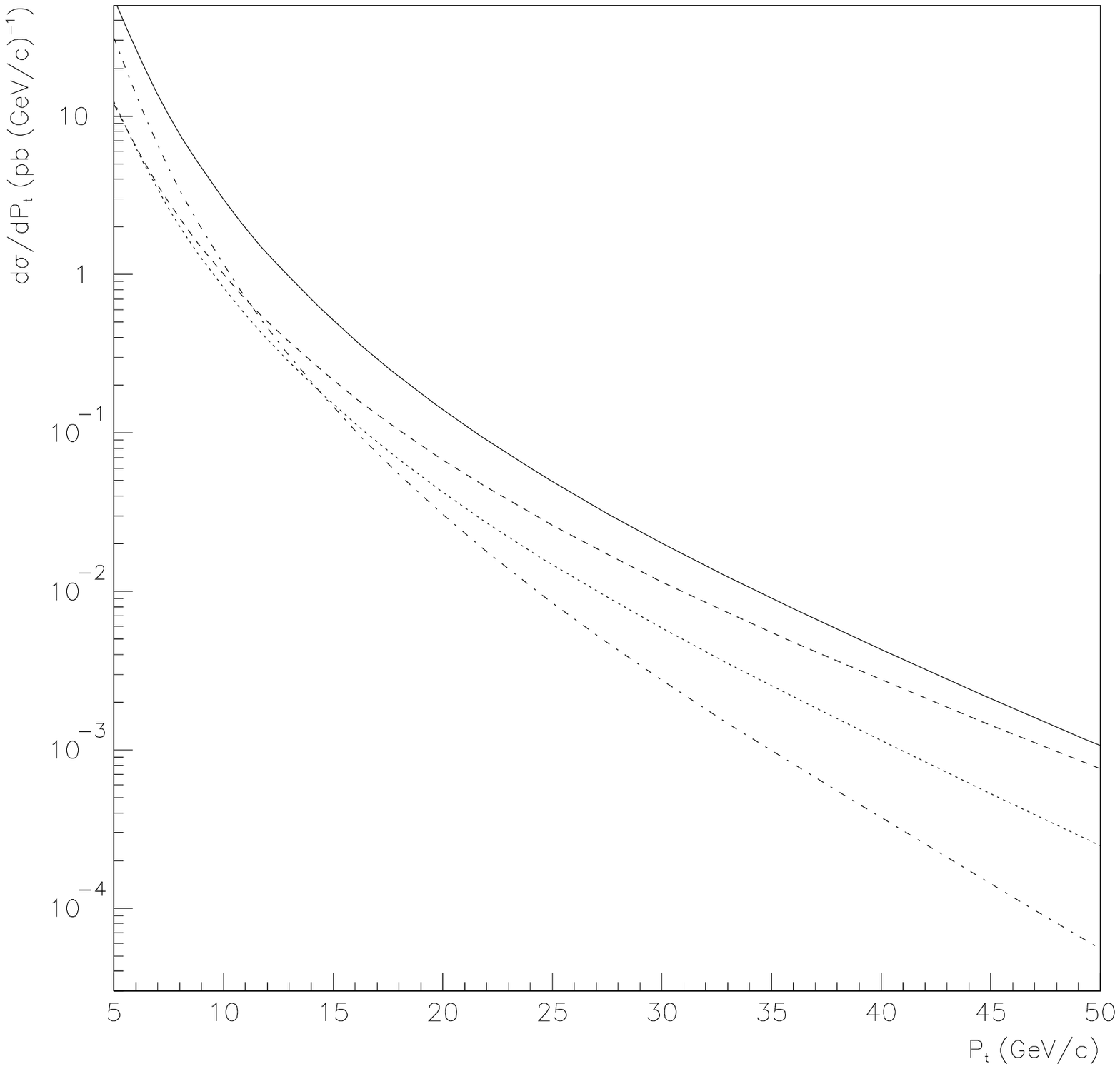}}
\vspace*{-.5cm}
\end{center}
\vspace*{-0.7cm}
\caption{\label{ptdetail}
{\em The various components of the jet $p_T$ spectrum.
The full cross section: full line; the DIR component: dashed;
the SF component: dotted; the DF component: dash-dotted.} }
\end{figure}
\vspace*{-.5cm}
\begin{figure}[htb]
\begin{picture}(484,227)(0,0)
\put(-10,0){\epsfig{file=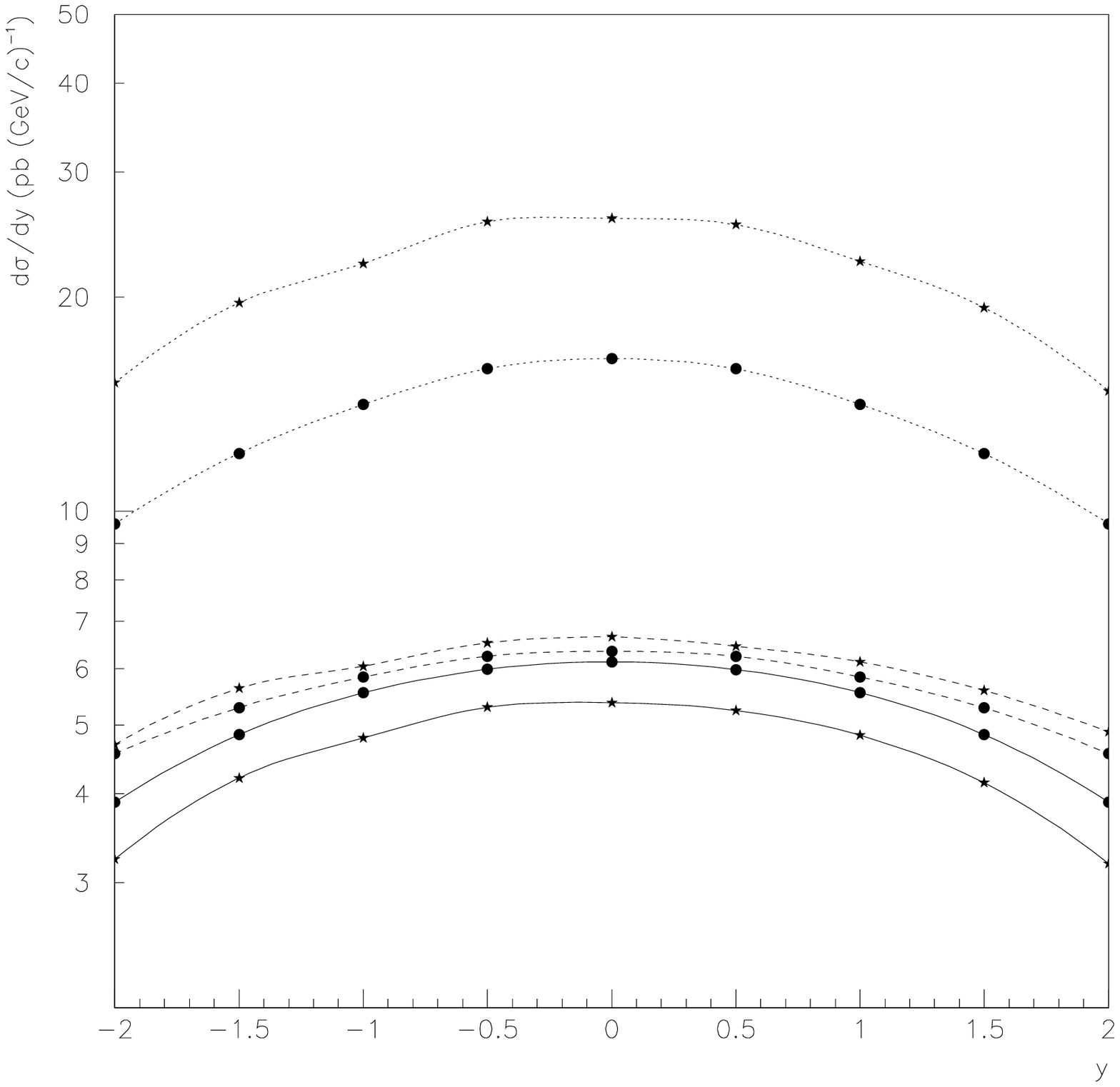,height=8.0cm,width=8.0cm}}
\put(230,0){\epsfig{file=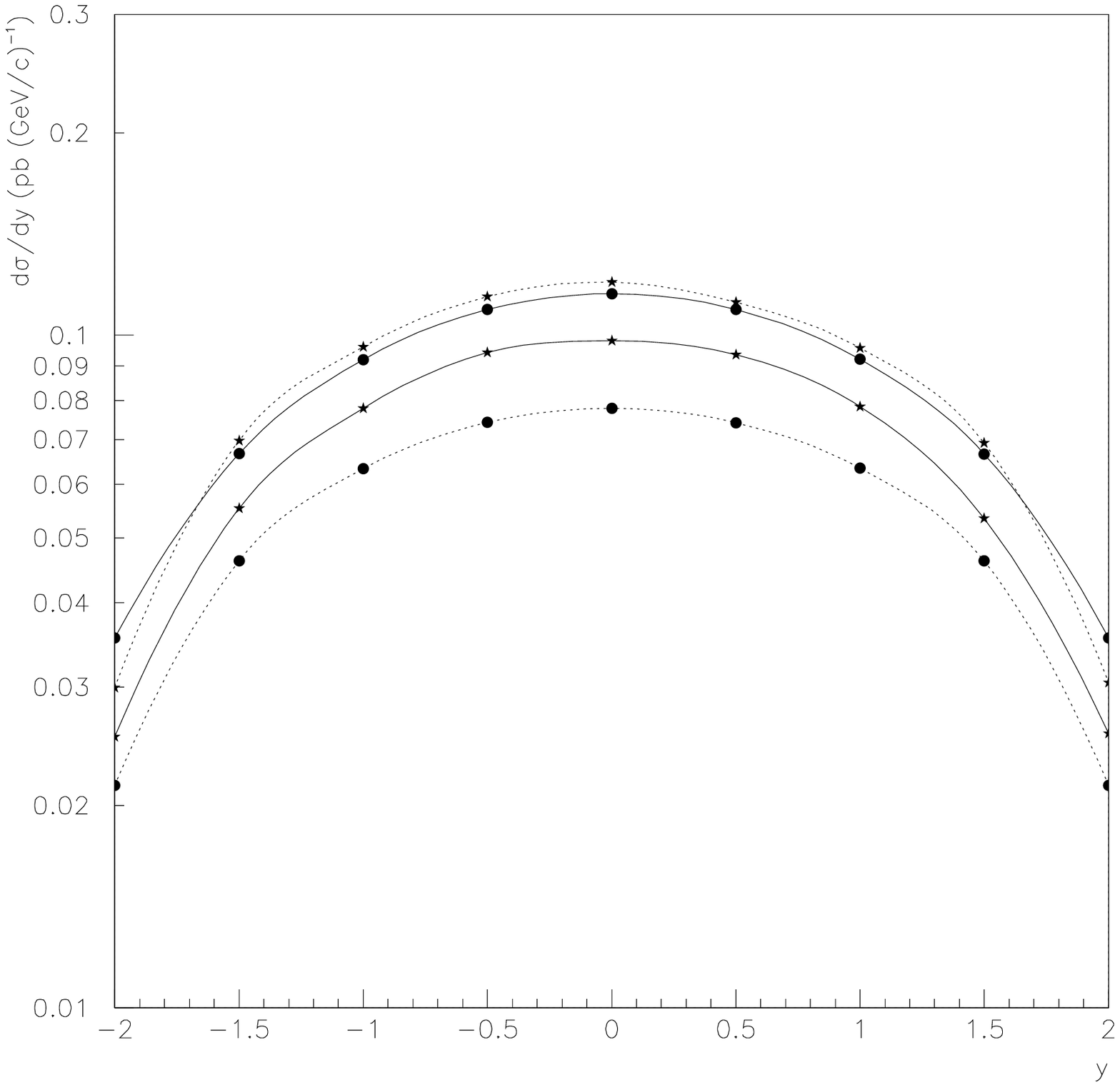,height=8.0cm,width=8.0cm}}
\end{picture}
\vspace*{-1.3cm}
\caption{ \em
The jet rapidity distributions at fixed $p_T$. The meaning of the curve
is as in the previous figure. The dotted lines are the LO
results while the starred line are the NLO results. Left: $p_T=5$ GeV/c;
right: $p_T=15$ GeV/c.
\label{rapdetail} }
\end{figure}
\vskip 0.2cm
 
Despite the warnings given above, we display in Fig. \ref{ptdetail} the
break-up of the single inclusive jet cross section, integrated in the
range $|\eta| <1$, into the DIR, SF and DF components. These curves
can serve useful purposes when comparing these analytic calculations
with those based on Monte-Carlo generators. Of course, the size of
the various components depend strongly on the choice of the
factorization scale (here we use $M=\mu=p_T$). In the figure, obtained
using the leading logarithmic expressions for the sub-processes we see
that DF component dominates the lower end of the distribution but is
less than $20\%$ at  $p_T= 20$ GeV/c. Nowhere does the SF terms
dominate. In Fig.~\ref{rapdetail} we show the rapidity dependence for
$p_T=5$ GeV/c and  $p_T=15$ GeV/c of the various components both in
the LO and the  NLO approximation.
One sees that the pattern of higher corrections is quite different for
the different subprocesses: the DF component is appreciably increased by
the higher-order corrections (about $40\%$) while the DIR component is
decreased and (roughly $15\%$)
the SF component remains stable. This pattern of higher order
corrections is independent of the transverse momentum or rapidity. It is
interesting to remark that at $p_T=15$ GeV/c the hierarchy of cross
sections is reversed when including the higher order corrections with
the DF term dominating over the DIR one despite the fact that the
overall cross section is not affected very much. In conclusion, one may
say that the higher-order corrections affect the structure of
the cross section more than its overall size. \\
\indent
It will be very useful to study a more exclusive observable
namely the di-jet cross section.  Quite interesting phenomenology
is coming out of HERA \cite{erdm}: it allows to separate on an
experimental basis events dominated by the direct component
from those dominated by the photon structure function.
For an application of this technique to $\gamma\gamma$ reactions we refer 
to the ''High-$p_T$'' section of \cite{gggen}. 
NLO calculations are in progress \cite{klei}. 
 
Recently data on jet distributions have been published by the TOPAZ
\cite{topjet} and AMY \cite{amyjet} collaborations. The above
calculations are in good agreement \cite{afg1} with the single jet
distribution of TOPAZ  when using the AFG parton distributions
\cite{AFG} while both the NLO  predictions \cite{klei} and the {\sc
Phojet} \cite{engran} results using the GRV parametrizations
\cite{GRVphot} fall below the data. In contrast, a rather good
agreement, at least at large $p_T$ values, is found with the two-jet
data of TOPAZ in \cite{klei} and \cite{engran}.
 
A word of caution should be said concerning large $p_T$ phenomenology.
Even at LEP2 we are dealing with rather low $p_T$ jets and the
comparison between the theoretical predictions, at the partonic level,
and the experimental distributions at the hadronic level may not be easy
because of the non negligible contribution  to the jet transverse
momentum from the ``underlying event" \cite{butt}. Good event generators
are required to understand this point and also a lot should be learnt
from present HERA studies. To avoid this difficulty it will be
extremely interesting to consider the single hadron inclusive
distribution which probes the same dynamics as jet production. New
parameters come into play through the fragmentation functions of partons
into hadrons but these distributions should soon be rather well
constrained as several groups are at present extracting NLO
parametrizations of fragmentation functions using data from both  $e^+
e^-$ colliders and $p p$ colliders \cite{frag}. Concerning lower energy
data, the situation concerning single hadron production is rather
confusing and paradoxical as the experimental results \cite{cords} are
much above the NLO theoretical predictions \cite{gord} at large $p_T$
(where the DIR term is predicted to dominate) while they agree with the
theory at lower $p_T$. Data are eargerly awaited to clarify this point.
 
\subsection{Three-jet events and the separation of gluon jets}
 
The first direct evidence for gluon jets was seen by the PETRA
experiments as three-jet events, $e^+e^-\to q\bar{q}g,$ where
all three partons have high energy and are well-separated in angle.
Ever since, one of the important aims of QCD studies has been to measure
the similarities and differences between quark and gluons jets, as well
as collective quark-gluon phenomena such as the string
effect\cite{MHSLEPrefs}.  One of the important questions is to what
extent such effects can be described by perturbative QCD, rather than
non-perturbative models.  However, many of the studies are inconclusive
on this issue, because at any given energy, most non-perturbative models
can be tuned to mimic the perturbative effects, and it is only in the
energy-dependence that definitive differences can be seen.  But
comparing experiments at different energies typically involves large
uncorrelated systematic errors.  In $\gamma\gamma$ collisions on the
other hand, one can study a wide range of energies in the same
experiment, and can thus study energy-dependent effects much more
reliably.
 
\begin{figure}[htb]
\begin{center}
\epsfig{file=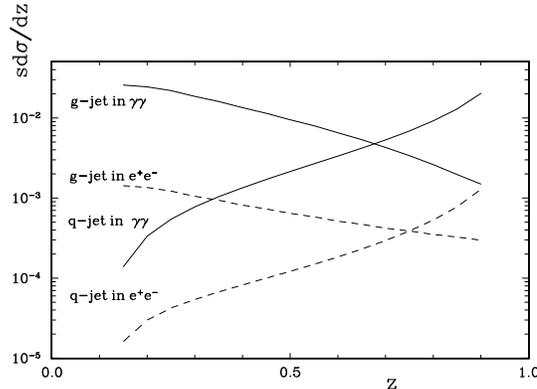,width=7.0cm}
\end{center}
\vspace*{-1.0cm}
\caption{
{\em Differential spectra $s d\sigma/dz$ of parent quark and gluon
 in the reactions $e^+ e^- \rightarrow q \bar q g$ and
 $\gamma \gamma \rightarrow q \bar q g$ }
\label{fig:dubin} }
\vspace*{-0.5cm}
\end{figure}

We can make a rough estimate of the three-jet rate by taking the leading
logarithmic approximation to the total two-jet rate\cite{MHSbible} and
simply multiplying it by a factor of $\alpha_s$ as an estimate of the
fraction of three-jet events.  We require each jet to have a transverse
momentum above $p_{t_{\mathrm{min}}}$ both with respect to the beam axis
and each other, and hence obtain
\[ \sigma(s,p_{t_{\mathrm{min}}}) =
   \frac43\left(\sum_qe_q^4\right)
   \frac{N_c\alpha^4}{\pi p_{t_{\mathrm{min}}}^2}
   \log\frac s{p_{t_{\mathrm{min}}}^2}\log^2\frac s{m_e^2}\times\alpha_s. \]
It is worth noting that the smallness of $\alpha_s$ cannot be
compensated by any large logarithms, because the appropriate logarithms
are $\log W^2/p_{t_{\mathrm{min}}}^2$ and the $W^2$ distribution is
dominated by $W^2\sim p_{t_{\mathrm{min}}}^2$.  Nevertheless, we obtain
around $10^4$, $10^3$ and $10^2$ events in 500 $\mathrm{pb}^{-1}$ for
$p_{t_{\mathrm{min}}}=5,\;15$ and 35 GeV respectively.
For such studies it is important to know which jet is the gluon.  The
cleanest way to do this is by flavour tagging the quark jets, but this
severly reduces the rate.  In $\mathrm{e^+e^-}$ annihilation, the fact
that the gluon's energy spectrum is softer than the quark's was used, by
always calling the softest of the three jets the gluon.  By explicit
analytical integration of the full five-dimensional matrix element for
$\gamma\gamma\to q\bar{q}g$ down to a distribution over the
energies of the quarks \cite{MHSDubinin}, we find that the same is also
true for \gaga collisions (Fig. \ref{fig:dubin}, see \cite{MHSDubinin}
for further details).  Thus the gluon jet in three-jet events
can be statistically tagged by calling it the softest jet.
 
Although a more complete study incorporating realistic detector cuts and
the effects of hadronization is clearly needed, it seems hopeful that a
sample of three-jet events could be isolated and a study of the
energy-dependence of quark-gluon jet effects made at LEP2.
\subsection{Role of the experimental cuts on the inclusive jet
spectrum}
We have seen
in sec.~\ref{ggtag}
that the radiative
production  of $Z^0$ bosons provides an important background to \gaga
physics when one looks for rare events such as large $p_T$ jets. 
\begin{figure}[htb]
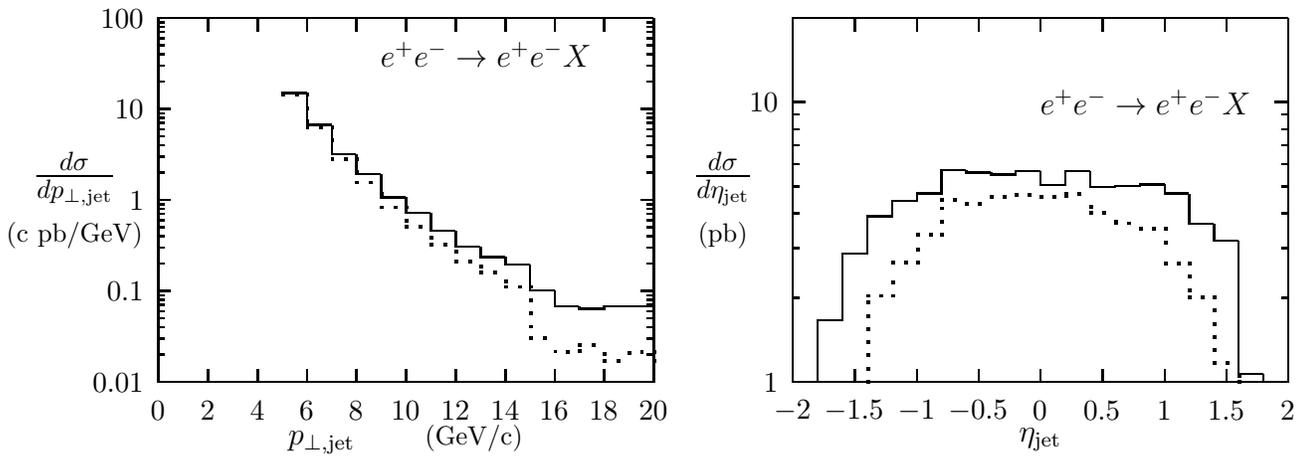

\begin{picture}(484,170)(0,0)
\put(-10,0){\input{jetptlep.pic}}
\put(230,0){\input{jeteta7.pic}}
\end{picture}
\caption{ \em
Jet cross section for events satisfying cut 1 (full line) and cut
1+2+3+4 (dotted line). Left: distribution in $p_T$ with
$|\eta_{jet}|<1$; right: distribution in $\eta_{jet}$ for fixed  $p_T
= 7$  GeV/c. (Calculated with {\sc Phojet}).
\label{fig:jeteta5} }
\end{figure}
Various cuts have been devised which considerably reduce the background. 
The study  of the effect of these cuts on the signal requires the use 
of a Monte-Carlo generator and we use here the {\sc Phojet} program
\cite{{Engel95a},{Engel95d}}. We have
checked that the unbiased jet  $p_T$ spectrum is in good agreement with
the analytical results  described above.  In Fig.~\ref{fig:jeteta5} we
see that, as expected, only the upper end of the spectrum is reduced by
the cuts (see sec.~\ref{ggtag} for the meanings of the cuts)
while at fixed
$p_T$  mainly the large $|\eta_{jet}|$ regions are reduced.
The net result is that, after cuts, the rate for producing jets
with $p_T \simeq 20\ \mbox{GeV}$ is comparable in the signal and in
background. Needless to
say that jets produced in single tag events are not affected by the
background.

\section[Heavy-quark physics]{Heavy-quark physics
\label{gghq1}
{\protect 
\footnote{ M.\ Cacciari, A.\ Finch, M.\ Kr\"{a}mer, E.\ Laenen, S.\
Riemersma, G.A.\ Schuler, S.\ S\"oldner-Rembold}}}
%
%
%
%
%


A favourable aspect of heavy flavour production in 
$\gamma\gamma$ collisions compared to other $\gamma\gamma$
processes is that the heavy quark mass ensures that the 
separation into a direct and resolved processes is, 
to next-to-leading order (NLO) in QCD, unambiguous, i.e.
does not depend on an arbitrary separation scale.
Experimentally one may perform this separation by
using single tag events (see below), by using 
non-diffractively produced $J/\psi$'s, or by using the
photon remnant jet, present in resolved processes, as
a separator. Therefore heavy flavour production at LEP2
provides a good opportunity for simultaneously testing
QCD (direct process) and measuring the poorly known 
gluon content of the photon (resolved processes).

\subsection{Theory}

\begin{figure}[hbt]
\begin{center}
\begin{tabular}{ccc}
\epsfig{file=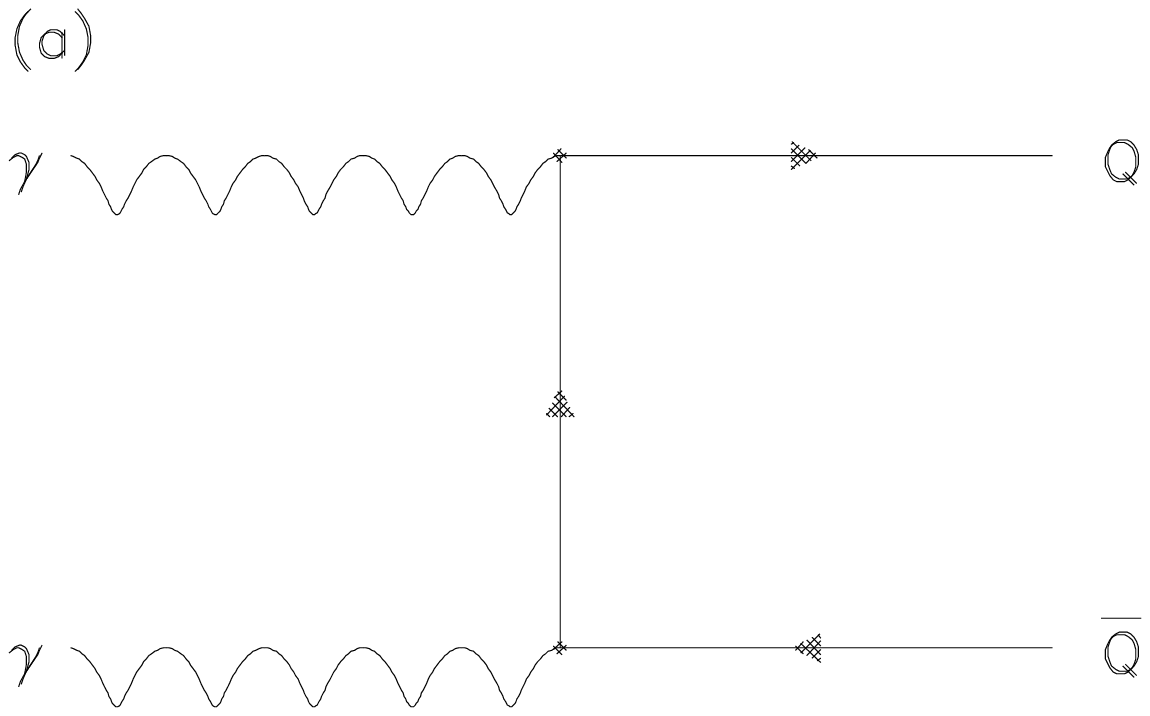,width=0.3\textwidth,height=3cm} &
\epsfig{file=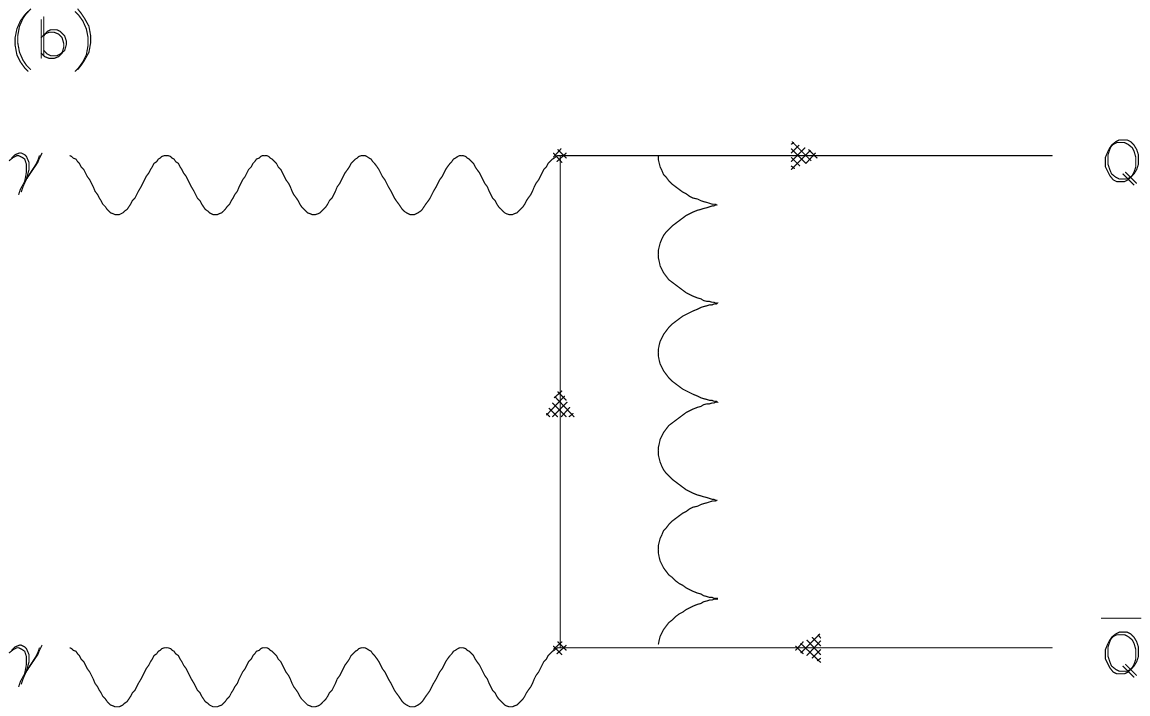,width=0.3\textwidth,height=3cm}&
\epsfig{file=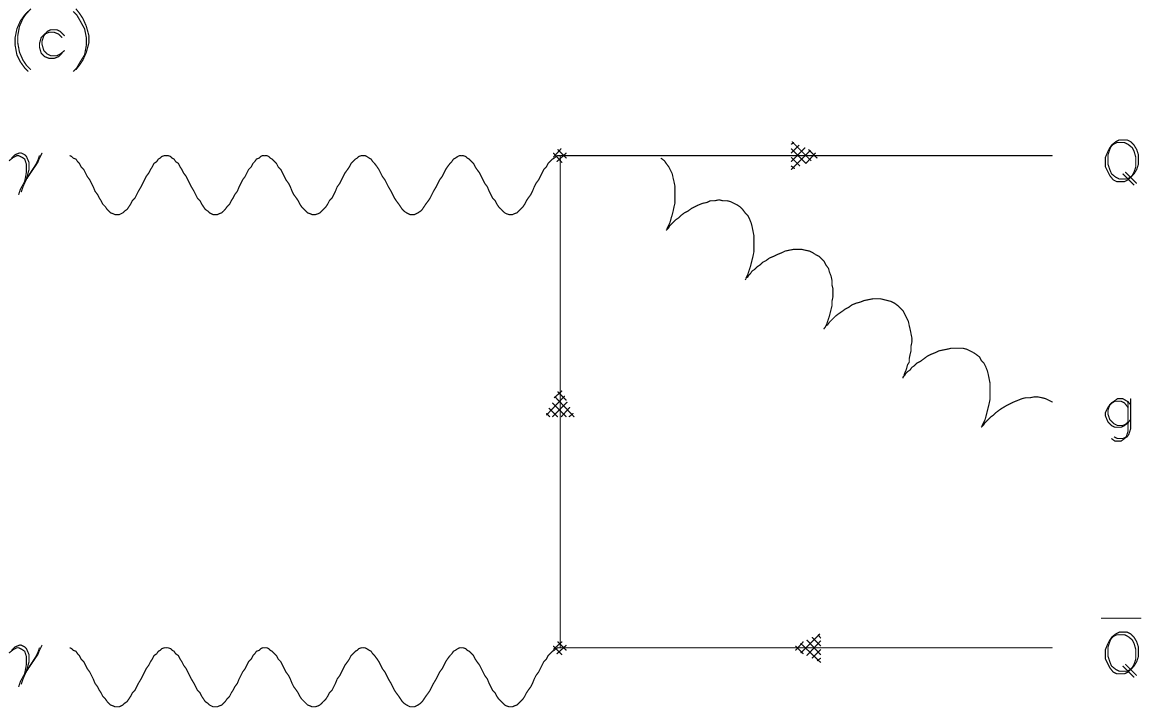,width=0.3\textwidth,height=3cm} \\
\epsfig{file=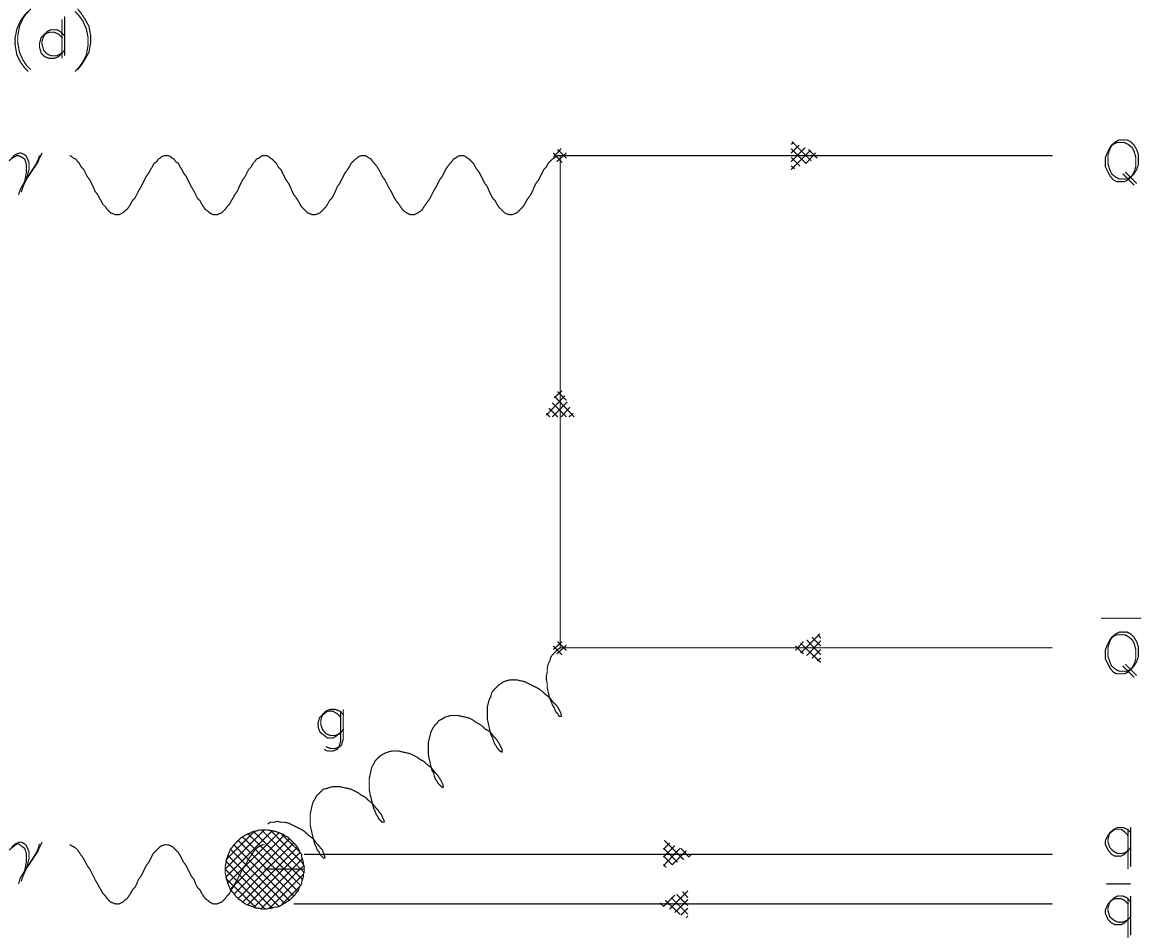,width=0.3\textwidth,height=3cm} &
\epsfig{file=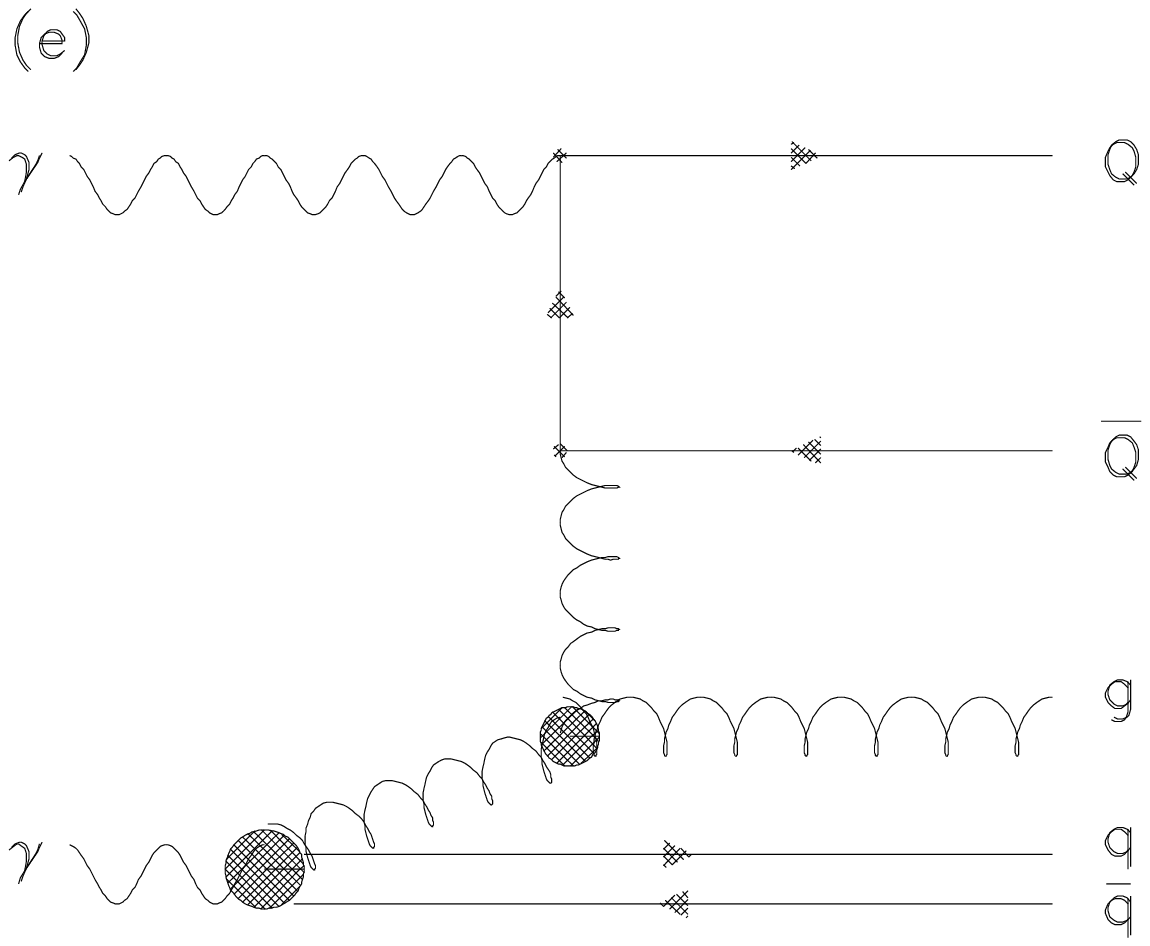,width=0.3\textwidth,height=3cm}&
\epsfig{file=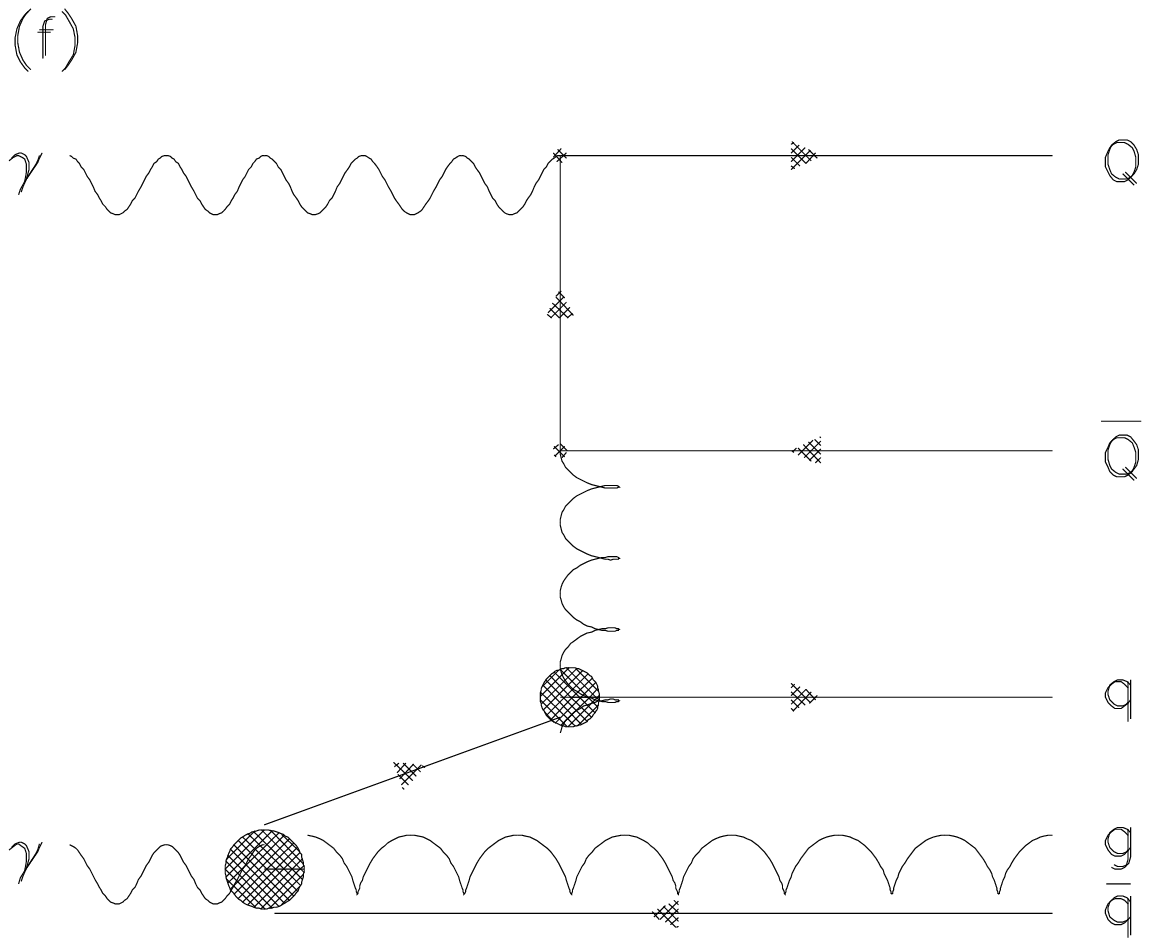,width=0.3\textwidth,height=3cm} \\
\end{tabular}
\end{center}
\caption{\em
Examples of diagrams contributing to heavy quark production in 
$\gamma \gamma$ collisions. (a) Direct contribution: Born term (QPM);
(b) virtual correction to direct term; (c) real correction to direct process;
(d-f) ``resolved'' contributions.}
\label{feynman}
\end{figure}

We first discuss the theoretical aspects of the reaction
$\gamma \gamma \rightarrow Q \bar Q $ where $Q$ ($\bar Q$) is a 
heavy (anti)-quark (charm or bottom).
In practice the cross section for bottom production is too 
small to be observed at LEP2 so, in what follows, only charm production 
will be considered.
 Figure \ref{feynman} shows some of the diagrams contributing to
heavy quark production in two-photon physics. 
Diagrams (a)-(c) are examples of the
so-called direct process in which the photon
couples directly to a quark. Diagram (a) is the Born term direct process which
is equivalent to the Quark Parton Model (QPM), (b) and (c) represent virtual and
real QCD corrections to the Born amplitude. 
At low beam energy the direct process is completely dominant, 
however at LEP2 the production of open charm in the
collision of two effectively on-shell equivalent
photons receives contributions in about equal amounts
from the direct- and once-resolved channels diagrams (d)-(f).
In Ref.\cite{DKZZ} this process was calculated to 
NLO in QCD, and all theoretical
uncertainties (due to scale choice, mass of the charm, and choice
of photonic parton densities) were thoroughly investigated.
The largest part of the resolved process is given by the
photon-gluon fusion process; this property offers the
possibility of measuring the gluon content of the photon which 
is currently poorly known. Doubly resolved processes have been
calculated to give a negligible contribution for presently available 
beam energies \cite{DKZZ}.
In Fig.~\ref{results} the total cross section and its theoretical
uncertainty is shown as a function of the center of mass energy,
together with some recent measurements.
LEP2 offers the possibility of a serious comparison
between a fairly well understood theory and experiment
with considerably more statistics than hitherto.
Given this larger statistics it will be interesting to make this
comparison not just for the total cross section, but
also for various differential distributions.
Single particle distributions in the transverse momentum
and rapidity of the heavy quark are given in \cite{DKZZ}.
Correlations between the heavy quarks in the direct
process, including NLO effects, have recently been studied in \cite{KL}, 
but will be difficult to observe at LEP2.

Furthermore, the NLO prediction for the one particle 
inclusive transverse momentum distribution contains 
potentially large terms $\sim \alpha_s\ln(p_t/m)$
which might spoil the convergence of the perturbation 
series at $p_t\gg m$. In Ref. \cite{LPT} 
the NLO cross section for large $p_t$ production
of heavy quarks in direct and resolved channels
has been calculated in the framework 
\cite{MeleNason}
of perturbative
fragmentation functions. This approach allows for
a resummation of the large logarithmic terms
and thereby reduces the scale dependence of the NLO 
prediction.

\begin{figure}[hbt]
\begin{center}
\epsfig{file=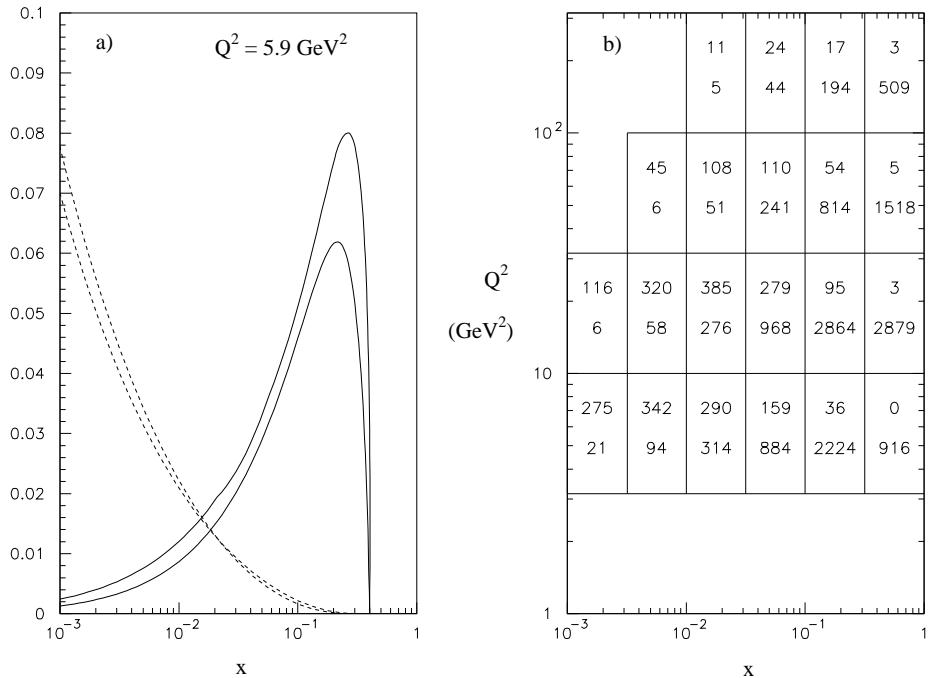,height=10cm} 
\caption{\em
a) $F_2^{\gamma}(x,Q^2)/\alpha$ as a function of $x$
at $Q^2 = 5.9 ({\rm GeV/c})^2$ for charm production at LO
and NLO in QCD. The dashed line is the resolved (or hadronic) 
contribution (lower curve at small $x$: NLO, upper: LO),
the solid the direct (or pointlike) (upper curve: NLO, lower: LO);\
b) Number of events in $x,Q^2$ bins. Upper number: resolved
contribution, lower number: direct.}
\label{erics}
\end{center}
\end{figure}

The process $\gamma^*\gamma \rightarrow c\cbar$ is considerably more 
suppressed than the previous one due to the 
photon being off-shell.
The structure functions $F_{2}^{\gamma}(x,Q^2)$ 
(and $F_{L}^{\gamma}(x,Q^2)$)
for open charm production in deep-inelastic single-tag
events were calculated to NLO in QCD in \cite{LSRN}. 
It was found that the contributions from the direct (or ``pointlike'')
and resolved (or ``hadronic'') separate in the variable
$x$: above $x \simeq 0.01$ $F_2^{\gamma}(x,Q^2)$
is almost exclusively due to pointlike photons, and hence
calculable, below this value it is mainly due to resolved
photons, and essentially proportional to the gluon density
in the photon, see Fig.~\ref{erics}a. Experimental
studies of the reaction 
$e^+e^- \rightarrow e^+e^- D^{*\pm} X$ with one outgoing
lepton tagged (``single-tag'') have been done 
by JADE \cite{jade} and by TPC/Two-Gamma\cite{tpcgg}
at low average value $\langle Q^2 \rangle$ of the 
momentum transfer squared of the tagged
lepton (below $1 \,\,({\rm GeV}/c)^2$)
and by TOPAZ\cite{topaz2} at somewhat larger $\langle Q^2 \rangle$.
The total number of events obtained was however very small
(about 30 for TOPAZ).

Theoretical uncertainties in these
quantities were investigated and are well under control.
Numbers of events expected per bin in $x,Q^2$ over the
lifetime of LEP2 are also given in Fig.~\ref{erics}b and in \cite{LR}.
One may conclude from these that with a not too pessimistic charm
acceptance (1-2\%) a measurement of $F_2^{\gamma}(x,Q^2)$
should be feasible. Some single particle differential 
distributions in the transverse momentum and rapidity
of the heavy quark have been studied in \cite{LR}.

Let us finally mention {\it onia} production. 
The radiative decay width of the charmonium states $\eta_c$, 
$\chi_{c0}$ and $\chi_{c2}$ can directly be measured in 
two-photon collisions at LEP2. These $\gamma\gamma$ partial 
widths provide an important test of the non-relativistic 
quarkonium model. 
We refer to the section on resonances and exclusive states for more details.
We comment here only further on the case of $J/\psi$.

Two-photon production of $J/\psi$ bound states is an attractive 
tool to determine the gluon distribution in the photon. 
In contrast to the case of open heavy 
flavour production, $J/\psi$ mesons are generated only via 
resolved photons and can be tagged in the leptonic decay 
modes. Higher order QCD corrections to $J/\psi$ production in 
photon-gluon fusion have been calculated in \cite{KZSZ}. 
Including NLO corrections, the cross section for 
inelastic $J/\psi$ production in $\gamma\gamma$ collisions
at LEP2 is predicted to be about 5 pb, suggesting that
a measurement of this reaction may be feasible.

\def\u{$\protect\displaystyle\left(\uparrow\right)$}
\def\d{$\protect\displaystyle\left(\downarrow\right)$}

\subsection{Experiment}
We review in this section the experimental
status of charm production in $\gamma \gamma $ collisions, list and 
evaluate presently available charm identification methods, and 
extrapolate published results to total charm cross sections for the
purpose of comparison. We think this is useful as no such review
is presently available in the literature.

\begin{figure}[hbt]
\begin{center}
\begin{tabular}{cc}
\epsfig{file=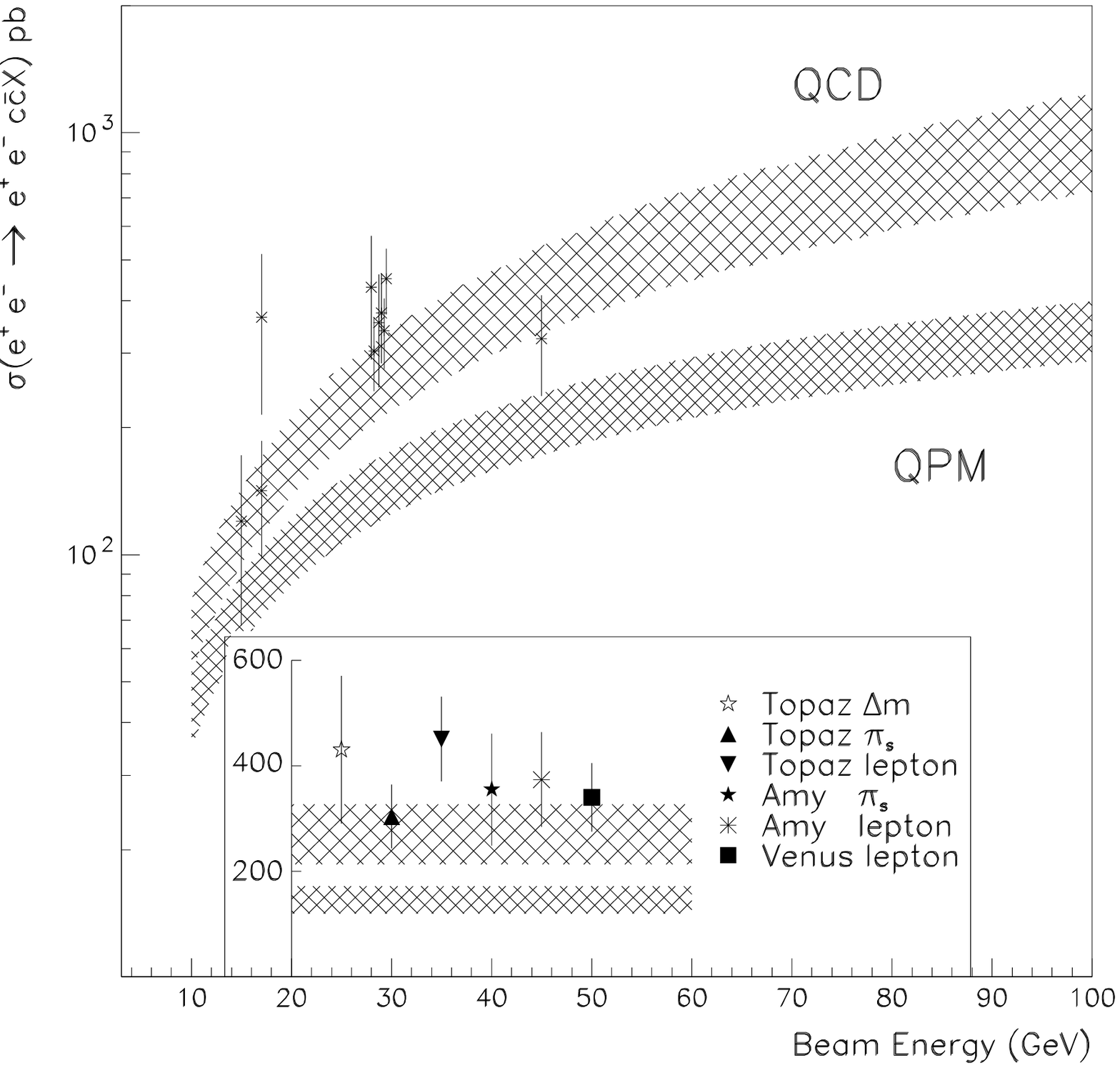,height=8cm} &
\epsfig{file=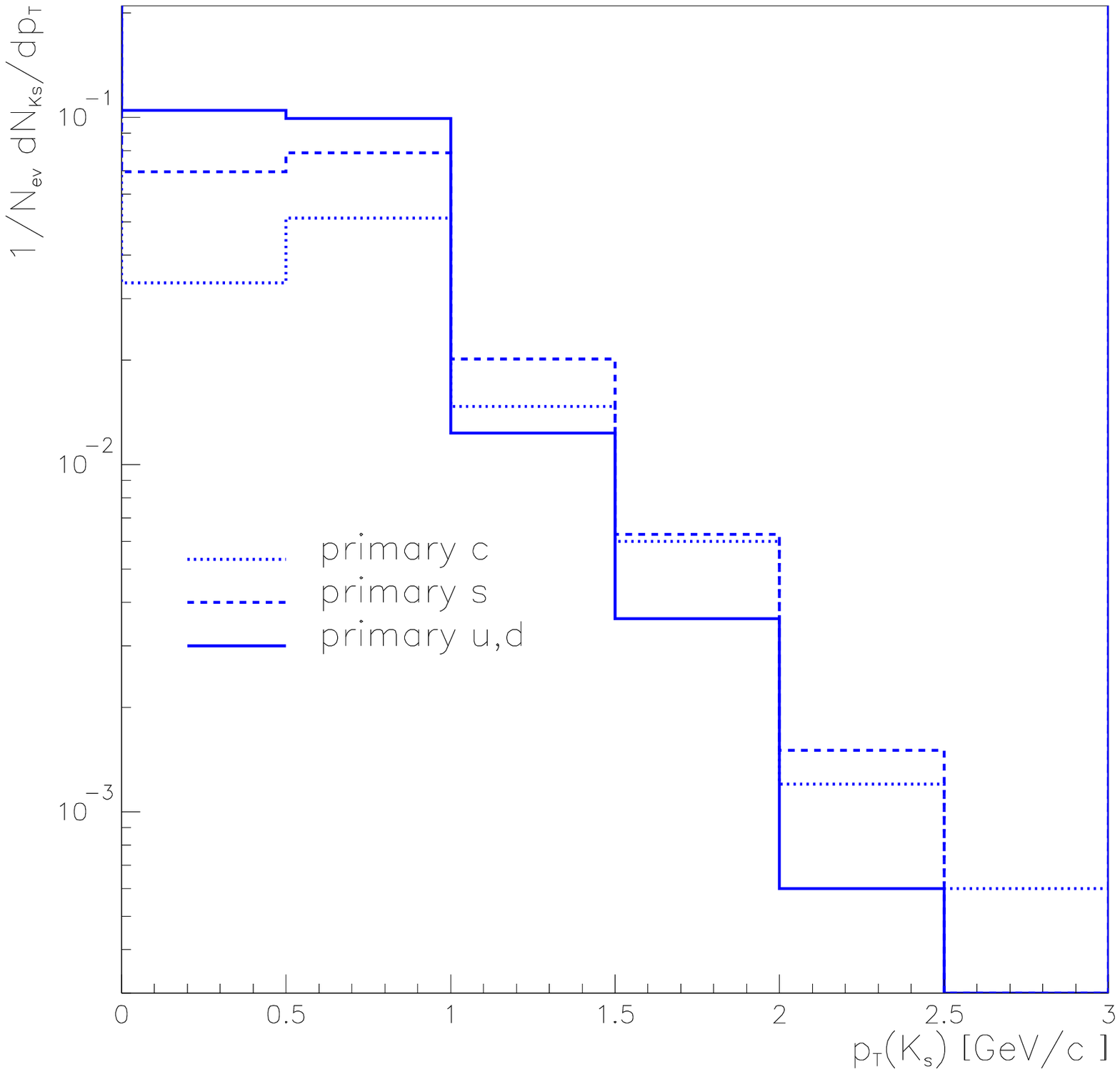,height=8cm}\\
\end{tabular}
\caption[]{\em
Comparison of theoretical predictions and experimental results for the
total charm cross section. The experimental results have been
 extrapolated to a total
cross section from published measurements. Insert shows the results from
Tristan experiments at a beam energy of 29 GeV. Results are tabulated in 
Table \ref{T.results}. The bands represent a range of theoretical predictions.
See table \ref{X-sect} for details.\label{results}}
\caption[]{\em
Distribution of transverse momentum $\protect p_t(K_s)$ of 
$\protect K_s$ simulated with {\sc Pythia} 5.7 
at $E_{\rm CM}=91$ GeV.\label{ks}}
\end{center}
\end{figure}

Experimental measurements of charm production in $\gamma \gamma $ physics 
require some form of tag to identify the presence of the charm quark. There are
a number of different techniques that have been used for this
in the past. They cover a spectrum in which generally the higher the 
selection efficiency of the tag, the larger the problems
due to backgrounds from non charm contributions. These different techniques are
described in the following and are summarised in Table~\ref{methods} and
Fig.~\ref{results}. In order to enable the different experimental results to be
compared we have attempted in this report to extrapolate the published results to a total
charm cross section. Some caution is in order however when comparing theory and
the results of different experiments in this way. This arises from the strong
dependence of the cross section at low p$_t$ on the choice of charm mass and
renormalization scale. As all experimental results are made above some
explicit or implicit p$_t$ cut, extrapolating back to a total cross section
increases the error on the measurement. 
This problem has been treated in
different ways by each experiment, e.g. TOPAZ\cite{topaz,topaz2} chose 
only to quote a cross section in a limited acceptance. Table ~\ref{T.mass}
summarizes the approach of each experiment.

\begin{table}[t]
\begin{center}
\begin{tabular}{|c|l|}
\hline
Experiment & Charm mass (GeV/c$^2$) \\
\hline
JADE\cite{jade}.          & 1.5 \\
TPC/Two-Gamma\cite{tpcgg} & 1.6  \\
TASSO \cite{tasso}         & 1.6 \\
TOPAZ D$^*$\cite{topaz}         & 1.3-1.8, $\mu = m/2$ - $2 m$\\
TOPAZ $\pi_s$\cite{topaz2}        & 1.3; iteratively tuned to the data \\
TOPAZ lepton \cite{Topaz_lepton}  & 1.3-1.5 (1.6 in fragmentation)\\
AMY $\pi_s$\cite{AMY_PHOTON95}    & 1.6, $\mu$ = $\sqrt 2 m$ \\
AMY lepton \cite{AMY_PHOTON95}    & 1.6, $\mu$ = $\sqrt 2 m$  \\
VENUS\cite{Venus}         & Direct $1.6\pm0.2$, Resolved $1.35\pm0.2$  \\
                          & (included in systematic error)          \\
ALEPH\cite{Aleph}         & 1.6  \\
\hline
\end{tabular}
\end{center}
\caption{\em
Treatment of charm mass ($m$)
 in extracting experimental results. $\mu$ is the renormalisation scale;
 unless otherwise stated it is equal to the charm mass.}
\label{T.mass}
\end{table}

\begin{table}[b]
\vspace{-0.5cm}
\begin{center}
\begin{tabular}{|c|l|l|l|}
\hline
Electron Energy & Direct (Born term) & Direct (NLO) & Resolved (NLO) \\
\hline
15 & 75  $\pm$ 14 & 103  $\pm$ 19 & 17  $\pm$  6\\
17 & 87  $\pm$ 16 & 120  $\pm$ 22 & 23  $\pm$  8\\
29 & 146 $\pm$ 26 & 200  $\pm$ 35 & 70  $\pm$  22\\
45 & 206 $\pm$ 34 & 258  $\pm$ 36 & 195 $\pm$ 86 \\
80 & 302 $\pm$ 49 & 416  $\pm$ 74 & 367 $\pm$ 120 \\
90 & 324 $\pm$ 53 & 444  $\pm$ 77 & 440 $\pm$ 152\\
\hline
\end{tabular}
\end{center}
\caption{\em
Cross Section (pb) for charm production in two photon physics based 
on Ref.~{\protect \cite{DKZZ}}. 
The highest values come from using a charm quark mass of $1.3 \,
{\rm GeV/c}^2$ with the scale equal to the charm mass. 
The lowest values use a charm mass of 
$1.7\, {\rm GeV/c}^2$ with the scale equal to $\protect \sqrt{2}$ times the charm mass.
The GRV {\protect \cite{GRVphot}} set of parton densities was used, and a lower cut of 3.8 GeV
was applied for the invariant mass of the $\gamma\gamma$ system.}
\label{X-sect}
\end{table}

\begin{table}[htb]
\begin{center}
\begin{tabular}{|c|c|c|c|}
 \hline
 Method & Experiments & N for  &  
 Comments\\
        &              & 500 pb$^{-1}$ &  
 \\  \hline \hline

 $\protect\displaystyle D^{*\pm}\rightarrow D^0\pi^{\pm}$ 
 & TPC/2$\protect\displaystyle \gamma$, TASSO, & 500  & 
 clean\u, \\
 & JADE, TOPAZ & &    
  no VMD contribution\u,  \\ 
 & AMY, ALEPH &  & $\protect\displaystyle  p_t^{D^*}$ cut\d\\\hline 

 Slow pions
 & TOPAZ, AMY & $\protect\displaystyle {\cal O} (10^3)$   & 
 large background \d\\ 
  & &  &   
  \\ \hline

 $\protect\displaystyle K^0_s\rightarrow \pi^+\pi^-$
 & TOPAZ & $\protect\displaystyle{\cal O}(10^4)$& 
 VMD contribution\d\\
  $\protect\displaystyle (e_c^4:e_s^4=16)$ &   &  &    
 \& fragmentation \d, s-quarks \d\\ \hline 

 Lepton Tags
 & TOPAZ, AMY & $\protect\displaystyle{\cal O} (10^2)$  & 
  background\d,\\
  & VENUS   &  &     fake leptons\d\\ 
    \hline 

 Secondary
 &  & ?  & 
 no $\protect\displaystyle b$ quarks\u \\ 
 Vertices & &  & boost ?  \\ \hline 
\end{tabular}
\end{center}
\caption{\em
Summary of experimental methods for tagging charm in 
$\gamma \gamma$
physics. In column 1 are listed the experiments which have published results in
each category. Column 3 is an estimate of the number of events likely to be
observed at LEP2 assuming similar selection efficiencies to those at current
experiments. Column 3 indicates advantages \u\, and disadvantages \d\, of the
different techniques. }
\label{methods}
\end{table}

We now discuss the various charm tagging techniques in more detail.
The {\it D$^*$ tagging} technique exploits the fact that
the available kinetic energy in the 
decay $\mathrm{D}^{\ast +}\rightarrow\mathrm{D}^{0}\pi^{+}$ is only 6 MeV.
  The signal is typically displayed by plotting 
$\Delta\mathrm{M} = \mathrm{M}_{\mathrm{D}^{\ast +}}-\mathrm{M}_{\mathrm{D}^{0}}$
  for all reconstructed decay product candidates. 
  A $\mathrm{D}^{0}$ decay mode can be used in this analysis if it has a reasonable
  branching ratio (at least of order 1\%).
Published results have included the decays 
$\mathrm{K}^{-}\pi^{+}$,  $\mathrm{K}^{-}\pi^{+}\pi^{0}$, 
 $\mathrm{K}^{-}\pi^{+}\pi^{+}\pi^{-}$.
  Having formed a candidate $\mathrm{D}^{0}$ meson, which is within
  the accepted mass range, tracks identified as pions are added in turn
  to form candidate $\mathrm{D}^{\ast +}$ mesons, $\Delta\mathrm{M}$ being
  determined in each case. 
For background
  tracks the spectrum rises from a kinematic lower limit of 139.6 MeV$/c^2$
  (M$_{\pi^{+}}$), whilst the signal produces a peak at 145.5 MeV$/c^2$, i.e.
  $\mathrm{M}_{\mathrm{D}^{\ast +}}-\mathrm{M}_{\mathrm{D}^{0}}$, which 
is a region where the background is 
  small.

Early  measurements were reported by JADE \cite{jade},
TPC/Two-Gamma\cite{tpcgg}, and TASSO\cite{tasso}. More recently results
have also been produced by TOPAZ\cite{topaz} and ALEPH\cite{Aleph}. 
These results are
summarised in Table~\ref{T.results}.
 The adjusted figure in column 4 takes account of various factors.
For TPC/Two-Gamma\cite{tpcgg}, JADE\cite{jade}, TASSO\cite{tasso}
 and the earlier TOPAZ result\cite{topaz}, the adjustment 
accounts for the latest
values for the $\mathrm{D}^{*+}$ and $\mathrm{D}^0$ branching 
ratios \cite{pdg}.
The published TOPAZ cross section is with the additional 
condition  $p_{\mathrm{t}}^{\mathrm{D}^{*+}} > 1.6 \,\mathrm{GeV}/c$. 
A total cross section was obtained  from the
published figures by multiplying the total QPM cross section by the 
ratio of the observed cross section to the QPM cross section for 
the same  acceptance.

\begin{table}
\begin{center}
\begin{tabular}{|c|c|c|c|c|}
\hline
Experiment & Beam & Published & $\sigma(e^+e^-~\!\rightarrow$ \\
          &  energy (GeV) & Measurement &  $~e^+e^-~c\bar{c} \mathrm{X})$ (pb)\\
\hline
TPC/Two-Gamma\cite{tpcgg} & 14.5 & 74 $\pm$ 32  pb  & 120 $\pm$\ 52\hspace{0.5em} \\
JADE\cite{jade}           & 17   & 20.5 events & 365 $\pm$\ 150  \\
TASSO\cite{tasso}         & 17   & 97  $\pm$ 29 pb  & 142 $\pm$\ 42\hspace{0.5em}   \\
TOPAZ\cite{topaz}         & 29   & 77  $\pm$ 25 pb  & 430 $\pm$\ 140  \\
ALEPH\cite{Aleph}         & 45   & 155 $\pm$ 39 pb  & 326 $\pm$\ 87\hspace{0.5em}  \\
\hline
TOPAZ\cite{topaz2}        & 29   & 23.5 $\pm$ 4.6 pb & 304 $\pm$\ 60\hspace{0.5em}   \\
AMY\cite{AMY_PHOTON95}    & 29   & 169  $\pm$ 48  pb & 355 $\pm$\ 106  \\
\hline
Venus\cite{Venus} & 29 & 68.4 $\pm$ 13 events             & 340 $\pm$ 65\hspace{0.5em}  \\
TOPAZ\cite{Topaz_lepton}       & 29   & 19.3 $\pm$ 3.4 pb & 451 $\pm$ 80\hspace{0.5em}  \\
AMY\cite{AMY_PHOTON95} & 29   & 1.0 $\pm$ 0.23 pb         & 374 $\pm$ 86\hspace{0.5em}  \\
\hline
\end{tabular}
\end{center}
\caption{\em
Measurements of charm production in two photon physics. 
The third column shows  the published measurement, either number of events or cross
section. As these numbers are not directly comparable we have extrapolated them
to a total charm cross section in the fourth column.
The upper 7 measurements are of $\mathrm{D}^{\ast +}$ production, and out 
of these
the first 5 measurement listed employed the `D$^*$ Trick', while 
the measurements in rows 6 and 7 used the `Soft Pion' method. The remaining 3
rows report measurements where lepton tagging was employed.}
\label{T.results}
\end{table}

The {\it soft pion method} is similar to the D$^*$ tagging method
in that it takes advantage of the small
kinetic energy available to the soft pion in the decay of a $\mathrm{D}^*$
 to a 
$\mathrm{D}^0$, but
avoids the reduction of statistics which results in looking for fully
reconstructed $\mathrm{D}^*$.
  It has been found that if one plots the transverse momentum
of all charged tracks in events with respect to the jet direction that there is
a small excess at very low values which is ascribed to these soft pions. This
excess sits on top of a background which is normally estimated by a fit to the
p$_t$ distribution in the nearby bins and by Monte Carlo studies. 
Measurements of this type have been made
by TOPAZ\cite{topaz2}, and AMY\cite{AMY_PHOTON95} and are  summarized  in 
Table~\ref{T.results}.
Note that TOPAZ published their result  for the restricted acceptance 
$p_{\mathrm{t}}^{\mathrm{D}^{\ast+}} > 1.6 \,\mathrm{GeV}/c $, 
$ |\cos(\theta)| < 0.77$.

In the {\it lepton tagging}
technique one uses the fact that there is roughly a 10\% branching
fraction for a charmed meson decay to include electrons or muons. However the
problem is that there are plenty of other sources of leptons in $\gamma\gamma$
events, so a measurement requires good 
modelling of the background. Results have been published by
TOPAZ\cite{Topaz_lepton},
 VENUS\cite{Venus}, and
AMY\cite{AMY_PHOTON95} and are summarised in Table~\ref{T.results}.

Furthermore, {\it kaons} may be used for charm tagging;
due to the quark charge, and neglecting quark masses, 
direct strange quark production is suppressed by a factor 16 
compared to direct charm quark production in $\gamma\gamma$
collisions. Therefore 
a large fraction of the $K_s$ observed come from decays of primary
charm quarks. However, a substantial number of kaons is
also produced from strange quarks in soft VMD process and
from $u,d$ quarks in the fragmentation. 
Modelling such production
introduces systematic errors. The statistical errors, however,
are expected to be very small due to the large number
of reconstructed $K$'s.
In Fig.~\ref{ks} the transverse momentum $p_t$ with respect
to the beam axis is plotted for $K_s$ from
$\gamma\gamma$ collisions reconstructed in
a typical LEP detector for $E_{\rm CM}=91$ GeV.
The events were simulated with {\sc Pythia} 5.7 \cite{jetset}
including the VMD, anomalous and direct photon components.
At $p_t(K_s)>1$ GeV/c the production from primary 
$s$ and $c$ quarks dominates over the production from $u,d$ quarks.
This effect will be even more pronounced at higher energies.
The still strong presence of primary $s$ quarks makes the 
measurement quite dependent on how this contribution is modeled.
The only published  measurement of 
$K_s$ production in $\gamma\gamma$ events 
to date is by TOPAZ~\cite{Topaz_Kaon}.

In principle it might be possible to observe the finite decay length of a
charm quark using techniques such as those used so succesfully 
to tag bottom quarks at LEP1 . These techniques take advantage of the vertex detectors installed on
LEP detectors and include impact parameter, secondary vertex finding and neural
nets. At present none of these techniques has been studied in detail. 
%

\subsection{Prospects for LEP2}

The cross section for charm production in both the direct and single resolved
mode grows with energy. The cross sections at Petra, Tristan, LEP1, and LEP2
energies are shown in Table~\ref{X-sect}.
Using these cross sections with the 500 pb$^{-1}$ promised for LEP2 and
assuming that selection efficiencies will be similar to those at present
experiments produces the estimated events rates given in column 3 of
Table~\ref{methods}. Note that the total number of $c\bar c$ events at LEP2
will be around $5\cdot 10^5$. 
It is clear that at LEP2 there is the prospect for high
statistics measurements of charm production. With these statistics it should be
possible to produce a clean measurement of the resolved and direct processes
separately. 
Recently the direct and resolved contributions to charm production 
were  measured \cite{Topaz_Kaon} by identifying the energy 
from the remnant jet, present in resolved processes, 
seen close to the beam pipe. There is good reason to believe that LEP
experiments will be at least as capable of observing this energy and thus
extend these studies to higher energies where the contribution of resolved
processes is larger.
As mentioned in the theory section there is also particular interest
in measuring charm production in $\gamma^*\gamma$ events, i.e. in events with a
tagged electron. The rates for this will be roughly 5-10\% of the untagged rates
given in Table~\ref{methods}.



\section[Exclusive channels]
{Exclusive channels
\label{ggex1}
{\protect 
\footnote{ A.\ Buijs, E.\ Boudinov, W.\ Da Silva, S.R.~Hou, M.N.\
Kienzle, P.\ Kroll, J.\ Parisi, T.\ van Rhee, G.A.\ Schuler, B.\ Wilkens}}}


The formation of light resonances by two-photon interactions is a powerful 
tool in understanding the hadron spectrum and the dynamics 
controlling the interaction of their constituents. 
A rigorous testing ground 
for nonperturbative QCD is provided by analyses of heavy quarkonia 
for which relativistic corrections and dynamical effects of gluons can be 
included in a systematic way. Similarly, the meson-photon transition 
form-factor at large $Q^2$ and exclusive (meson and/or baryon) pair 
production at large $p_T$ can reliably be calculated in QCD. Last but not 
least, high-energy $\gamma^\star(Q^2) + \gamma \rightarrow 
V_1 +  V_2$ and/or $V_1 + X_2$ reactions, 
where either $Q^2$, the resonance mass or the momentum transfer is
large,  allow us to enter a new domain of perturbative QCD. Prospects
for LEP2 will be discussed in the following.

\subsection{Resonance production by quasi-real photons}
Two-photon couplings provide a useful probe of the internal structure 
of mesons. Two quasi-real photons couple in a selective way to 
$C=+1$ states, thus simplifing the analysis of mass spectra
where many hadrons are superimposed. The accessible resonances with 
spin $\leq 2$ have $J^{PC} = 0^{-+}$ (${}^1S_0$), 
$2^{-+}$ (${}^1D_2$), 
$0^{++}$ (${}^3P_0$), and
$2^{++}$ (${}^3P_2$) where we have given in parentheses 
the $q\bar{q}$ quark-model assignements in the spectroscopic notation. 
In particular the ``classical'' resonances have been observed 
at $e^+e^-$ machines with $\gamma\gamma$ partial widths consistent 
with quark-model predictions \cite{twophotconf}. 
These are the light pseudoscalar 
$0^{-+}$ and tensor $2^{++}$ states $\pi^0$, $\eta$, $\eta'$, 
$f_2(1270)$, $a_2(1320)$, $f_2'(1525)$,  
and, although with poor statistics, the $c\bar{c}$ states 
$\eta_c$, $\chi_{c0}$, and $\chi_{c2}$. 

Particularly interesting are, of course, those resonances whose 
$\gamma\gamma$ couplings are not consistent with quark-model
estimates. On the one hand, conventional $\gamma\gamma$-width calculations 
might not be reliable enough. As an example let us mention 
the $\pi_2(1670)$, thought to be the ${}^1D_2$ $(u\bar{u}
 - d\bar{d})/\sqrt{2}$ quarkonium state. 
Here the discrepancy between the measured $\gamma\gamma$ 
partial width and the non-relativistic calculation \cite{and:91}
could be due to large relativistic corrections \cite{Barnes1}.
On the other hand, mesons thought to be 
non-$q\bar{q}$ states generally have $\gamma\gamma$ widths far from 
expectations for a $q\bar{q}$ state. These anomalous states include the 
$f_0(1500)$, $f_{0/2}(1720)$, $f_J(2230)$, the $\eta(1410)$, $\eta(1460)$,
and the $f_1(1420)$ (for a recent review see e.g.\ \cite{Nils}). 
Measurements of their $\gamma\gamma$ widths have the potential to 
resolve the enigma of these mesons. 

Exploring the exclusive channels at LEP2 has 
pros and cons compared to lower-energy machines. 
There are two advantages.
\begin{table}[htb]
\begin{center}
\begin{tabular}{|c|r|r|r|r|r|} 
\hline
Experiment&$\sqrt{s}$ & $\int L$dt   & $\sigma$ & A&$\sigma \cdot$A
\\
 &(GeV) & pb$^{-1}$  &(pb) & & (pb)
\\ \hline
 CLEO II \cite{CLEO1}& 11 & 3000  & 17 &0.54 & 9.2 
\\ \hline 
 TPC/2$\gamma$ \cite{TPC1} & 29 & 69 & 48 & 0.35& 16.8 \\
   PLUTO ~\cite{PLUTO}& 35 & 45 & 56 & 0.33& 18.5 
\\ \hline
  L3 \cite{L3etac}& 91 & 30  & 104&0.25 & 26.0 
\\ \hline
  LEP2 & 175 & 500  & 147 &0.21 & 30.2 
\\ \hline
\end{tabular}
\caption{\it Comparison of existing data samples
for $\sigma (\epem\ra\epem\eta_{c})$ 
with estimates 
for LEP2. The geometrical acceptance $A$ is defined in the text.
The fourth column gives the full resonance production cross-section  
calculated with exact kinematics and a $J/\psi$ form factor.}
 \label{tab:expts}
 \end{center}
 \end{table}
First, the signal cross section 
$\sigma (\epem\ra\epem\mbox{R})$ 
rises\footnote{In the Low approximation, the two-photon cross-section 
rises as $\ln^3{s}$ for a $4\pi$ detector and as 
$\ln^2{s}$ for a realistic, limited angular acceptance detector 
\cite{ex:Courau}; compare with Table~\ref{tab:expts}.}
with the $e^+e^-$ centre of mass (cm) energy $\sqrt{s}$, 
while the background  from the $s$-channel
annihilation reaction decreases as $\approx 1/s$. Second, since the 
energy released in annihilation events rises with $\sqrt{s}$, 
the two contributions become more separated at higher energy \cite{x2}.

There are also two disadvantages compared to experiments at lower
energies. First, the detector acceptance is reduced since 
the photon-photon system receives on average a larger Lorentz boost. 
Table~\ref{tab:expts} displays the acceptance as a function of 
$\sqrt{s}$ for the case of $\eta_c$ production 
\cite{PLUTO,TPC1,TASSO1,CLEO1,L3etac,ARGUS1}
in the decay channel $\eta_c \rightarrow 4\pi$. 
All LEP detectors have a good solid angle coverage. For our studies 
we \cite{L3note} use the following geometrical acceptance:
$20^{\circ}\leq \theta \leq 160^{\circ}$ for tracks
with a  $p_{t} \geq 0.1\,$GeV 
and  $15^{\circ}\leq \theta \leq 165^{\circ}$ for photons
with  E $\geq 0.1\,$GeV. 
As can be seen from Table~\ref{tab:expts}
the acceptance slowly decreases with the centre of mass energy. 
Due to the increase in the cross section, 
we find still a net gain 
on the number of observed 
events.
For lighter resonances the acceptance decreases, 
from $\sim 20$\% at $3\,$GeV to $\sim 4$\% at $1.3\,$GeV.  

The second disadvantage concerns the triggers. Their efficiencies are
more difficult to evaluate.
Since the interest of the LEP experiments is centered on the maximum
available energy, the two-photon events are mainly seen by triggers
based on tracks, thus excluding the observation of purely neutral decays of
resonances. The combined effect of the trigger and analysis cuts reduces
the efficiency by factors varying from about two at a mass of $3\,$GeV to
about 15 at $1.3\,$GeV. 
\begin{table}[htb]
\begin{center}
%
%
\begin{tabular}{|c|c|c|c|c|c|c|c|} 
\hline   
Resonance & decay & cross sections& A &$\epsilon_{untag}$&Events
 & $\epsilon_{tag}$   &Events \\ 
  & & (pb) & & & untagged& & tagged
\\ \hline
$\pi ^0$ & & no trigger   & & & & &  \\
$\eta $& $\pi ^{+} \pi ^{-} \pi ^{0} $& no acceptance& & & & &\\ 
$\eta ^{'} $& $\pi ^{+} \pi ^{-} \gamma $&3381$\pm$23&0.133 
  &0.0188 &9534&0.0016&811\\
 &$\pi ^{+} \pi ^{-} \eta$ & &0.025 & $\le 10^{-4}$& & & 
\\ \hline 
$f_2 $& $\pi ^{+} \pi ^{-}$ &4463 $\pm$ 44  &0.264 &0.125 & 159000
  &0.003 &3815 \\
$a_2 $& $\pi ^{+} \pi ^{-} \pi ^{0}$&1436 $\pm$ 5.5 & 0.0416
& 0.0033& 1658&0.0006  &301 \\
$f_{2}^{'} $& $K_{s}K_{s} $&82.04 $\pm$ 0.37 & 0.0935&0.0564 
& 190&0.0011 & 4
\\ \hline
\end{tabular}
\caption[]{\it Examples of low-mass resonances at LEP2  
($e^{+}e^{-} \rightarrow e^{+}e^{-} R$ at $\sqrt{s}=175\,$GeV 
for ${\cal L}=500\,$pb$^{-1}$).
A is the geometrical acceptance as described in the text and 
$\epsilon$ the efficiency
including trigger and analysis cuts: $\epsilon_{untag}$ for untagged 
and $\epsilon_{tag}$ for tagged events ($0.2 \leq Q^{2} \leq 0.8\,$GeV$^{2}$,
 and $ Q^{2} \geq 7\,$GeV$^{2}$). The third column gives the full 
resonance production cross-section, while the number of events is 
calculated taking into account the branching ratio.
 \label{tab:resex}}
 \end{center}
 \end{table}
As examples we show in Table~\ref{tab:resex} 
the expectations for the lightest $0^{-+}$ and $2^{++}$ states. With current 
triggers, only the $\eta^{'}$ can be studied in
the pseudoscaler octet. The expected rates for the tensor octet
are rather good, an analysis of these data has already started at
LEP1 \cite{L3a2}.

Let us emphasize the generally very low efficiencies, 
for example, the $a_2$ in Table~\ref{tab:resex}. Its trigger 
could easily be improved to reach the acceptance limit \cite{L3note}. 
We conclude that, 
if specialized triggers are installed, good results on 
light resonance physics can be achieved except for purely neutral decays. 
Hence even searches for glueballs might be within reach. 
Since their two-photon widths are expected to be at least one order 
of magnitude below that of normal $ q \bar q $ states
\cite{Goun2,Kada}, searches 
for associated glueball ($G$) production, 
$ \gamma \gamma \rightarrow \pi^0 G$ are welcome: 
of the order of 10 to 100 events should be produced 
above $p_T = 1\,$GeV at $500\,$pb$^{-1}$ \cite{Ichola}.

\subsection{Resonance production with one off-shell photon}
Resonances can also be studied in 
two-photon events in which one 
photon is far off the mass shell. 
Usually, this is the domain of single-tag events where 
the virtuality $Q^2$ is determined from a measurement 
of the scattered electron. The $Q^2$ range can be extended by 
including no-tag events, in which case $Q^2$ is reconstructed 
by the measurement of the $p_T$ of the resonance. 
In particular one may cover $Q^2$ ranges where 
the $Q^2$ determination from the electron is not possible (i.e.\ 
from about $0.8\,$GeV$^2$ to $7\,$GeV$^2$ at LEP2).
Note that resonance production in single-tag events 
is just the exclusive limit of the photon structure 
function (i.e.\ $e\gamma \rightarrow e M$). The interest here is twofold.
First, the meson--photon transition form factor can be measured, 
and secondly, spin-$1$ states can be produced (the Landau--Yang theorem 
forbids their production from two on-shell photons). 
Measurements of the pseudoscalar--photon transition form factors 
are now becoming quite precise \cite{sav:95}. 
Also the $1^{++}$ (${}^3P_1\ q\bar{q}$) state $f_1(1285)$ has been observed 
\cite{spin1}
with a $\gamma\gamma$ coupling consistent with quark-model
estimates \cite{Cahn} for $Q^2 \neq 0$ photons. 

\begin{figure}[htb]
\[
    \epsfig{figure=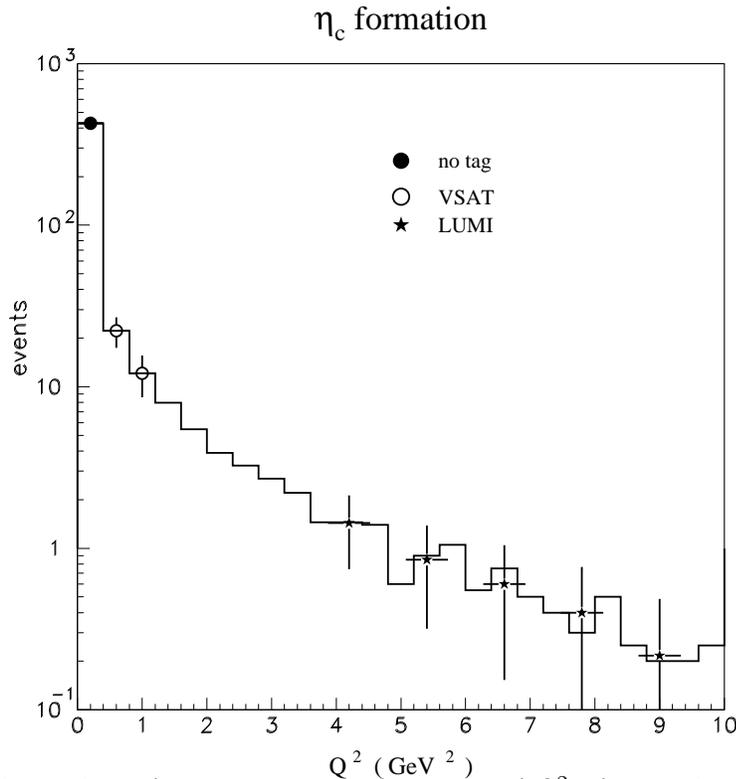,%
        width=10cm}
\]
\vspace*{-1.5cm}
\caption[dummy2]{\it 
Expected number of $\eta_{c}$  events as a function
of $Q^2$, 
for an integrated
 luminosity of 500  pb$^{-1}$. All observable decay channels are included.}
\label{fig:etac}
\end{figure}

The LEP experiments can cover a large
$Q^2$ range with various tagging systems, $e.g.$ for L3:
VSAT (very small angle tagger) ($0.2 \leq Q^{2} \leq 0.8\,$GeV$^{2}$),
LUMI (luminosity monitor ) ($ Q^{2} \geq 7\,$GeV$^{2}$), and 
ECAL (the endcap electromagnetic calorimeter)
($ Q^{2} \ge 40\,$GeV$^{2}$). Given the high cm energy, high $Q^2$ values 
should be reachable. We estimate that form-factor measurements will,
even with current triggers, be possible for at least the $\eta'$,  
the $f_2$, and the $\eta_c$, see Table~\ref{tab:resex} and
Fig.~\ref{fig:etac}.
\begin{figure}[htb]
\centerline{
\epsfig{file=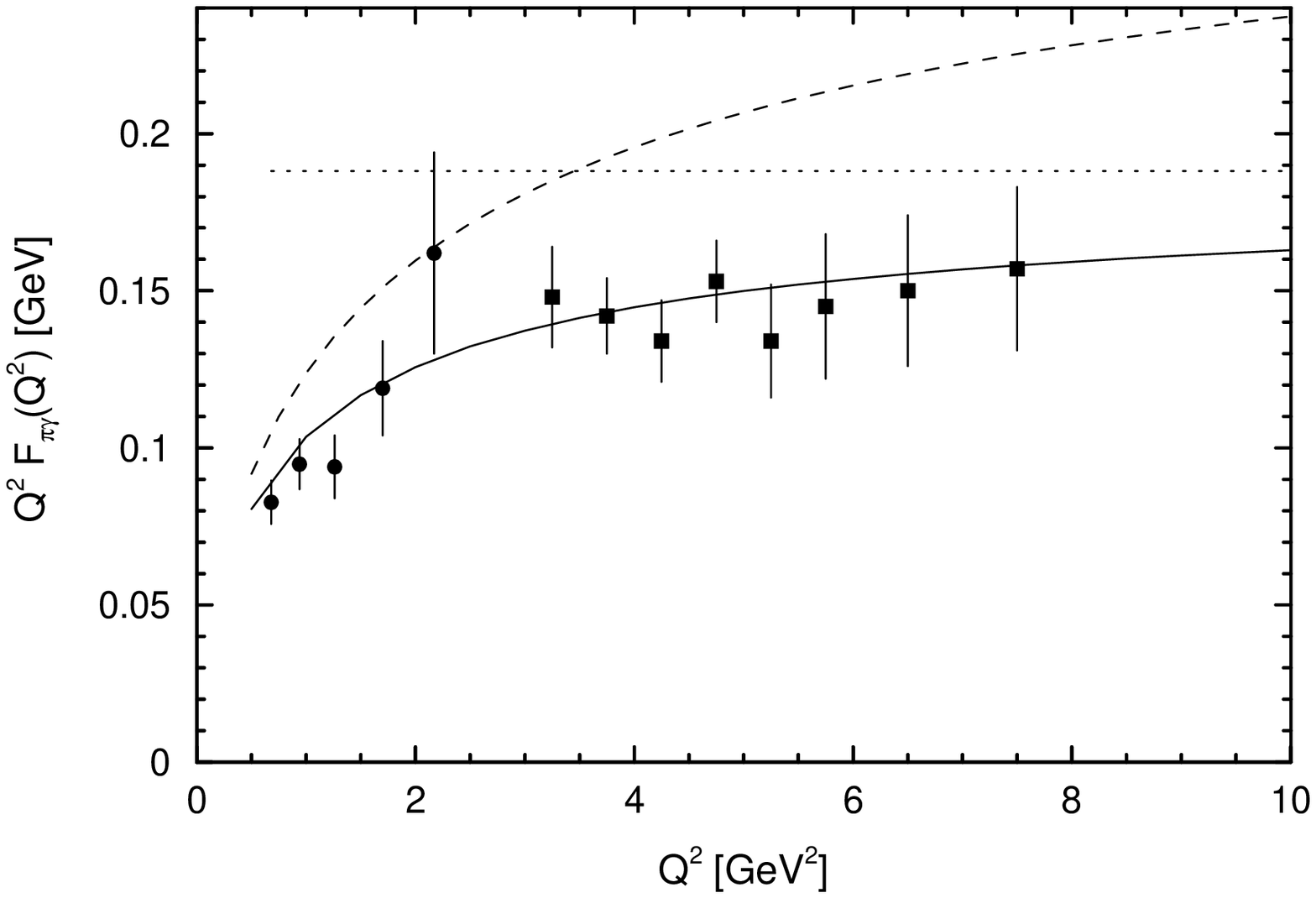,width=0.48\textwidth}
\hfill\epsfig{file=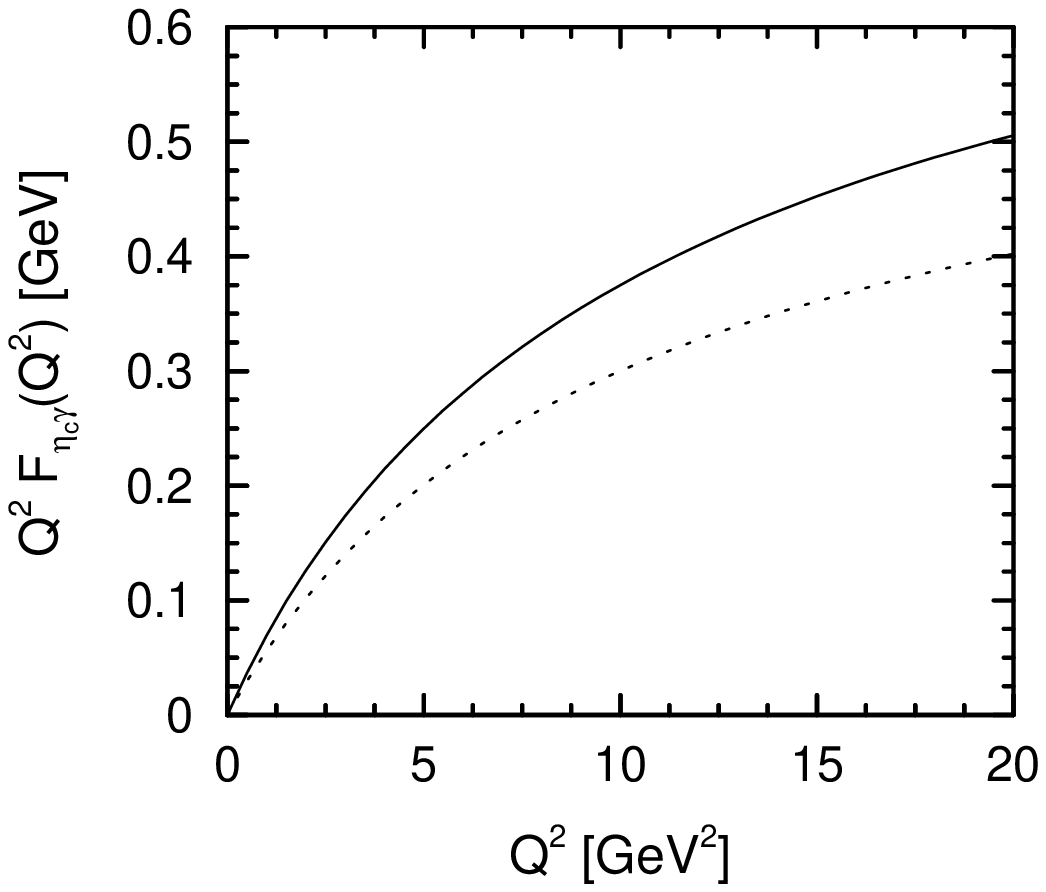,width=0.48\textwidth}}
\caption[dummy1]{\it Left: The $\pi\gamma$ transition form factor vs.~$Q^2$.
The solid (dashed) line represents the prediction obtained with the
modified HSA using the asymptotic (CZ) wave function \cite{jak:94}. 
The dotted line
represents the results of the standard HSA for the asymptotic wave function.
Data are taken from \cite{beh:91},\cite{sav:95}.   
Right: The $\eta_c\gamma$ transition form factor calculated 
assuming for the $\eta_c$ decay constant
$383\,$MeV (solid line) and $284\,$MeV (dotted line).}
\label{fig:pigaff}
\end{figure}
Concerning the spin-$1$ mesons, 
the $f_1$ and $\chi_{c1}$ will certainly be observed. Otherwise 
the same remark as above applies: with more dedicated triggers more 
resonances will become accessible.  

Theoretical predictions for the $\eta_c$-photon transition form factor 
$F_{\eta_c \gamma}(Q^2)$ \cite{kro:95}
are shown in Fig.~\ref{fig:pigaff}, as well as a similar 
calculation \cite{jak:94}
for $F_{\pi\gamma}(Q^2)$ compared to recent data \cite{beh:91,sav:95}. 
For low $Q^2$, models based on vector-meson-dominance (VMD), 
constituent quarks, QCD sum rules or chiral perturbation
theory work, in general, successfully for pseudoscalar mesons (P), 
see e.g.\ ref.~\cite{ame:92}. The $Q^2$ dependence can be parametrized by
\begin{equation}
  F_{P\gamma}(Q^2) = \frac{A_P}{1 + Q^2 / \Lambda_P^2}
\ .
\label{Q2depe}
\end{equation}
For example, in VMD $A_P$ is related to the $V\pi\gamma$ and 
$Ve^+e^-$ coupling constants (here $V$ is the corresponding vector 
state(s), i.e.\ $\rho$ and $\omega$ for the $\pi^0$, and $J/\psi$ for
the $\eta_c$), and $\Lambda_P$ is given by the vector-meson mass.

Form factors are particularly interesting at high $Q^2$ since 
a factorization formula holds \cite{lep:80}, which expresses large-$Q^2$ 
exclusive reactions as a product of process-independent meson distribution 
amplitudes and perturbatively calculable short-distance coefficients. 
In fact, asymptotically, i.e.\ for $\ln Q^2/\mu_0^2 \rightarrow \infty$ 
($\mu_0$ a typical 
hadronic scale $\sim 0.5$--$1\,$GeV), the distribution amplitudes 
are known and one derives the parameter-free result
\cite{lep:80,wal:74}
\begin{equation}
  F_{P\gamma}(Q^2) \rightarrow \frac{\sqrt{2} f_P}{Q^2}
\label{Q2asympto}
\end{equation}
where $f_P$ is the meson's decay constant (i.e.\ $130.7\,$MeV for the pion). 
A simple all-$Q^2$ formula is arrived at 
\cite{lep:80} by assuming (\ref{Q2depe}) and fixing the two parameters
from (\ref{Q2asympto}) ($\Lambda_P=2\pi f_{P}$) 
and the PCAC value of the form factor at 
$Q^2=0$ ($A_P=1/(2\sqrt{2}\pi^2 f_{P})$). 

For finite $Q^2$, a comparison of the full calculation and data allows the 
determination of the distribution amplitude. 
The calculations shown in Fig.~\ref{fig:pigaff} are based on a modified 
hard-scattering-approach (HSA) to exclusive reactions, in which
transverse degrees of freedom, representing higher twist effects, and
Sudakov suppression are taken into account. 
The pion data nicely agree with the calculation if one uses
a wave function $\sim \exp [ -a^2 k_{\perp}^2/x(1-x) ]$, which 
leads, after $k_{\perp}$-integration, 
to the asymptotic distribution amplitude $\sim x(1-x)$ 
($x$ is the momentum fraction carried by the quark inside the pion).
In contrast, the use of a wave function implying the 
Chernyak-Zhitnitsky distribution amplitude \cite{che:82},
leads to results in severe conflict with the data, see Fig.~\ref{fig:pigaff}. 
That observation may have far-reaching consequences for our understanding
of other large momentum transfer exclusive reactions, as for instance
$\gamma\gamma\to \pi\pi$.
 
The $\eta_c$-photon transition form factor has not yet been measured. 
The predictions shown in Fig.~\ref{fig:pigaff} are obtained using  
the Bauer-Stech-Wirbel wave function \cite{bau:85}
\begin{equation}
\Psi_0(x,{\bf k_\perp})=N x(1-x)\exp \left[ -a^2M_{\eta_c}^2\,(x-1/2)^2 \right]
                      \exp \left[ -a^2 k^2_\perp \right]
\ .
\label{WSBwave}
\end{equation}
Its two parameters, namely the transverse
size parameter $a$ and the normalization constant $N$ cannot be
fully fixed from the $\eta_c\to\gamma\gamma$ decay width. To illustrate 
the parameter dependence we have taken 
the valence quark Fock state probability to be $0.8$ and
required a value of either $383$  MeV or $284$ MeV
for the $\eta_c$ decay constant. Referring to Fig.~\ref{fig:etac},
a measurement of $F_{\eta_c \gamma}(Q^2)$ up to $Q^2 \sim 10\,$GeV$^2$ 
seems feasible.

\subsection{Charmonium}
The heavy quark systems can, to first approximation, 
be described by nonrelativistic quark-potential models 
and many of their properties are expected to 
reflect the underlying dynamics of QCD.
%
For example (see e.g.\ \cite{x1}), 
\begin{equation}
 \Gamma(\eta_c \rightarrow \gamma \gamma)  =  \frac{64 \pi \alpha^2}{27 M_c^2}
|\Psi_{\eta_c}(0)|^2 \left [1-\frac{3.4}{\pi} \alpha_s (M_c)\right]   
\ .
\end{equation}
Here $\Psi_{\eta_c}(0)$, the non relativistic wave function at 
the origin, contains all non-perturbative QCD effects. 
Replacing the two photons by two gluons gives 
$\Gamma(\eta_c \rightarrow gg)$ which, in the potential model, 
is the total width $\Gamma_{\eta_c}$ of the $\eta_c$. 
It turns out \cite{x4} that the value of $\alpha_s(M_c)$ determined 
from the relation, 
 \begin{equation}
\frac{\Gamma_{\eta_c}}{\Gamma(\eta_c \rightarrow \gamma \gamma)} =
\frac{9 \alpha_s(M_c)^2}{8 \alpha^2}
 \left [1+\frac{8.2}{\pi}\alpha_s(M_c) \right] 
\end{equation}
is about 3 standard deviations below the value expected from other 
QCD measurements: $\alpha_s(M_c) = 0.33 \pm 0.02$ ~\cite{x2}.  
This indicates that corrections to the non-relativistic quark-potential 
model are non-negligible. 
Recently, a rigorous factorization of heavy quarkonium decays has 
been developed \cite{Bodwin,MS:94}. 
Both relativistic corrections and effects of dynamical 
gluons (those associated with the binding) can be calculated in
a systematic way. 
Phenomenological studies \cite{x4,GS:93,x7} show
results that depend crucially on the input data. 

These problems can be solved by
high precision LEP measurements of the two-photon
widths $\Gamma_{ \gamma \gamma }$ for charmonium states.
In Table~\ref{tab:charm} we estimate the expected number of events
for the charmonium states decaying into 4$\pi$.
For tagged events, the $Q^2$ dependence is modelled by a 
$J/\psi$ form-factor.
\begin{table} [ht]
\begin{center}
\begin{tabular}{|c|r|c|c|l|c|c|} 
\hline    
Resonance& cross sections& A & $\epsilon_{untag}$&Events
& $\epsilon_{tag}$ &Events 
\\ 
  & (pb) &  & &untagged& & tagged 
\\  \hline
$\eta_{c}$   & 146.8$\pm$1.5~    & 0.21 & 0.11 & 97 &0.01 &9. \\
$\chi_{c0}$  & 25.9$\pm$0.3~   & 0.22 & 0.12 & 57 &0.01 & 6.  \\
$\chi_{c2}$  & 18.2$\pm$0.11   & 0.22 & 0.12 & 17.5 &0.01 & 2.   
\\ \hline
\end{tabular}
\caption[]{\it Example of charmonium production expected at LEP2; 
given is the channel $e^{+}e^{-} \rightarrow 
e^{+}e^{-} R$, $ R \rightarrow 4 \pi$
for ${\protect\cal L}=500\,$pb${}^{-1}$ at $\protect\sqrt{s} = 175\,$GeV.
A and $\epsilon$ as in Table~\ref{tab:resex}.}
\label{tab:charm}
 \end{center}
 \end{table}
Since the charmonia have very small branching ratios into $4\pi$ 
\cite{pdg}, the number of events can be increased substantially 
(factors $5$--$6$) by measuring also other decay channels 
\cite{L3etac,L3note}.
Hence good measurements of the charmonium states, including the
$\chi_{c1}$ are expected.

\subsection{Exclusive pair production}
Two-photon exclusive meson and baryon production provides particularly 
clean tests of QCD. At large angles, a factorization holds 
\cite{lep:80} which 
tells us that the exclusive scattering amplitude is given as the product 
of a hard scattering amplitude and soft distribution amplitudes
$\Phi_i(x,p_T^2)$. 
Although the latter are nonperturbative quantities, they are subjected
to several constraints (evolution equation in $p_T$, asymptotic ($p_T 
\rightarrow \infty$) form, normalization in the case of mesons). 
This leads to a parameter-independent prediction for the 
fall-off with $\sgaga$ at fixed angle ($d\sigma/d t \propto 
\sgaga^{-4}$, $\sgaga^{-6}$ for meson and baryons, respectively) and 
essentially parameter-independent predictions for the 
angular distribution at moderate $\sgaga$ (see e.g.\ \cite{Aachen}).
Current experiments reach $\sqrt{\sgaga}$ values up to about $3\,$GeV, 
where a transition from the VMD-like angular distribution to the
one predicted by the HSA just becomes visible. Production of
pseudoscalar- and vector-meson pairs at LEP2 has been estimated
during the last workshop: sufficient counting rates can be expected 
to test the predictions up to $\sqrt{\sgaga} \approx $5--$6\,$GeV 
\cite{Aachen}. 

Calculations within the HSA have been extended in various ways. 
On the one hand, the potentially dangerous endpoint regions 
in the $x$ integration have been shown to be suppressed by Sudakov 
form factors \cite{Li:92}. On the other hand, predictions now exist  
for the pair production of baryons \cite{Cher:89},
heavy mesons ($D$, $D^\star$) \cite{Kiselev:95}, 
and (light) mesons with non-zero orbital angular momenta \cite{Ichola:94}. 
Among the $L > 0$ mesons, $a_2^+\, a_2^-$ might have 
a rate large enough to be observable at LEP2, namely about 
$500$ ($50$) events for a $p_T$-cut of $1\,$GeV ($2\,$GeV) 
(without taking into account experimental cuts).

The standard (or possibly improved) HSA gives the leading-twist 
cross section at fixed cm scattering angle or, equivalently, at fixed
$t/\sgaga$. At LEP2, a new domain of perturbative QCD may be entered. 
This is the region $\mu_0^2 \ll | t | \ll \sgaga$ (so-called
semi-hard region) which is discussed in sec. \ref{ggso1} where
numerical estimates are presented.


\subsection{Summary}
In summary, many light resonances can be studied with good counting rates 
at LEP2, provided the triggers of the experiments will be extended 
to low-momentum charged particles. The trigger efficiency 
and acceptance generally increases 
with the resonance mass. Good results can be expected for 
the four lowest-lying $C=+1$ charmonium states. Bottomonium resonances, 
however, are beyond the reach of LEP2. Measurements of spin-$1$ 
mesons (notably the $f_1$ and the $\chi_{c1}$) and various meson--photon 
transition form factors will be possible both in single-tag events 
and in untagged events through $Q^2$ reconstruction from the hadronic 
final state. Measurements of conventional pair production of $S$-wave mesons 
should be possible up to $5$--$6\,$GeV CM energy, allowing for definite
tests of the HSA. Effects of QCD in a new perturbative domain are
expected to become accessible in semi-hard reactions.

\vfill
\noindent
{\large \it Acknowledgements:} We warmly thank J.\ Butterworth (ZEUS), 
M.\ Erdmann (H1), and H.~Hayashii (TOPAZ) for interesting 
discussions about recent HERA and TRISTAN results.

\end{document}